\documentclass[fullpage,a4paper,10pt, sort&compress]{elsarticle}
\journal{Information Systems}

\usepackage{amsthm}

\def\url#1{#1}{}%

\usepackage[utf8x]{inputenc}
\usepackage{graphicx}
\usepackage{multirow}
\usepackage{subfigure}
\usepackage{color}
\usepackage{enumitem}

\theoremstyle{definition}
\newtheorem{example}{Example}[section]

\newenvironment{code}
{ % \scriptsize
    \fontsize{8}{10}\selectfont
  \begin{tabbing}
\rule{\textwidth}{0.25mm} \\
xxxx\=xxxx\=xxxx\=xxxx\=xxxx\=xxxx\=xxxx\=xxxx \kill}
{\rule{\textwidth}{0.25mm}
\end{tabbing}}

\newcommand{\csa}{\mathsf{CSA}}
\newcommand{\icsa}{\mathsf{iCSA}}
\newcommand{\wm}{\mathsf{WM}}
\newcommand{\wt}{\mathsf{WT}}
\newcommand{\wtht}{\mathsf{WTHT}}

\newcommand{\tcsa}{\mathsf{CTR}}    %%UNIFICAR AL FINAL
\newcommand{\repres}{\mathsf{CTR}}   %%UNIFICAR AL FINAL
\newcommand{\represName}{{\em Compact Trip Representation}}   %%UNIFICAR AL FINAL

\newcommand{\Sswx}{\texttt{starts-with-x}}
\newcommand{\Sewx}{\texttt{ends-with-x}}
\newcommand{\Sfxty}{\texttt{from-x-to-y}}
\newcommand{\Sux}{\texttt{uses-x}}
\newcommand{\Stk}{\texttt{top-k}}
\newcommand{\Stkseq}{\texttt{top-k-seq}}
\newcommand{\Stkbin}{\texttt{top-k-bin}}
\newcommand{\Stdiez}{\texttt{top-10}}
\newcommand{\Stcien}{\texttt{top-100}}
\newcommand{\Stdiezseq}{\texttt{top-10-seq}}
\newcommand{\Stcienseq}{\texttt{top-100-seq}}

\newcommand{\Stks}{\texttt{top-k-starts}}

\newcommand{\Stksseq}{\texttt{top-k-starts-seq}}

\newcommand{\Tswx}{\texttt{starts-with-x}}
\newcommand{\Tewx}{\texttt{ends-with-x}}
\newcommand{\Tux}{\texttt{uses-x}}
\newcommand{\Tfxty}{\texttt{from-x-to-y}}
\newcommand{\Tfxtys}{\texttt{from-x-to-y-strong}}
\newcommand{\Tfxtyw}{\texttt{from-x-to-y-weak}}

\newcommand{\STtk}{\texttt{top-k}}
\newcommand{\STtks}{\texttt{top-k-starts}}

\newcommand{\Tst}{\texttt{starts-t}}
\newcommand{\Tut}{\texttt{uses-t}}
\newcommand{\Ttt}{\texttt{trips-t}}

\newcommand{\Tet}{\texttt{ends-t}}

\begin{document}
\begin{frontmatter}		
	
\title{A Compact Representation for Trips over Networks built on self-indexes\tnoteref{t1}}

\tnotetext[t1]{\footnotesize 	
	Funded in part by European Union's Horizon 2020 research and innovation programme under the Marie Sklodowska-Curie grant agreement No 690941 (project BIRDS). 
	The Spanish group is also partially funded by Xunta de Galicia/FEDER-UE [CSI: ED431G/01 and GRC: ED431C 2017/58]; by MINECO-AEI/FEDER-UE [Datos 4.0: TIN2016-78011-C4-1-R; Velocity: TIN2016-77158-C4-3-R; and ETOME-RDFD3: TIN2015-69951-R]; and by MINECO-CDTI/FEDER-UE [CIEN: LPS-BIGGER IDI-20141259 and INNTERCONECTA: uForest ITC-20161074].
	M. A. Rodr\'iguez is partially funded by Fondecyt [1170497] and and by the Millennium Institute for Foundational Research on Data. 	
}

\tnotetext[]{\footnotesize An early partial version of this article appeared in {\em Proc SPIRE'16} \cite{bfgrspire16}. 
}

	\author[udc]{Nieves R. Brisaboa}
	\ead{brisaboa@udc.es}
	
	\author[udc]{Antonio Fari\~na\corref{cor1}}
	\ead{fari@udc.es}

	\author[udc]{Daniil Galaktionov}
	\ead{d.galaktionov@udc.es}
	
	%\author[udec]{M. Andrea Rodr\'iguez}
	\author[udec]{M. Andrea Rodriguez}
	\ead{andrea@udec.cl}

	\cortext[cor1]{Corresponding author}

\address[udc]{University of A Coru\~na, Database Laboratory, Spain.\\}
\address[udec]{University of Concepci\'on, Department of Computer Science, Chile. \\ Millennium Institute for Foundational Research on Data, Chile}

%\authorrunning{Brisaboa et al.}
%\authorrunning{N. Brisaboa, A. Fari\~na, D. Galaktionov, and A. Rodr\'iguez}

%\institute{University of A Coru\~na, Spain \and % University of Chile \and
%    University of Concepci\'on, Chile}

%\maketitle

%%%%%%%%%%%%%%%%%%%%%%%%%%%%%%%%%%%%%%%%%%%%%%%%%%%%%%%%%%%%%%%%%%%%%%%%%%%%%%%%%%%%%%%

%%%%%%%%%%%%%%%%%%%%%%%%%%%%%%%%%%%%%%%%%%%%%%%%%%%%%%%%%%%%%%%%%%%%%%
%%%%%%%%%%%%%%%%%%%%%%%%%%%%%%%%%%%%%%%%%%%%%%%%%%%%%%%%%%%%%%%%%%%%%%
\begin{abstract}

Representing the movements of objects (trips) over a network  in 
a compact way while retaining the capability of exploiting such data
effectively is an important challenge of  real applications. 
We present a new {\em Compact Trip Representation} ($\repres$) that 
handles the spatio-temporal data associated with users' trips over transportation networks. Depending on the network and types of queries, nodes in the network can represent intersections, stops, or even street segments.

$\repres$ represents separately  sequences of 
nodes and the time instants when users traverse these nodes. 
The spatial component is handled with a data structure based on the well-known
Compressed Suffix Array (CSA), which provides both a compact
representation and interesting indexing capabilities. The temporal component is
self-indexed with either a Hu-Tucker-shaped Wavelet-tree  or a 
Wavelet Matrix  that solve range-interval queries efficiently.  We show how $\repres$\ can solve relevant  counting-based spatial, temporal, and spatio-temporal queries over large sets of trips.
Experimental results show the space requirements (around 50-70\% of the space
needed by a compact non-indexed baseline) and query
efficiency (most queries are solved in the range of $1$-$1000$ microseconds) of $\repres$.

%We present a new {\em Compact Trip Representation} ($\repres$) that
%allows us to manage users' trips (moving objects) over networks.
%These could be public transportation networks (buses, subway,
%trains, and so on) where nodes are stations or stops, or road
%networks where nodes are intersections. $\repres$ represents the
%sequences of nodes and time instants in users' trips. The spatial
%component is handled with a data structure based on the well-known
%Compressed Suffix Array ($\csa$), which provides both a compact
%representation and interesting indexing capabilities. We also
%represent the temporal component of the trips, that is, the time
%instants when users visit nodes in their trips. We create a sequence with these
%time instants, which are then self-indexed with a balanced Wavelet Matrix ($\wm$). This
%gives us the ability to solve range-interval queries efficiently. We
%show how $\repres$\ can solve relevant spatial and spatio-temporal
%queries over large sets of trajectories. Finally, we also provide
%experimental results to show the space requirements and query
%efficiency of $\repres$.

\end{abstract}


\end{frontmatter}

%%%%%%%%%%%%%%%%%%%%%%%%%%%%%%%%%%%%%%%%%%%%%%%%%%%%%%%%%%%%%%%%%%%%%%
\section{Introduction}
%%%%%%%%%%%%%%%%%%%%%%%%%%%%%%%%%%%%%%%%%%%%%%%%%%%%%%%%%%%%%%%%%%%%%%

Due to the current advances in sensor networks, wireless technologies, and RFID-enabled ubiquitous 
computing, data about moving-objects (also referred to as trajectories) is an example of massive 
data relevant in many real applications. Think in the notion of Smart Cities, where the implementation of new technologies in public transportation systems has become more widespread all around the world in the last decades. For instance, nowadays many cities -from London to Santiago- provide
the users of the public transportation with smartcards that help in making the payment to access buses
or subways an easier task. Even though  smartcards may only collect  data  when users enter to the transportation system, it is possible to derive the users' trip (when they enter and leave the system) using historical data and  transportation models~\cite{Munizaga20129}.  %\marginpar{\tiny quizas podamos quitar las referencias a privacy; en este caso quitar ...and considering that for privacy issues the individual trajectory must not be revealed... }
When having this data,
% and considering that for privacy issues the individual trajectory must not be revealed, 
 counting or aggregate queries of trajectories become useful tools for traffic monitoring, road planning, and road navigation systems.

New technologies and devices generate a huge amount of highly detailed,
real-time data.  Several research exists about moving-object databases (MODs)~\cite{DBLP:conf/chorochronos/GutingBEJLNSV03,DBLP:journals/geoinformatica/Spaccapietra01,DBLP:conf/sigmod/ForlizziGNS00,DBLP:journals/geoinformatica/ErwigGSV99} and indexing structures~\cite{DBLP:books/sp/PelekisT14,DBLP:conf/vldb/PfoserJT00,DBLP:conf/vldb/PapadiasT01,DBLP:conf/ssd/Frentzos03,DBLP:journals/geoinformatica/AlmeidaG05,DBLP:journals/vldb/PopaZOBV11,DBLP:conf/icde/Cudre-MaurouxWM10}.  They, however, have addressed typical spatio-temporal queries such as
time slice or time interval queries that retrieve trajectories or
objects that were in a spatial region at a time instant or during a
time interval. They were not specially designed to answer queries that are
based on counting, such as the number of  distinct trips,
which are more meaningful queries for public-transportation or
traffic administrators.  This problem was recently highlighted in~\cite{DBLP:conf/cikm/LiCDYZZ0Z15}, where authors describe an approximate query processing of aggregate queries that count the number of distinct trajectories within a region. In this work, we concentrate on counting-based queries on a network, which  includes the number of trips  starting
or ending at some time instant in specific stops (nodes) or the
top-k most used stops of a network during a given time interval.

The work in this paper proposes  a new structure named \represName\ ($\repres$) that answers  counting-based queries and uses compact self-indexed data structures to represent the large amount of trips in a compact space.
$\repres$ combines two well-known data structures. The first one,
initially designed for the representation of strings, is
Sadakane's Compressed Suffix Array ($\csa$) \cite{Sad03}. The second
one is the Wavelet Tree \cite{WT03} ($\wt$). To make the
use of the $\csa$ possible, we define a trip or trajectory of a moving object
over a network as the temporally-ordered sequence of the nodes the trip
traverses.  An integer $ID$ is assigned to each node such that a trip
is a string with the $ID$s of the nodes. Note that this representation avoids the cost of storing coordinates to represent the locations users pass through during a trip. It is just enough to identify the stops or nodes and when necessary to map these nodes to geographic locations. Then a $\csa$, over the concatenation of
these strings (trips), is built with some adaptations for this
context. In addition, we discretize the time in periods of fixed
duration (i.e. timeline split into 5-minute intervals) and each time
segment is identified by an integer $ID$. In this way, it is possible
to store the times when trips reach each node by associating the
corresponding time $ID$ with each node in each trip. The sequence of
times for all the nodes within a trip is self-indexed with a $\wt$
to efficiently answer temporal and spatio-temporal queries.

We experimentally tested our proposal using two sets of %synthetic
data representing trips over two different real public
transportation systems. Our results are promising because the
representation uses around  $50$\% of its original size and
answers most of our spatial, temporal,  and spatio-temporal queries within $1\!-\!1000$ microseconds. 
In addition, since $\repres$ implicitly keeps all the original trajectories in a compact and self-indexed
way, it would permit us to extend its functionality with additional operations that could benefit
from the indexed access provided both by the underlying $\csa$ and $\wt$ structures.
 %\marginpar{\tiny Daniil, ibamos a mirar la representacion que usaba oracle?}
No experimental comparisons with classical spatial or spatio-temporal
indexing structures were possible, because none of them were designed to
answer the types of queries in this work. Our approach can  be
considered as a proof of concept that opens new application
domains for the use of well-known compact data structures such as the
$\csa$ and the $\wt$, creating a new strategy for
exploiting trajectories represented in a self-indexed way. 

The organization of this paper is as follows. Section~\ref{sec:prevwork} reviews previous works on
trip representations. It also presents 
$\csa$ and $\wt$ upon which we develop our proposal. We pay special attention to
show the internals of those structures and discuss also their properties and functionality.
Then, in Section~\ref{sec:wt}, we also present the {\em wavelet matrix} ($\wm$) and show how to create
a {\em Hu-Tucker-shaped $\wt$} ($\wtht$). These are the two variants of $\wt$ we use to represent temporal data. 
In Section~\ref{sec:queries}, we present the main counting-based queries that are of
interest for a transportation network.
In Section~\ref{sec:ctr}, we present $\repres$ and show how to reorganize  
a dataset of trips to allow a $\csa$ to handle the spatial data and a $\wt$-based structure
to manage the temporal data. 
Section~\ref{sec:transnet_repr} shows how  $\repres$ represents the spatial 
component and how spatial queries are dealt with. In Section~\ref{sec:time_repr}, we focus on
how to represent the temporal component of trips and how to answer temporal queries. We also include
a brief comparison of the space/time trade-off of $\wm$ and $\wtht$.
In Section~\ref{sec:stq}, we show how spatio-temporal queries are solved by $\repres$, and
Section~\ref{sec:experiments} includes our experimental results.
%gives the experimental evaluation of $\repres$., showing  its efficiency in space and time.
Finally, conclusions and future work are discussed in Section~\ref{sec:conclusions}.

%%%%%%%%%%%%%%%%%%%%%%%%%%%%%%%%%%%%%%%%%%%%%%%%%%%%%%%%%%%%%%%%%%%%%%%%%%%%%%%%%%

%%%%%%%%%%%%%%%%%%%%%%%%%%%%%%%%%%%%%%%%%%%%%%%%%%%%%%%%%%%%%%%%%%%%%%
\section{Previous Work} \label{sec:prevwork}
%%%%%%%%%%%%%%%%%%%%%%%%%%%%%%%%%%%%%%%%%%%%%%%%%%%%%%%%%%%%%%%%%%%%%%
\subsection{Models of trajectory and types of queries}
There  is a large amount of work on  data models for  moving-object data~\cite{DBLP:conf/ssdbm/WolfsonXCJ98,DBLP:conf/icde/SistlaWCD97,DBLP:journals/tods/GutingBEJLSV00,DBLP:conf/chorochronos/GutingBEJLNSV03,DBLP:journals/geoinformatica/Spaccapietra01,DBLP:conf/sigmod/ForlizziGNS00,DBLP:journals/geoinformatica/ErwigGSV99,DBLP:books/mk/GutingS2005}.   Basically, a moving-object data model represents the continuous change of the location of an object over time, what is called the trajectory of the object.

Moving-object data is an example of big data that differ  in the representation of  location, contextual or environmental  information where the movement takes place, the time dimension that can be continuous or  discrete, and the level of abstraction or granularity on which the trajectories are described~\cite{DBLP:journals/sigspatial/DamianiIGV15}.   A common classification of trajectories  distinguishes free from network-based trajectories.  \textit{Free trajectories} or Euclidean trajectories are a sequence of GPS points represented by an ad-hoc data type of moving points~\cite{DBLP:conf/ssdbm/WolfsonXCJ98,DBLP:conf/icde/SistlaWCD97,DBLP:journals/tods/GutingBEJLSV00}. \textit{Network-based} trajectories are a temporal ordered sequence of locations on networks.  This trajectory model includes a data type for representing  networks and  for representing the relative location of static and moving  points on the network~\cite{DBLP:journals/vldb/GutingAD06}. In a recent work, \textit{network-matched trajectories} are defined to  avoid the need of a mobile map at the moving-object side~\cite{DBLP:journals/tits/Ding0GL15}.

The definition of trajectories at an abstract level must be materialized in an internal representation with access methods for query processing. An early and broad classification of spatial-temporal queries for \textit{historical positions} of moving objects \cite{DBLP:conf/vldb/PfoserJT00} identifies coordinate- and trajectory-based queries. Coordinate-based queries include the well-known  {\it time-slice}, {\it time-interval} and \textit{nearest-neighbor queries}. Examples are \textit{find objects or trajectories in a region at a  particular time instant or during some time interval}. Another important example of range-based queries is \textit{find the k-closest objects with respect to a given point at a given time instant}. Trajectory-based queries  involve topology of trajectories (e.g., overlap and disjoint) and information (e.g., speed, area, and heading) that can be derived from the combination of time and space. An example of such queries would be  \textit{find objects or trajectories that satisfy a spatial predicate (eg., leave or enter a region)  at a particular time instant or time interval}. There also exist combined queries addressing information of particular objects: \textit{Where was object X at a particular time instant or time interval?}. In all previous queries, the results are individual trajectories that satisfy the query constraints.

%\marginpar{\tiny quizas podamos quitar las referencias a privacy issues de nuevo y %simplemente decir que: ~~~~ ~~~~~~~~  ``when dealing with large datasets of trajectories we %can find scenarios where answering counting based or aggregated queries are typically of %concern. This is for example the case of network management applications or when there are %privacy issues that prevent us from revealing the original individual trajectories."  }
%
%When dealing with large datasets of trajectories, and due to privacy issues, the individual %trajectory cannot be revealed, then anonymized and aggregated trajectories are of concern.

When dealing with large datasets of trajectories we can find scenarios where answering counting based or aggregated queries are typically of concern. This is for example the case of network management applications, those for mobility analysis, or when there are privacy issues that prevent us from revealing the original individual trajectories. 
In this context, we can  further distinguish range- from trajectory-based queries. Range queries impose constraints in terms of a spatial location  and temporal interval.  Examples of these queries are  to retrieve the number of distinct trajectories  that intersect a spatial region or spatial location (stop) at a given time instant or time interval, retrieve the number of distinct trajectories that start at a particular location (stop) or in a region and/or end in another particular location of region, retrieve the number of trajectories that follow a path, 
%\marginpar{\tiny Daniil, entiendo que el pattern debe ser exacto: si pongo P=ABC, no sirve una trayectoria ABxC. Esta seria otra consulta interesante? \DAGAL{Delego esta decisión en Andrea :-)} \ART{Yo creo que es %interesante esa consulta pero en este paper se estan enfatizando consultas de agregation o conteo. De hecho se destaca esto para diferenciar con NETTRA. Ojo, yo diria tambien que estas path queries caen en la categoria de %trajectory-based queries.  } } 
%
 and retrieve the  top-k locations (stops) or regions  with the larger number of  trajectories that  intersect  at a given time instant or time interval. Trajectory-based queries require not only to use the spatio-temporal points of trajectories  but also the sequence of these points. Examples of such queries are  to find the number of trajectories that are heading (not necessarily ending at) to a spatial location during a time interval, find the destination of trajectories that are passing through a region during a time interval, find the number of starting locations of  trajectories that go or pass through a region during a time interval.%, and find the number of trajectories that intersect during  a time interval.

\subsection{Trajectory indexing} Many data structures have been proposed to support efficient 
query capabilities on collections of
trajectories. We refer to~\cite[Chapter 4]{DBLP:books/sp/PelekisT14} for a comprehensive and
up-to-date survey on data management techniques for trajectories of moving objects. We can broadly
classify these data structures into two groups: those that index trajectories in free space and
those that index trajectories that are constrained to a network.

In free space, it is common to see spatial indexes that extend the {\em R-Tree} %spatial
index~\cite{DBLP:conf/sigmod/Guttman84} beyond a simple {\em 3D R-Tree} where the time is one of the dimensions.
Two examples of such indexes can be found in~\cite{DBLP:conf/vldb/PfoserJT00}
where the authors present two fundamental variations of the {\em R-Tree}: the {\em STR-Tree} and the {\em TB-Tree}.
Both indexes modify the classical construction algorithm for the {\em R-Tree},
where the nodes are not only grouped by the spatial distance among the indexed objects,
but also by the trajectories they belong to. In the {\em MV3R-tree}~\cite{DBLP:conf/vldb/PapadiasT01},
the construction takes into account temporal information of the moving objects,
adapting ideas from the Historical {\em R-Tree}~\cite{nascimento1998towards}. Another
interesting approach is described in~\cite{chakka2003indexing}, where the authors
split trajectories of moving objects across partitions of space, indexing each partition
separately. This improves query efficiency, as only the partitions that intersect
a query region are accessed.

{\em R-Tree} adaptations can also be useful when the trajectories are constrained to a network.
They exploit the constraints imposed by the topology of the network to optimize the data structure.
This is the case of the {\em FNR-tree}~\cite{DBLP:conf/ssd/Frentzos03}, which consists of a 
{\em 2D R-Tree} to index the %line 
segments of the trajectories over the network,
and a forest of {\em 1D R-Trees} used to index the time interval when each trajectory is moving through
each segment of the network.
The {\em MON-Tree}~\cite{DBLP:journals/geoinformatica/AlmeidaG05} can be seen as an
improvement over the {\em FNR-Tree}, saving considerable space by indexing MBRs of larger
network elements (edge segments or entire roads) and reducing the number of disk accesses at query time.
Both indexes are outperformed by the {\em TMN-Tree}~\cite{chang2010tmn} in query time,
which indexes whole trajectories of moving objects with a {\em 2D R$*$-Tree} and
indexing the temporal component with a B$⁺$-Tree, which proves to be more efficient for
that application than the {\em R-Tree}.

%\marginpar{\tiny\DAGAL{No sé si TMN-Tree es más o menos eficiente que PARINET. No he podido encontrar un paper que los compare}}
%PARINET is an interesting alternative for the network bound trajectories~\cite{DBLP:journals/vldb/PopaZOBV11}. %
{\em PARINET} is another interesting alternative to represent trajectories constrained to a network~\cite{DBLP:journals/vldb/PopaZOBV11}. %
%
% It divides trajectories into segments associated to the
% underlying roads in a network, and for each segment keeps:
% the trajectory-id, the road-id for the segment, and the position and times of the starting and ending points. Both roads and
% trajectory segments are stored in a data base.
% At query time, the segments of the trajectories can be easily grouped by road, and a temporal
% B$⁺$-tree %(relying on the DBMS)
% index (there is one index for each road)
% is used  to filter out candidate segments matching time constraints.
%
%
It partitions trajectories into segments from an underlying road network using a
complex cost model to minimize the number of disk accesses at query time. It takes into
account the spatial relations among the indexed network elements, as well as some statistics
of the data to index. Then it adds a temporal B$⁺$-tree to index the trajectory segments from each road.
Those indexes permit {\em PARINET} to filter out candidate trajectory segments matching time constraints at query time.
The same ideas were used in {\em TRIFL}~\cite{that2015trifl}, where the cost model is adapted
for flash storage.
\medskip

All previous data structures were designed to answer spatio-temporal queries, where the space, in particular geographic coordinates, and time are the main
filtering criteria. Examples of such queries are: {\em retrieve trajectories that crossed a region
within a time interval}, {\em retrieve trajectories that intersect}, or {\em retrieve the $k$-best connected
trajectories} (i.e., the most similar trajectories in terms of a distance function).
Yet, they could not easily support queries such as {\em the number of trips starting in X and ending at Y}. 
A recent work in~\cite{DBLP:conf/cikm/LiCDYZZ0Z15} proposes  a method to compute the approximate number of distinct trajectories that cross a region. Note that computing aggregate queries of trajectories 
in the  hierarchical structure of classical
spatio-temporal indices is usually done by aggregating the
information maintained in index nodes at the higher levels to avoid
accessing the raw spatio-temporal data.
 However, for a trajectory
aggregate query, maintaining the statistical trajectory information
on index nodes does not work because what matters for these queries is to determine the number of \textit{distinct} trajectories in
a spatio-temporal query region. 
 
% \ART{Extendi el parrafo anterior para enfatizar el tipo de consulta con una cita que  no teniamos incluida}
 
%Although existing data structures for trajectories support efficient
%query processing on  large datasets, they can have  limitations to
%deal with current massive data.
 %In this context, data compression techniques has been explored in the past to help in handling  the  problem of massive data.
The application of data compression techniques has
been explored in  the context of massive data about trajectories. The work by Meratnia and de
By~\cite{DBLP:conf/edbt/MeratniaB04} adapts a classical
simplification algorithm by Douglas and Peucker to reduce the number
of points in a curve and, in consequence, the space used to represent
trajectories. Potamias \textit{et
al.}~\cite{DBLP:conf/ssdbm/PotamiasPS06} use concepts, such as speed
and orientation, to improve compression. It is also possible~\cite{cao2006spatio}
to compress a trajectory in a way that the maximum error at query time is
deterministic, although the method greatly depends on the distance function to be used.
%When we know that object movements are constrained to a
%network, we can do even better. This aspect has been
%explored
%in~\cite{DBLP:journals/josis/RichterSL12,DBLP:journals/jss/KellarisPT13,DBLP:conf/w2gis/FunkeSSS15}.
%Those works focus mainly in how to represent trajectories and in how to gather the location of one or more
%given  moving objects from those trajectories. Yet, they would poorly support more complex aggregated queries.
%\marginpar{porfa daniil, mira se o anterior é correcto}
%

In~\cite{DBLP:journals/josis/RichterSL12,DBLP:journals/jss/KellarisPT13,DBLP:conf/w2gis/FunkeSSS15},
they focus mainly on how to represent trajectories constrained to
networks, and in how to gather the location of one or more given
moving objects from those trajectories.
%Yet, these works are also out of our scope as they would poorly support queries related to an
%unbounded set of objects, or more complex aggregated queries
%oriented to exploit the data about the network usage.
Yet, these works are out of our scope as they would poorly support
queries oriented to exploit the data about the network usage such as
those that compute the number of trips  with a specific  spatio-temporal
pattern (e.g. {\em Count the trips starting at stop $X$ and ending at stop $Y$ in
working days between 7:00 and 9:00}).

A recent work~\cite{DBLP:conf/gis/KroghPTT14} proposed an indexing structure called  {\em NETTRA} to answer  
{\em strict}  and {\em approximate path queries} that can be implemented in standard SQL using $B^+$-trees and self-JOIN operations.  
For each trajectory, {\em NETTRA} represents the sequence of adjacent network edges touched by the trajectory as  
entries in a table  with four columns:  id, entering and leaving time, and a hash value 
 of the entire path up to and including the edge itself.
Using the hash value  for the first and last
edge on a query path, {\em NETTRA}  determines whether
the trajectory followed a specific path between these edges. 
Also for {\em strict path queries}, Koide et al.~\cite{DBLP:conf/gis/KoideTY15} proposed a spatio-temporal
index structure  called {\em SNT-Index} that is based on  the integration of a  FM-index~\cite{DBLP:conf/focs/FerraginaM00}  
to store spatial information with a forest of B+trees that stores temporal information.
%\OJOFARI{ Os parece bien dejar la siguiente frase? ~~~~~~
To the best of our knowledge, this makes up the first technique using compact data structures to
handle spatial data in this scenario. Yet, in our opinion, {\em strict path queries} have little interest in the context of exploiting
data to analyze the usage of a transportation network.
%}
%\ART{Solo edite levemente esto}
	%While strict path queries have little interest in the context of a transport network,
	%they could also be answered with our proposal in case of need.
	%\marginpar{\tiny Aqui teniamos una linea diciendo que strict path queries podian ser respondidas con $\repres$. Sin embargo creo que es mejor no decir nada
	%salvo que se hiciese una comparacion contra NETTRA o SNT-Index.}
	%\medskip

%\ART{Quizas sea bueno que esta estructura ha sido  pensada esencialmente para consultas de conteo que son las que se resuelven finalmente en la parte experimental.}
Unlike previous works, we designed an in-memory representation, that targets at solving
counting-based queries, and is completely
based on the use of compact data structures (discussed in the next section)
to make it successful not only in time but also in space needs. 
Since $\repres$ keeps data in a compressed way, it  will permit to handle larger sets 
of trajectories entirely in memory and consequently to avoid costly disk accesses. 

%Data structures envisioned in our work must support the
%representation of trajectories on a network. Indeed, we are not
%interesting in codifying trajectories as a sequence of locations on
%a free space, but as a sequence of nodes visited on a transportation
%network.
%Also, the recent work
%in~\cite{DBLP:conf/w2gis/FunkeSSS15} encodes  a
%trajectory as a concatenated string  and  queries match
%patterns within a large concatenation of trajectories.

\subsection{Underlying Compact Structures of $\repres$}
$\repres$ relies on two components: one to handle the spatial information and
another to represent temporal information.
The spatial component is based on the well-known a Compressed Suffix
Array ($\csa$)~\cite{Sad03}. The temporal component can be implemented with
either a Wavelet Tree ($\wt$)~\cite{WT03} or a Wavelet Matrix
($\wm$)~\cite{CNO15}. The latter is a variant of $\wt$ that performs better 
when representing sequences built on a large alphabet as we see below.

%%%%%%%%%%
\subsubsection{Sadakane's Compressed Suffix Array (CSA)} \label{sec:csa}

%%%%%%%%%%%%%%%%%%%%%%%%%%%%%%%%%%%%%%%%%%%%%%%%%%%%%%%%%%%%%
%
Given a sequence $S[1,n]$ built over an alphabet $\Sigma$ of length
$\sigma$, the {\em suffix array} $A[1,n]$  built on $S$ \cite{MM93}
is a permutation of $[1,n]$ of all the suffixes $S[i,n]$ so that
$S[A[i],n] \prec S[A[i+1],n]$ for all $1 \le i < n$, being $\prec$
the lexicographic ordering. Because $A$ contains all the suffixes of $S$ in lexicographic order,
this structure permits to search for any pattern $P[1,m]$ in time
$O(m \log n)$ by simply binary searching the range $A[l,r]$ that
contains pointers to all the positions in $S$ where $P$ occurs.
%For the construction of the $A$ a linear time construction
%algorithm such as the SA-IS~\cite{nong2011two} could be used.
%which is based on induction sorting.

To reduce the space needs of $A$, Sadakane's CSA \cite{Sad03} uses
another permutation $\Psi[1,n]$ defined in \cite{GV00}. For each
position $j$ in $S$ pointed from $A[i]=j$, $\Psi[i]$ gives the
position $z$ such that $A[z]$ points to $j+1$. There is a special
case when  $A[i]=n$, in this case $\Psi[i]$ gives the position $z$
such that $A[z]=1$.
 %
%\DAGAL{Fari, que es $\sigma'$?}
In addition, we could set up a vocabulary array $V[1,\sigma'], (\sigma' \leq \sigma)$ with
all the different symbols from $\Sigma$ that appear in $S$, and a bitmap $D[1,n]$
aligned to $\Psi$ so that $D[i] \leftarrow 1$ if  $i=1$ or if
$S[A[i]] \neq S[A[i-1]$ ($D[i]\leftarrow 0$; otherwise). Basically, a
$1$ in $D$ marks the beginning of a range of suffixes pointed from
$A$ such that the first symbol of those suffixes coincides. That is,
if the $i^{th}$ and ${(i+1)}^{th}$ one in $D$ occur in $D[l]$ and
$D[r]$ respectively,
%select_1(D,i)=l$ and $select_1(D,i+1)=r$
we will have $V[rank_1(D,l)] = V[rank_1(D,x)]~ \forall x \in [l,r-1]$. Note that $rank_1(D,i)$
returns the number of 1s in $D[1,i]$ and can be computed in constant time using $o(n)$ extra bits~\cite{Jac89,Mun96}.

By using $\Psi$, $D$, and $V$, 
%it is possible to perform binary search without the need of accessing $A$ nor $S$. 
it is possible to simulate a binary search for the interval
$A[l,r]$ where a given pattern $P$ occurs ($[l,r]~\leftarrow~bsearch(P)$) without keeping $A$ nor $S$. 
Note that, the symbol $S[A[i]]$ pointed by $A[i]$ can be obtained as
$V[rank_1(D,i)]$, and we can obtain the following symbol from the source sequence  $S[A[i]+1]$ as
$V[rank_1(D, \Psi[i])]$, $S[A[i]+2]$ can be obtained as $V[rank_1(D, \Psi[\Psi[i]])]$, and so on.
Therefore, $\csa$\  replaces $S$, and it does not need $A$ anymore to perform searches.

However, in principle, $\Psi$ would have the same space requirements
as $A$. Fortunately, $\Psi$ is highly compressible. It was shown to
be formed by $\sigma$ subsequences of increasing values \cite{GV00} so that
it can be compressed to around the zero-order entropy of $S$~\cite{Sad03} and, by using $\delta$-codes to represent the
differential values, its space needs are $nH_0+O(n\log\log\sigma)$ bits.
In \cite{NM07}, they showed that $\Psi$ can be split into
$nH_k+\sigma^k$ (for any $k$) {\em runs} of consecutive values so
that the differences within those runs are always $1$. This
permitted  to combine $\delta$-coding of gaps with run-length
encoding (of $1$-$runs$) yielding higher-order compression of $\Psi$.
In addition, to maintain fast random access to $\Psi$, absolute
samples at regular intervals (every $t_{\Psi}$ entries) are kept. Parameter
$t_{\Psi}$ implies a space/time trade-off. Larger values lead to better compression of $\Psi$ but
slow down access time to non-sampled $\Psi[i]$ values.

In \cite{FBNCPR12}, authors adapted Sadakane's CSA to deal with large (integer-based) alphabets
and created the {\em integer-based CSA} ($\icsa$). They also showed that, in their scenario (natural language
text indexing), the best
compression of $\Psi$ was obtained by combining differential encoding of runs with Huffman and
run-length encoding.

\subsubsection{The Wavelet Tree (WT)}\label{sec:wt}

\begin{figure}[ht]
  %\vspace{-0.4cm}
  \begin{center}
  {\includegraphics[width=0.49\textwidth]{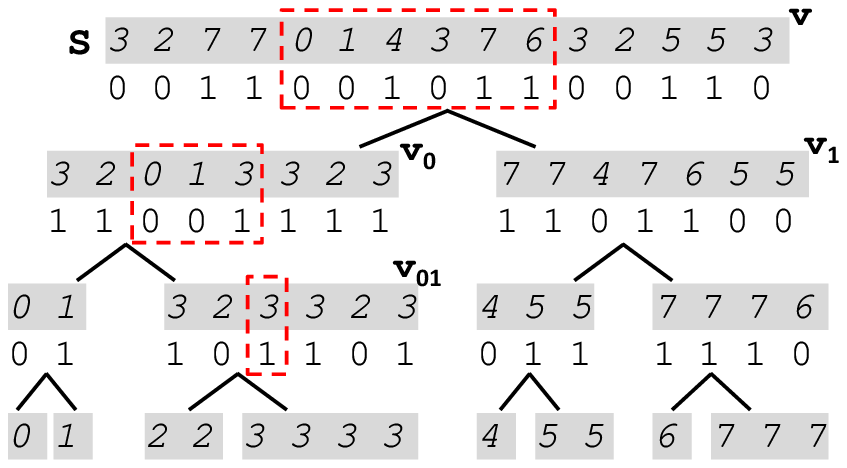}}  %% probar se funciona en local
  {\includegraphics[width=0.49\textwidth]{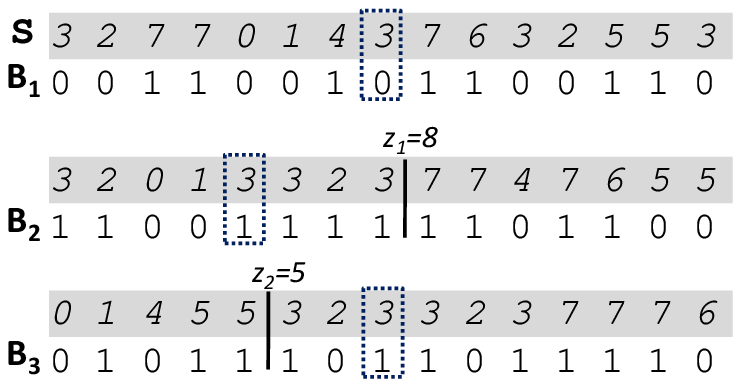}}  %% probar se funciona en local
  \end{center}
  %\vspace{-0.4cm}
  \caption{$\wt$ (left) and $\wm$ (right) for sequence $S=\langle 3\, 2\, 7\, 7\, 0\, 1\, 4\, 3\, 7\, 6\, 3\, 2\, 5\, 5\, 3\rangle$. 
  Shaded areas are neither included in the $\wt$ nor in the $\wm$.}
  \label{fig:wt}
  %\vspace{-0.5cm}
\end{figure}

Given a sequence $S[1,n]$ built on an alphabet $\Sigma$ with $\sigma$ symbols that 
are encoded with a fixed-length binary code $[0,\sigma)$,  
a $\wt$~\cite{WT03} built over $S$ is a balanced binary tree where leaves are labeled with the different
symbols in $S$, and each internal node $v$ contains a bitvector $B_v$.
The bitvector in the root node contains the first bit from the codes of all the $n$ symbols in $S$. 
Then symbols whose code starts with a 0 are assigned to
the left child, and those with codes starting with a 1 are assigned to the right child.
In the second level, the bitvectors contain the second bits of the codes of their assigned
symbols. This applies recursively for every node, until a leaf node is reached.
Leaf nodes can only contain one kind of symbol. The height of the tree is $\log\sigma$, and 
since the bitvectors of each level contain $n$ bits, 
the overall size of all the bitvectors is $n\log\sigma$ bits. To calculate the total size of
the $\wt$ we also need to take into account the space needed to store pointers from each symbol
to its corresponding tree node which is $O(\sigma\log n)$ bits. In addition, as we see below
the $\wt$ reduces the general problem of solving $access(i)$, $rank_{c}(i)$, and $select_{c}(i)$ operations to
the problem of computing $access$, $rank$ and $select$ on the bitvectors. Therefore, additional structures to
efficiently support those operations add up to $o(n\log\sigma)$ space. The overall size of the $\wt$ is
$n \log\sigma  (1 +o(1))$ + $O(\sigma\log n)$. Figure~\ref{fig:wt}.(left) shows a $\wt$ built on the
sequence $S=\langle 3\, 2\, 7\, 7\, 0\, 1\, 4\, 3\, 7\, 6\, 3\, 2\, 5\, 5\, 3\rangle$  
assuming we use a 3-bit binary encoding for the symbols in $\Sigma=[0..7]$. Shaded areas are not 
included in the $\wt$ but help us to see the subsequences handled by the children of a given node. 

Among others, the $\wt$ permits to answer  the following queries in $O(\log\sigma)$ time:

\begin{itemize}
	\item $access(i)$ returns $S[i]$.
	\item $rank_{c}(i)$ returns the number of occurrences of symbol $c$ in $S[1,i]$.
	\item $select_{c}(i)$ returns the position of the i-th occurrence of  symbol $c$ in $S$.
	\item $count(i,j,\alpha,\beta)$ described in~\cite{gagie2012new}, returns the
	number of occurences in $S[i,j]$ of the symbols between $\alpha$ and $\beta$.
\end{itemize}

To solve $access(i)$ and $rank_c(i)$ operations we traverse the $\wt$ from the root 
until we reach a leaf. In the case of $rank_c(i)$ we descend the tree taking into account the
encoding of $c=c_1c_2...$ in each level.  Being $B$ the bitmap in the root node, if $c_1=0$ we move
to the left child and set $i \leftarrow rank_0(i) $; otherwise we move to the right child and
set $i \leftarrow rank_1(i) $. We proceed recursively until we reach a leaf where we return $i$. 
$access(i)$ is solved similarly, but at each level, we either move left or right depending on 
if $B[i] = 0$ or $B[i] = 1$ respectively. The leaf where we arrive corresponds to the symbol 
$c=S[i]$ which is returned. 
To solve $select_c(i)$ we traverse the tree from the leaf corresponding to symbol $c$ until the root. 
At level $j$, we look at the value of the $j$-th bit of the encoding of $c$ ($c_j$). 
If $c_j=0$ we set $i \leftarrow select_0(B_{j-1})$ (where $B_{j-1}$ is the bitmap of the parent
of the current node), otherwise we set $i \leftarrow select_1(B_{j-1})$. 
Then we move to level $j-1$ and proceed recursively until the root, where the final value of $i$ is returned.

%Using the classical binary $rank$ and $select$ over the bitvectors,
%it is possible to navigate up and down the tree and answer all the queries
%described above in $O(\log\sigma)$ time. As in this work we are only interested
%in $count(i,j,\alpha,\beta)$, we provide a small example of $count(6,7,2,4)$ over
%the $\wt$ from the Figure~\ref{fig:wt}:

In this work we are also interested in operation $count(i,j,\alpha,\beta)$ that allow us to
count the number of occurrences of all the symbols $c \in [\alpha,\beta]$ within $S[i,j]$.
Assuming the encodings of the symbols $c \in [\alpha,\beta]$ form also a contiguous range 
this can be solved in $O(\log \sigma)$ \cite{gagie2012new,CNO15}. The idea is to traverse 
the tree from the root and descend through the nodes that cover the leaves in $[\alpha,\beta]$. 
At each node $v$ (whose bitmap is $B_v$ and range $B_v[i,j]$ is considered), 
that covers symbols in range $[a,b]$ we check whether 
$[a,b] \subseteq [\alpha,\beta]$. In that case we sum $j-i+1$ occurrences. If both ranges are disjoint we found
a node not covering the range $[\alpha,\beta]$ and stop the traversal. Similarly, if
range $[i,j]$ becomes empty traversal stops on that branch.  Otherwise, we recursively descend 
from node $v$ to their children $v_l$ and $v_r$ where we map the interval $B_v[i,j]$ into $B_{v_l}[i_l,j_l]$
and $B_{v_r}[i_r,j_r]$ with $rank_0$ operation. In practice, $i_l\leftarrow rank_0(i-1)+1$, 
$j_l\leftarrow rank_0(j)$ and $i_r \leftarrow i-i_l+1$, $j_r \leftarrow j-j_l$. For more details and 
pseudocodes see \cite{Nav16, gagie2012new, CNO15}. 
In Figure~\ref{fig:wt}.(left) we can see the nodes ($\mathbf{v}$, $\mathbf{v_0}$, and $\mathbf{v_{01}}$) that
must be traversed, and the ranges within the bitmaps in those nodes, to solve $count(5,10,3,7)$.
Therefore, we want to compute the number of occurrences of symbols between $3$ and $7$ that occur
within $S[5,10]$. We start in the root node $\mathbf{v}$, where $B_v[5,10]$ contains $3$ zeroes 
and $3$ ones. We compute $i_l=rank_0(4)+1=3, j_l=rank_0(10)=5$, and $i_r=5-3+1=3, j_r=10-5=5$.
At this point we could move to $\mathbf{v_1}$, but we can see that all the encodings of the 
symbols $4, 5, 6, 7$ start by $1$ and they are covered by $\mathbf{v_1}$. Therefore, we report
$i_r-i_l+1 = 3 $ occurrences of symbols in range $[4,7]$ and no further processing is done in the
subtree whose root is $\mathbf{v_1}$. However, we descent to $\mathbf{v_0}$ since $0s$ in $B_v[5,10]$
could belong to any symbol in range $[0,3]$, and we have to track only occurrences of symbol $3$.
We check $B_{v_0}[3,5]$ and compute $i_l=rank_0(3)+1=1, j_l=rank_0(5)=2$, and $i_r=3-1+1=3, j_r=5-2=3$.
Since the second bit of the encoding of symbol $3$ is a $1$ (as for symbol $2$), we can discard descending
on the left child of $\mathbf{v_0}$ and move only to its right child $\mathbf{v_{01}}$ where we are 
interested in the range $B_{v_{01}}[3,3]$. Since $\mathbf{v_{01}}$ covers both symbols $2$ and $3$, and the
third bit of the encoding of $2$ is a zero whereas it is a one for $3$, we do only need to count the
number of ones in $B_{v_{01}}[3,3]$. After computing $i_l=rank_0(2)+1=2, j_l=rank_0(3)=1$, 
and $i_r=3-2+1=2, j_r=3-1=2$, we report $j_r -i_r+1 =1$ occurrence of symbol $3$. 
Therefore, we conclude $count(5,10,3,7) = 3+1=4$.

%
%We provide a small example of $count(6,7,2,4)$ over
%the $\wt$ from the Figure~\ref{fig:wt}:
%
%We start by calculating $rank_0(5)~=~5$ and $rank_0(7)~=~6$ on the bitvector of the root node.
%$6-5=1$, which means that in $S[6,7]$ there is one symbol between 1 and 2,
%and we need to check the left child at the position 6, done with
%$rank_1(5)~=~1$ and $rank_1(6)~=~2$\footnote{This could be done with an $access$ in this particular case}
%and as $2-1=1$ we can say that the symbol in that position is a 2 and not a 1.
%
%In the case of the symbols from 3 to 4, we calculate $rank_1(5)~=~0$ and $rank_1(7)~=~1$
%back at the root node. In this case we don't need to check anything at the right
%child because any 1 bit in the root comes from either a 3 or a 4, both of them being
%valid for our query. This is what makes this operation to be $O(\log\sigma)$
%instead of $O(p\log\sigma)$ where $p~\leftarrow~\beta-\alpha+1$: there is no
%need to reach the leaves when the queried codes are contiguous.
%Finally, we can say that in $S[6,7]$ there is one 2 and one
%symbol that is either a 3 or a 4, so $count(6,7,2,4)~=~2$.

One way of reducing the space needs of a WT consists in compressing its bitvectors 
\cite{Navjda13}. Among others (i.e. Golynski et al. \cite{Golynski2007} which is better theoretically), 
Raman el al. technique \cite{Raman:2002:SID:545381.545411} (RRR) is, in practice, 
one of the best choices. The overall size of the $\wt$ becomes 
$nH_0(S)+ o(n \log \sigma) + O(\sigma \log n)$, whereas operations still require $O(\log \sigma)$ time.

Another way of compressing a $\wt$ is to use a prefix-free variable-length encoding for the symbols. 
For example,  Huffman~\cite{huffman1952method} code can be used to build a
Huffman-Shaped $\wt$~\cite{ferragina2009compressed}, where the tree is not balanced
anymore. The size reduces to $n(H_0(S)+1)+ o(n(H_0(S)+1)) +  O(\sigma \log n)$,\footnote{$O(\sigma \log n)$ term 
includes both the tree pointers and the size of the Huffman model.}   and average time becomes
$O(H_0(S))$ for $rank$, $access$, and $select$  (worst-case time is still $O(\log\sigma)$ 
\cite{Barbay:2013:CPA:2562345.2562626}). By using compressed
bitvectors \cite{{CNO15}} space can be reduced even further to $nH_0(S) +o(n(H_0(S)+1)) +  O(\sigma \log n)$.
Unfortunately, the Huffman codes given to
adjacent symbols are no longer contiguous, and it is not possible to give a $O(\log\sigma)$ bound for
$count(i,j,\alpha,\beta)$ anymore, even if the code is canonical. Hu-Tucker codes~\cite{hu1971optimal}
can be used instead.\footnote{Hu-Tucker ~\cite{hu1971optimal}
	 is an optimal prefix code that preserves the order of the input vocabulary.
	This means that the lexicographic order of the output variable-length binary codes is the same as the order of the input symbols.}
Compression degrades slightly with respect to using Huffman coding,\footnote{Being $L_h$ and $L_{ht}$ the average codeword 
	length of Huffman coding and Hu-Tucker codes respectively, it holds: $H_0 \leq L_h \leq H_0+1$ and $H_0 \leq L_{ht} \leq H_0+2$
	(see \cite{Cover:2006:EIT:1146355} (pages 122-123), or \cite{HORIBE1977148, GilbertandMore1959}).} but the
codes for adjacent symbols are lexicographically contiguous. This permits to solve $count$ efficiently. 
%The size of a Hu-Tucker-shaped $\wt$ ($\wtht$) can be bounded to $n(H_0(S)+2) + o(n(H_0(S)+2)) +  O(\sigma \log n)$ and can be reduced
%to  $n(H_0(S)+1) + o(n(H_0(S)+2)) +  O(\sigma \log n)$ using compressed bitvectors as well.
%\ART{Por que en negritas?}
The size of a {Hu-Tucker-shaped $\wt$} ($\wtht$)  can be bounded to $n(H_0(S)+2) + o(n(H_0(S)+1)) +  O(\sigma \log n)$ and can be reduced
to  $nH_0(S) + o(n(H_0(S)+1)) +  O(\sigma \log n)$ by using compressed bitvectors as well.

%%%%%%%%%%%%%%%%%%%%%%%%%%%%%%%%%%%%%%%%%%%%%%%%%%%%%%%
\paragraph{The Wavelet Matrix (WM)}\label{sec:wm}

For large alphabets, the size of the $\wt$ is affected by the term $ O(\sigma \log n)$. A {pointerless}
$\wt$ \cite{CNspire08.1} permits to remove\footnote{In a pointerless Huffman-shaped $\wt$ a
term $O(\sigma \log\log n)$ still remains due to the need of storing the canonical Huffman model.} 
that term by concatenating all the bitmaps level-wise 
and computing the values of the pointers during the $\wt$ traversals. 
The operations on a pointerless $\wt$ have the same time complexity but become slower in practice. 

By reorganizing the nodes in each level of a pointerless $\wt$, the {\em Wavelet Matrix} 
($\wm$)~\cite{CNO15} obtains the  same space requirements ($n \log\sigma  (1 +o(1))$ bits), 
yet its performance is very close 
to that of the regular $\wt$ with pointers.
Figure~\ref{fig:wt}.(right) shows an example.

As in the $\wt$, the $i$-th level stores the $i$-th bits of the encoded symbols. 
A single bitvector $B_i$ is kept for each level. In the first level, $B_1$ stores the
$1$-st bit of the encoding of the symbols in the order of the original sequence $S$.
From there on, at level $i$, symbols are reordered according to the $(i-1)$-th bit of their encoding; 
that is, according to the bit they had in the previous level.
Those symbols whose encoding had a zero at position $i-1$ must be arranged before those that
had a one. After that, the relative order from the previous level is maintained. That is, if 
a symbol $\alpha$ occurred before other symbol $\beta$, and the $(i-1)$-th bit of their encoding
coincides, then $\alpha$ will precede $\beta$ at level $i$. %Note that, following this rule,
%the symbols in the last level shall be ordered by their reversed bit encoding.

If we simply keep the number of zeros at each level ($z_l\leftarrow rank_0(B_l,n)$), we can easily see that the $k$-th zero
at level $i-1$ is mapped at position $k$ within $B_i$, whereas the $j$-th one at level $i-1$ is 
mapped at position $z_l +j$ within $B_i$. This avoids the need for pointers and permits to retain
the same time complexity of the $\wt$ operations. For 
implementation details see \cite{CNO15,ordonez2015statistical}. For example to solve $access(S,8)$, we
see that $B_1[8]=\mathbf{0}$ and $rank_0(B_1,8)=5$. We move to the next level where we check position $5$; 
we see that $B_2[5]=\mathbf{1}$ and $rank_1(B_2,5)=3$. We move to next level and check position $3+z_2 = 3+5 = 8$,
where we finally see $B_3[8]=\mathbf{1}$. Therefore, we have decoded the bits $\mathbf{011}$ that correspond
to symbol $3 = access(S,8)$.

To reduce the space needs of $\wm$ we could use compressed bitvectors as for $\wt$s. Space needs become $nH_0(S)+ o(n \log \sigma) $ bits. Yet, 
compressing the $\wm$ by giving either a Huffman or Hu-Tucker shape is not possible as the reordering 
of the $\wm$ could lead to the existence of holes in the structure that would ruin the process of 
tracking symbols during traversals. To overcome this issue an optimal Huffman-based coding was 
specifically developed for wavelet matrices \cite{CNO15, Farina2016}. This allows to obtain space
similar to that of a pointerless Huffman-shaped $\wt$ but faster $rank$, $select$, and $access$ operations.
Unfortunately, since the encodings of consecutive symbols do not form a contiguous range, $count(i,j,\alpha,\beta)$ is 
no longer supported in $O(\log\sigma)$ time and computing $rank_c(S,j)-rank_c(S,i)+1$ is required 
for each $c$ in $[\alpha,\beta]$.

As indicated before, since in $\repres$ we need efficient support for $count$ operation, we 
will try (see Section~\ref{sec:time_repr}) %both the uncompressed and 
the Hu-Tucker-shaped $\wt$ as well as the uncompressed $\wm$.
In addition, we will couple them with both uncompressed and RRR compressed bitvectors.

% %%%%%%%%%%%%%%%%%%%%%%%%%%%%%%%%%%%%%%%%%%%%%%%%%%%%%%%%%%%%%%%%%%%%%%%%%%%%%%%%%%

%%%%%%%%%%%%%%%%%%%%%%%%%%%%%%%%%%%%%%%%%%%%%%%%%%%%%%%%%%%%%%%%%%%%%%
\section{Counting-based queries} \label{sec:queries}
%%%%%%%%%%%%%%%%%%%%%%%%%%%%%%%%%%%%%%%%%%%%%%%%%%%%%%%%%%%%%%%%%%%%%%

%\ART{En otros paper se habla de aggregate queries  y hablamos de counting queries. Lo que si es que %aggregate queries son el numero de trayectorias distintas que cruzan una region (especificada como %ventanas). Las nuestras son mas especificas al definir comienzo o fin de un viaje. Mejor dejar como %counting-based queries, pero quizás hacer alguna referencia. Algo deje en la intro de eso. }

In  transportation systems, new technologies such as automatic fare collection (e.g., smartcards) and automatic passenger counting have made possible to generate a huge amount of highly detailed,
real-time data useful to define measures that characterize  a transportation network. This data is particularly useful because it actually consists of real trips, combining  implicitly the service offered by a public transportation system with  the demand for the system. When having this data,  it is not the data about individual trajectories but measures of the use of the network what matters for traffic monitoring and road planning tasks. Examples of useful measures are  accessibility and centrality indicators, referred to how easy  is to reach locations or how important certain stops are within a network~\cite{Morency2007193, El-Geneidy2011, Wang2015335,gmtnNAS15}. All these measures are based on some kind of counting queries that determine the number of \textit{distinct} trips that occur within a spatial and/or temporal window.   

Among other types of queries, in this work we focus on the following counting queries, which to the best of our knowledge have not been  addressed by previous proposals. In general terms, we define two general queries, number-of-trips queries and top-k queries, upon which we apply spatial, temporal or spatio-temporal constraint when useful.

%\vspace{-2mm}
\begin{itemize} [leftmargin=6mm]
	\setlength{\itemindent}{0mm}
	\item[(a)] {\em Number-of-trips queries.} This is a general type of queries that counts the number of distinct trips. When applying spatial, temporal or spatio-temporal constraints, it can specialized in the following queries:
	
	\begin{enumerate}
		[leftmargin=3mm]
		\setlength{\itemindent}{0mm}
		\item Pure spatial queries:
		\begin{itemize}
			[leftmargin=3mm]
			\setlength{\itemindent}{0mm}
			\item[-] {\em Number of trips starting at node $X$ (\Sswx).}
			\item[-] {\em Number of trips ending at node $X$ (\Sewx).} 
			\item[-] {\em Number of trips starting at $X$ and ending at $Y$ (\Sfxty).}
			\item[-] {\em Number of trips using or passing through node $X$ (\Sux)}
			%	\medskip
		\end{itemize}
		
		\item Spatio-temporal queries:
		\begin{itemize}[leftmargin=3mm]
			\setlength{\itemindent}{0mm}
			\item[-] {\em Number of trips starting at node $X$ during time interval $[t_1,t_2]$ (\Tswx).}
			\item[-] {\em Number of trips ending at node $X$ during the time interval $[t_1,t_2]$ (\Tewx). }
			\item[-] {\em Number of trips starting at $X$ and ending at $Y$ occurring during  time interval $[t_1,t_2]$ (\Tfxty).} This type of queries is further classified into: (i)  \Tfxty\ with strong semantics (\Tfxtys), which considers trips that completely occur within interval $[t_1,t_2]$. (ii) \Tfxty\ with weak semantics (\Tfxtyw), which considers trips whose life time overlap $[t_1,t_2]$.
			\item[-] {\em Number of trips using node $X$ during the time interval $[t_1,t_2]$ (\Tux).}
		\end{itemize}
		
		\item Pure temporal queries:
		\begin{itemize}	[leftmargin=3mm]
			\setlength{\itemindent}{0mm}
			\item[-] {\em Number of trips starting during the time interval $[t_1,t_2]$ (\Tst). } 
			\item[-] {\em Total usage of network stops during the time interval $[t_1,t_2]$ (\Tut).} 
			\item[-] {\em Number of trips performed during the time interval $[t_1,t_2]$ (\Ttt).} 
		\end{itemize}
		
	\end{enumerate}
	
	\setlength{\itemindent}{0mm}
	\item[(b)] {\em Top-k queries.} In this type of queries we want to retrieve the $k$ nodes with the highest number of trips. In this case, depending on having a temporal constraint or not we include the following queries:
	\begin{enumerate}
		[leftmargin=3mm]
		\setlength{\itemindent}{0mm}
		\item Pure spatial {\em Top-k} queries:
		\begin{itemize}	[leftmargin=3mm]
			\setlength{\itemindent}{0mm}
			\item[-] {\em Top-k most used nodes (\Stk)} that returns the nodes with the largest number of trips passing through. %\ART{Incluye a las trayectorias que parten en esos nodos?}
			\item[-] {\em Top-k most used nodes to start a trip (\Stks)} that returns the nodes with the largest number of trips that start at that node.
			%\item {\em Top-k most used times (\Ttk)} that returns the discrete time units with largest number of trips.
			\medskip
		\end{itemize}
		
		\item Spatio-temporal {\em Top-k}  queries:
		\begin{itemize}[leftmargin=3mm]
			\setlength{\itemindent}{0mm}
			\item[-] {\em Top-k most used nodes during time interval $[t_1,t_2]$(\STtk)} that returns the nodes with the largest number of trips passing through within time interval $[t_1,t_2]$. 
			\item[-] {\em Top-k most used nodes to start a trip during time interval $[t_1,t_2]$(\STtks)} that returns the nodes with the largest number of trips starting there within time interval $[t_1,t_2]$ at that node.
		\end{itemize}
		
	\end{enumerate}
	%\item[(c)] {\em Number of nodes.} 
	%\ART{A mi me parece que esta ultima no es tan relevante. Yo me concentraria en las anteriores y todos sus versiones y ya esta.}
	%
	%\begin{itemize}
	%\item {\em Number of nodes used between $t_1$ and $t_2$ (\Tut)}. 
	%\end{itemize}
	%
\end{itemize}

% %%%%%%%%%%%%%%%%%%%%%%%%%%%%%%%%%%%%%%%%%%%%%%%%%%%%%%%%%%%%%%%%%%%%%%%%%%%%%%%%%%

 \section{Compact Trip Representation ($\repres$)} \label{sec:ctr}

If we consider a network $\mathcal{N}$ with $\sigma_s$ nodes, 
%and we assume time is discretized into  $\sigma_t$ time instants, 
we can see a dataset of trips $\mathcal{T}$ over $\mathcal{N}$ as 
a set of $z$ trips, where for each trip $\mathcal{T}_i$, we represent a list with the $l_i$ 
temporary-ordered nodes it traverses and the corresponding timestamps: 
$\mathcal{T}= \{ \langle (s^i_1, s^i_2, \dots,  s^i_{l_i}),(t^i_1, t^i_2, \dots,  t^i_{l_i}) \rangle\}$, $i\in[1,z]$, 
$s^i_j \in [1,\sigma_s]$, and $t^i_{x} \leq t^i_y, \forall x < y$. 
Note that every node in the network can be identified with an integer ID ($s^i_1$) and that, if we are interested in
analyzing the usage patterns of the network, we will probably be interested in discretizing time into 
time intervals (i.e. 5-min, 30-min intervals). Therefore,
we will have $\sigma_t$ different time intervals that can also be identified with an
integer ID ($t^i_j \in [0,\sigma_t)$).

The size of the time interval is a parameter for the time-discretizing process
that can be adjusted to fit the required precision in each domain.
For example, in a public
transportation network where we could have data including five years of trips, one
possibility would be to divide that five-years period into
10-minute intervals hence obtaining a
vocabulary of roughly $\sigma_t=5\times 365 \times 24 \times 60/10 = 262,800$ different intervals. 
Other possibility would
be to use cyclically annual 10-minute periods resulting in $\sigma_t=262,800 / 5 = 52,560$. 
However,  in public transportation networks, queries such
as \textit{``Number of trips using the stop X on May 10 between 9:15 and 10:00''} may be not 
as useful as queries such as \textit{``Number of trips using stop X on Sundays between 9:15 and
	10:00''}.
% Therefore, it is more useful
%to encode with the same codes the hours in working days on one hand and
%hours in weekend in the other.
For this reason, $\repres$ can adapt how the
time component is encoded depending on the queries that the system must answer.

\begin{figure}[ht]
	%\vspace{-0.4cm}
	\begin{center}
		% {\includegraphics[width=0.5\textwidth]{figures/wt.png}}  %% probar se funciona en local
		{\includegraphics[width=1\textwidth]{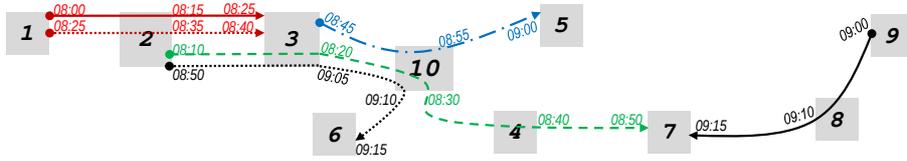}}  %% probar se funciona en local
	\end{center}
	\vspace{-0.2cm}
	\caption{A set of trips over a network with 10 nodes.}
	\label{fig:network}
	%\vspace{-0.5cm}
\end{figure}

\begin{example} \label{exp:ctr}
Figure~\ref{fig:network} shows a network that contains $\sigma_s=10$ nodes 
numbered from $1$ to $10$. Over that network we have six trips ($z=6$),
and, for each of them, we indicate the sequence of nodes it traverses
and the time when the trip goes through those nodes. If we discretize time into
5-minute intervals, starting at 08:00h, and ending at 9:20h, we will have
have $\sigma_t=16$ different time intervals. Any timestamp within
interval $\mathit{[08\!:\!00,08\!:\!05)}$ will
be assigned time-code $0$, those within $\mathit{[08\!:\!05,08\!:\!10)}$ code $1$, and so on until
times within $\mathit{[09\!:\!15,09\!:\!20)}$ that are given time-code $15$.  
Therefore, our dataset of trips will be: 
$\mathcal{T}$: $\{$%
$\langle (\mathbf{1,2,3     })$, $(\mathit{5,7,8})                     \rangle$, 
$\langle (\mathbf{2,3,10,6  })$, $(\mathit{10,13,14,15})           \rangle$, 
$\langle (\mathbf{1,2,3     })$, $(\mathit{0,3,5})                     \rangle$, 
$\langle (\mathbf{2,3,10,4,7})$, $(\mathit{2,4,6,8,10}) \rangle$, 
$\langle (\mathbf{3,10,5    })$, $(\mathit{9,11,12})                     \rangle$, 
$\langle (\mathbf{9,8,7     })$, $(\mathit{12,14,15})                    \rangle$$\}$, 
where bold numbers indicate node IDs and slanted ones indicate times. \qed
\end{example}

In $\repres$ we represent both the spatial and the temporal component of the trips using well-known
self-indexing structures in order to provide both a compact representation and the ability to 
perform fast indexed searches at query time. In Section~\ref{sec:transnet_repr} we focus on the
spatial component and discuss how we adapted  $\csa$ to deal with trips. We also
show how we support spatial queries. Then, in Section~\ref{sec:time_repr} we show that the times,
which are kept aligned with the spatial component of the trips, can be handled with   
a $\wt$-based representation. Actually we study two alternatives (a $\wtht$ and a $\wm$) 
and show how temporal and spatio-temporal (Section~\ref{sec:stq}) queries are supported by $\repres$.

\section{Spatial component of $\repres$}
\label{sec:transnet_repr}
%%%%%%%%%%%%%%%%%%%%%%%%%%%%%%%%%%%%%%%%%%%%%%%%%%%%%%%%%%%%%%%%
%As show in the previous section, the spatial component of each trip $\mathcal{T}_i, (1\leq i< z)$ consists
%in a sequence of integers (node IDs) $s^i_1, s^i_2, \dots,  s^i_{l_i}$. In addition recall that for 
%each node we keep the time $t^i_j, (1\leq j < l_i)$. 

We use a $\csa$ to represent the spatial component of our dataset of trips within $\repres$. Yet,
we perform some preprocessing on $\mathcal{T}$ before building a $\csa$ on it. Initially, we 
sort the trips by their first node ($s^i_1$), then by the last node ($s^i_{l_i}$), then by the starting time
($t^i_1$), and finally, by its second node ($s^i_2$), third node ($s^i_3$), and successive nodes ($s^i_j, 3<j\leq l_i$). 
Note that the start time ($t^i_1$) of the trip does not belong to the spatial component, 
but it is nevertheless used for the sorting.\footnote{This initial sorting of the trips will allow us 
to answer some useful queries very efficiently  (i.e., count trips starting at $X$ and ending at $Y$).} 

Following with Example~\ref{exp:ctr}, after sorting the trips in $\mathcal{T}$ with the criteria above, 
our sorted dataset $\mathcal{T}^s$ would look like: 
$\mathcal{T}^s$: $\{$%
$\langle (\mathbf{1,2,3     })$, $(\mathit{0,3,5})                     \rangle$, 
$\langle (\mathbf{1,2,3     })$, $(\mathit{5,7,8})                     \rangle$, 
$\langle (\mathbf{2,3,10,6  })$, $(\mathit{10,13,14,15})           \rangle$, 
$\langle (\mathbf{2,3,10,4,7})$, $(\mathit{2,4,6,8,10}) \rangle$, 
$\langle (\mathbf{3,10,5    })$, $(\mathit{9,11,12})                     \rangle$, 
$\langle (\mathbf{9,8,7     })$, $(\mathit{12,14,15})                    \rangle$$\}$. 
Note that  $ (\mathbf{2,3,10,6  })$ appears before $(\mathbf{2,3,10,4,7})$ because
during the sorting process we compare $ (\mathbf{2,6,\mathit{2},3, 10,6  })$ with $ (\mathbf{2,7,\mathit{10},3, 10,4,7})$;
that is, we compare the starting nodes ($\mathbf{2}$ and $\mathbf{2}$) and then the ending nodes ($\mathbf{6}$ and $\mathbf{7}$).
If needed  (not in this example) we would have also compared the slanted values ($\mathit{2}$ and $\mathit{10}$) 
that are the starting times of the trips, and finally the rest of nodes  ($ \mathbf{3, 10,6  }$ and $ \mathbf{3, 10,4,7}$).
Similarly, the two trips containing nodes $ (\mathbf{1,2,3})$ are sorted by the starting times ($\mathit{0}$ and $\mathit{5}$).

In a second step, we enlarge all the trips $\mathcal{T}^s_i, 1\!\leq\!i\!<\!z$ with a fictitious terminator-node $\$_i$ whose
timestamp is set to that of the initial node of the trip. We choose terminators such that $\$_i \prec \$_j, \forall i<j$; 
that is the lexicographic value of $\$_i$ is smaller for smaller $i$ values. In addition, the lexicographic value
of any terminator must be lower than the ID of any node in a trip. Therefore, an enlarged trip $\mathcal{T}^s_i$
would become $\mathcal{T}'_i =  \langle (s^i_1, s^i_2, \dots,  s^i_{l_i}, 
\mathbf{\$_i}),(t^i_1, t^i_2, \dots,  t^i_{l_i}, \mathbf{t^i_1}) \rangle$. 

\begin{figure}[h!]
  \vspace{-0.1cm}
  \begin{center}
  {\includegraphics[width=1.00\textwidth]{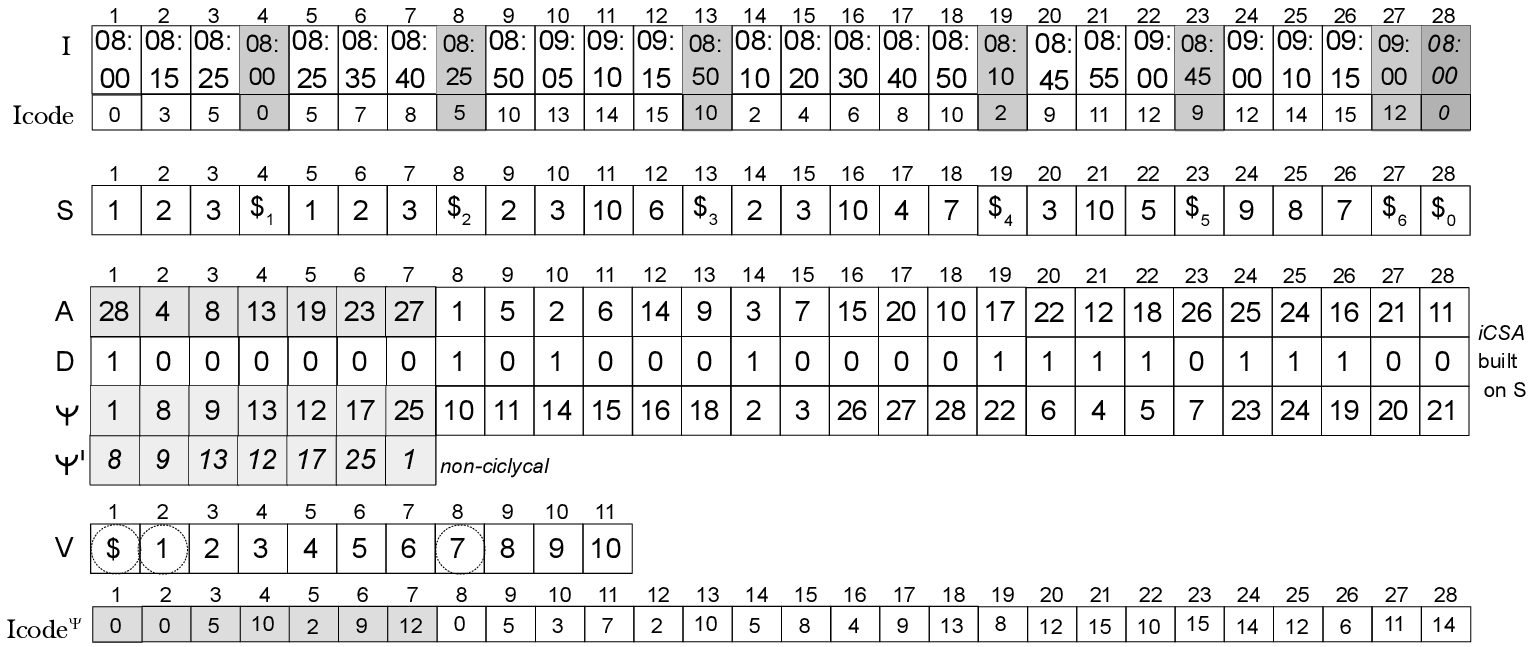}}
  \end{center}
  \vspace{-0.3cm}
  \caption{Structures involved in the creation of a $\tcsa$.}
  \label{fig:tcsa}
  \vspace{-0.2cm}
\end{figure}

The next step involves concatenating the spatial component of all the enlarged trips and to add an 
extra trailing terminator $\$_0$ to create a sequence $S[1,n]$. $\$_0$ must be  lexicographically 
smaller than any other entry in $S$ (then it also holds $\$_0 \prec \$_i$ $\forall i \in [1,z]$). In the top part of
Figure~\ref{fig:tcsa}, we can see array $S$ for the running example, as well as the corresponding time-IDs that
are regarded in sequence $Icode$  ($I$ shows the original times).

Finally, we build a $\csa$ on top of $S$ to obtain a self-indexed representation of the spatial component in $\repres$.
Figure~\ref{fig:tcsa} depicts the structures $\Psi$ and $D$ used by $\tcsa$ built over $S$. There is also a vocabulary
$V$ containing a $\$$ symbol and the different node IDs in lexicographic order.

Note that the use of different values $\$_i$ as terminators ensures that our sorting criteria are kept even if we follow the
standard suffix-sort procedure\footnote{Suffix $S[i,n]$ is compared with suffix $S[j,n]$.} 
required to build suffix array $A$ during the creation of $\csa$. Yet, when we finish that
process, we can replace all those $\$_i$ terminators  by a unique $\$$. This is the reason why 
there is only one $\$$ symbol in $V$. 
 
%while in $V$ there is only one single entry for all the $\$$. 
%The use of different values for $\$_i$ during the sorting explains why $A[22] = 18$ is placed before $A[23]= 26$. Note that
%the suffix starting at $S[18]$ is ``$7 \cdot \$_4 \cdot 2 \cdot 3 \dots$'' and that suffix at
%$S[26]$ is ``$7 \cdot \$_6 \cdot 9 \cdot \dots$''. Therefore, it holds that $A[22] \prec A[23]$. However, considering
%the traditional definition of a {\em suffix}, these suffixes would be ``$7 \cdot \$_4 \cdot 3 \cdots $''
%and ``$7 \cdot \$_6 \cdot \$_0 \cdots $'' respectively,  and  $A[22] \prec A[23]$ would not hold.

Although they are not needed in $\repres$, we show also suffix array $A$ and $\Psi$' for clarity reasons in Figure~\ref{fig:tcsa}. 
$\Psi'$  contains the first entries of $\Psi$ from a regular $\csa$, whereas we introduced a small variation
in $\repres$ for entries $\Psi[1,z+1]$.  %just to explain the difference of how we build $\Psi$. 
For example, $A[8]=1$ points to the first node of the first trip $S[1]$.
$\Psi[8]=10$ and $A[10]=2$ point to the second node.  $\Psi[10]=14$ and $A[14]=3$ point to the third node.
$\Psi[14]=2$ and $A[2]=4$ point to the ending $\$_1$ of the first trip. Therefore, in the standard 
$\csa$, $\Psi'[2]=9$ and $A[9]=5$  point to the first node of the second trip. 
However, in  $\repres$, $\Psi[2]=8$ and $A[8]=1$ point
to the first node of the first trip. With this small change, subsequent applications of $\Psi$ will allow 
us to cyclically traverse the nodes of the trip instead of accessing the following entries of $S$.

%Finally, note that aligned with sequence $S$, we could keep the times associated with the nodes in each
%trip with the structures $I$ and $Icode$, which are explained in the following subsection. \marginpar{\tiny YA NO ES SUBSECTION}

%In  Figure~\ref{fig:tcsa}, $\Psi'$  contains
%the first entries of $\Psi$ from a regular \csa. For example, $A[8]=1$ points to the first stop of the first trip $S[1]$.
%$\Psi[8]=10$ and $A[10]=2$ point to the second stop.  $\Psi[10]=14$ and $A[14]=3$ point to the third stop.
%$\Psi[14]=2$ and $A[2]=4$ point to the ending $\$$ of the first trip. Therefore, $\Psi'[2]=9$ and $A[9]=5$ point
%to the first stop of the second trip. However, we managed to modify $\Psi$ so that $\Psi[2]=8$ and $A[8]=1$ point
%to the first stop of the first trip. Thus, subsequent applications of $\Psi$ will allow us to cyclically
%traverse the stops of each trip. This will be interesting at query time.

Another interesting property arises from the use of a cyclical $\Psi$ on trips, and from using trip terminators.
Since the first entries in $\Psi[2,z+1]$ correspond the $\$$ symbols that 
mark the end of each trip in $S$ (remember that $\Psi[1]$ corresponds the $\$_0$), we
can see that the $j^{th}$ node of the $i^{th}$ trip can
be obtained as $V[rank_1(D, \Psi^j[i+1])]$, (where $\Psi^3[x]= \Psi[\Psi[\Psi[x]]]$). This property
makes it very simple to find starting nodes for any trip.
For example, if we focus on the shaded area $\Psi[2,7]$, we can find the ending terminator $\$_4$ of the
fourth trip at the $5^{th}$ position (because the first $\$_0$
corresponds to the final $\$$ at $S[28]$). Therefore, its starting node can be found 
as $V[rank_1(D, \Psi[4+1])]$. Since $\Psi[5] = 12$ and $rank_1(D,12)= 3$, 
the starting node is $V[3]=\mathbf{2}$. For illustration purposes note that it would correspond to $S[A[12]]$.
By applying $\Psi$ again, the next node of that trip would be obtained by computing $\Psi[12] = 16$, 
$rank_1(D,16)=4$, and accessing $V[4]=\mathbf{3}$  (that is, we have obtained 
 $V[rank_1(D, \Psi[\Psi[4+1]])]=\mathbf{3}$, and so on.

%In addition, note that in the shaded range $\Psi[1,7]$, the first entry is
%related to terminator $\$_0$, whereas the next six entries  correspond
%to the $\$$ symbols that mark the end of each trip in $S$. 
%%(sorted by the starting stop, then by the ending stop, then by their initial time, 
%%and finally by the second, third and following up stops).
%This property makes it very simple to find
%starting %and ending
%nodes. For example, the ending $\$_4$ of the
%$4^{th}$ trip is at the $5^{th}$ position (because the first $\$_0$
%corresponds to the final $\$$ at $S[28]$). Therefore, its starting
%node can be obtained by  $\Psi[5] = 12$ and $rank_1(D,12)= 3$;
%that is, the starting node is the $3^{th}$ entry in the vocabulary.
%The next node of that trip would be obtained by $\Psi[12] = 16$
%and $rank_1(D,16)=4$, and so on. 
%In practice, the $j^{th}$ node of the $i^{th}$ trip can
%be obtained as $rank_1(D, \Psi^j(i+1))$. \marginpar{\tiny verificar que formula este ok}

Regarding the space requirements of the $\csa$ in $\repres$, we can expect to obtain a good compressibility
due to the structure of the network, and the fact that trips that start in a given node or simply
those going through that node will probably share the same sequence of ``next'' nodes. This will
lead us to obtaining many {\em runs} in $\Psi$~\cite{NM07}, and consequently, good compression.

\subsection{Dealing with Spatial Queries}
\label{sec:sq}

With the structure described for representing the spatial component of the trips,
the following queries can be solved.

%\vspace{-2mm}
\begin{itemize}[leftmargin=3mm]
\setlength{\itemindent}{0mm}
\item {\em Number of trips starting at node $X$ (\Sswx).}
Because $\Psi$ was cyclically built in such a way that every $\$$ symbol is followed by the first node 
of its trip, this query is solved by $[l,r]~\leftarrow~bsearch(\$X)$ over the $\csa$, 
which results on a binary search for the pattern $\$X$ over the section $\Psi[2,z+1]$ corresponding to $\$$ symbols. 
Then $r-l+1$ gives the number of trips starting at $X$.

\item {\em Number of trips ending at node $X$ (\Sewx).} In a similar way to the previous query, 
this one can be answered with $bsearch(X\$)$.

\item {\em Number of trips starting at $X$ and ending at $Y$ (\Sfxty).}
Combining both ideas from above, and thanks to the cyclical construction of $\Psi$, this query is solved 
using $bsearch(Y\$X)$.

\item {\em Number of trips using node $X$ (\Sux).}
%Instead of performing a binary search over $\Psi$, we can operate on bitmap $D$. 
Even though we could solve this query with $bsearch(X)$, it is more efficient to solve it by directly operating on $D$. 
 Assuming that $X$ is
at position $p$ in the vocabulary $V$ of  $\repres$ ($V[p]=X$), its total frequency is obtained by
$occs_X \leftarrow select_1(D,p+1) - select_1(D,p)$. %That is, we find the range in $D$ corresponding to the node $X$.
If $p$ is the last entry in $V$, we set $occs_X \leftarrow n+1-select_1(D,p)$.

\setlength{\itemindent}{0mm}
\item {\em Top-k most used nodes (\Stk).}
We provide two possible solutions for this query named: %These queries have two possible implementations:
sequential and binary-partition approaches. 

\begin{itemize}
\item To return the $k$ most used nodes using {\em sequential approach (\Stkseq)}. The idea is
to apply  $select_1$ operation sequentially for every node from $2$ to $|V|$ to compute the 
frequency of each node and to return the $k$ nodes with highest frequency.
We use a min-heap that is initialized with
the first $k$ nodes, and for every node $s$ from $k+1$ to $|V|$, 
we compare its frequency with that of the minimum node (the root) from
the heap. In case the frequency of $s$ is higher, the root of the heap is replaced by $s$ and
then moved down to comply with the heap ordering. At the end of the process, the heap
will contain the top-k most used nodes $\langle p_1\:,\:p_2,\:\dots,\:p_k \rangle$, which can be 
sorted with the heapsort algorithm if needed. Finally, we return $\langle V[p_1],V[p_2],\dots,V[p_k]\rangle$.
Note that this  approach always performs $|V|$ $select_1$ operations on $D$.

\item The {\em binary-partition (\Stkbin)} approach takes advantage of a skewed 
distribution of frequency of the nodes that trips traverse.  Working over $D$ and $V$, we 
recursively split $D$  into two segments after each iteration. 
If possible, we leave the same number of different nodes in each side of the partition. 
Initially, we start considering the range in $D[l,r] \leftarrow D[select_1(D,2),n]=D[z+2,n]$ 
which corresponds to the nodes that appear in 
$V$ from positions $i=2$ to $j=|V|$.\footnote{We skip the $\$$ at the first entry of $V$ and its corresponding 
entries in $D$; that is, $D[1,select_1(D,2)-1]$.}
%	$D$ (without its initial range 	%of $|V|$ $\$$ symbols).
We use a priority queue that is initialized as $Q \leftarrow (\langle i,j\rangle, \langle l,r\rangle)$.
Then, assuming $m=i + \frac{j-i+1}{2}$ and $q=select_1(D,m)$, we create two partitions 
$D[l, q-1]$ and  $D[q, r]$, which correspond respectively to the nodes in $V[i,m-1]$ and $V[m, j]$.
These  segments created after the partitioning step are
pushed into  $Q$. %, storing the initial and the final positions of the segment in $D$,
%and also the initial and final corresponding positions in $V$. 
The pseudocode can be found in  Figure~\ref{fig:topk_nieves}.

The priority of each segment in $Q$ is
directly the size of its range in $D$ ($r-l+1$). 
%The priority queue $Q$ is initialized with a segment covering the whole $D$ (without its initial range
%of $|V|$ $\$$ symbols). 
When a segment extracted from $Q$ represents the instance of only one node ($(\langle i,j\rangle, \langle l,r\rangle)$, with $i=j$),
that node is returned as a result of the top-k algorithm (we return $V[i]$). The algorithm stops when the first $k$
nodes are found.

For example, when searching for the top-1 most used nodes in the example from Figure~\ref{fig:tcsa}, $Q$ is initialized with
the segment $[8, 28]$, corresponding to nodes from~1~to~10 (positions from~2~to~11 in $V$). Note
that the entries of $D$ from~1~to~7 and $V[1]$ represent the $\$$ symbol. Since it is not an actual node, it
 must be skipped. Then $[8, 28]$ is split producing the segments $[8, 20]$ for nodes 1~to~5 ($V[2,6]$)
and $[21, 28]$ for nodes~6~to~10 ($V[7,11]$). After three more iterations, we extract
$(\langle 3,3\rangle, \langle 14,18\rangle)$, hence obtaining the segment $[14, 18]$ for
the single node 3 (position $4$ in $V$), concluding that the  {\em Top-1 most used node} is 
$\mathbf{3}=V[4]$ with a frequency equal to $5=18-14$.

%\marginpar{\tiny algorithm ten que comezar metendo $<2,|V|>$, non $1,V$}
%\marginpar{\tiny $q$ debe ser $select_1(D,m)$, non $select_1(D,m+1)$}

%	\begin{figure}[t]
%	%\vspace{-0.5cm}
%	\begi%n{center}
%	\begin{minipage}{0.5\textwidth}
%	\begin{code}
%	\textbf{GetTopK} $(k)$: \\
%	 \> $ Q ~\leftarrow$ \textbf{new PriorityQueue()}; \\
%	 \> $ Q .$\textbf{push$(<1, |V|>, <$select$_1(D,2), n-1>)$}; \\
%	 \> $ current\_k ~\leftarrow 0$; \\%\\
%	
%	 \> \textbf{while }$current\_k < k$: \\
%	 \> \> $ (<i,j>, <l,r>) ~\leftarrow ~Q.$\textbf{pop$()$}; \\%\\
%	
%	  \> \> \textbf{if} $i = j$: \\
%	 \> \> \> $ topK[current\_k] ~\leftarrow i$; \\
%	 \> \> \> $ current\_k ~\leftarrow current\_k + 1$; \\
%	 \> \> \textbf{else}: \\
%	 \> \> \> $ m ~\leftarrow i + \frac{j - i + 1}{2}$; \\
%	 \> \> \> $ q ~\leftarrow$ \textbf{select$_1(D,m + 1)$}; \\
%	 \> \> \> $ Q .$\textbf{push$(<i, m-1>, <l, q-1>)$}; \\
%	 \> \> \> $ Q .$\textbf{push$(<m, j>, <q, r>)$}; \\%\\
%	 %
%	 \> \textbf{return} $topK$; \\
%	\end{code}
%	\end{minipage}
%	\end{center}
%	\vspace{-0.2cm}
%	\caption{Top K most used nodes implementation using binary partitions}
%	\label{fig:topk_nieves}
%	%\vspace{-0.5cm}
%	\end{figure}

\end{itemize}

\begin{figure}[t]
	%\vspace{-0.5cm}
	\begin{center}
		\begin{minipage}{0.70\textwidth}
			\begin{code}
				\textbf{GetTopK\_most\_used\_nodes} $(k)$: \\
				\>(l.1~) \> ~$ Q ~\leftarrow$ \textbf{new PriorityQueue()}; \\
				\>(l.2~) \> ~$ Q .$\textbf{push$(\langle2, |V|\rangle, \langle$select$_1(D,2), n\rangle)$}; \\
				\>(l.3~) \> ~$ current\_k ~\leftarrow 0$; \\%\\
				
				\>(l.4~) \> ~\textbf{while }$current\_k < k$: \\
				\>(l.5~) \> ~\> $ (\langle i,j\rangle, \langle l,r\rangle) ~\leftarrow ~Q.$\textbf{pop$()$}; \\%\\
				
				\>(l.6~) \> ~\> \textbf{if} $i = j$: \\
				\>(l.7~) \> ~\> \> $ topK[current\_k] ~\leftarrow V[i]$; \\
				\>(l.8~) \> ~\> \> $ current\_k ~\leftarrow current\_k + 1$; \\
				\>(l.9~) \> ~\> \textbf{else}: \\
				\>(l.10) \> ~\> \> $ m ~\leftarrow i + \frac{j - i + 1}{2}$; \\
				\>(l.11) \> ~\> \> $ q ~\leftarrow$ \textbf{select$_1(D,m + 1)$}; \\
				\>(l.12) \> ~\> \> $ Q .$\textbf{push$(\langle i, m-1 \rangle, \langle l, q-1 \rangle)$}; \\
				\>(l.13) \> ~\> \> $ Q .$\textbf{push$(\langle m, j \rangle, \langle q, r \rangle)$}; \\%\\
				\>(l.14) \> ~\textbf{return} $topK$; \\
			\end{code}
		\end{minipage}
	\end{center}
	\vspace{-0.3cm}
	\caption{Algorithm {\em Top-k most used nodes} using binary-partition approach.}
	\label{fig:topk_nieves}
	%\vspace{-0.5cm}
\end{figure}

\item {\em Top-k most used nodes to start a trip (\Stks).}
Both \Stk\ approaches above can be adapted for answering \Stks.
However, unlike its simpler variant, it requires performing $bsearch(\$X)$ over $\Psi$ (rather than a $select$ on $D$) at
each iteration, hence increasing the temporal complexity of the operation.

The implementation of the linear approach is straightforward. The binary-partition approach differs slightly 
from the algorithm in Figure~\ref{fig:topk_nieves}: in (l.2) we insert $(\langle 2, |V| \rangle, \langle 2,z+1 \rangle)$ into $Q$, and we 
replace (l.11) with $[x,y]~\leftarrow~bsearch(\$V[m])$; $q \leftarrow x$.

%interval covers only the range corresponding to $\$$ symbols 
%(therefore $\langle l,r\rangle \leftarrow \langle 1,z+1\rangle$), and $q$ is computed as $[q,\_]~\leftarrow~bsearch(\$V[m])$.

%Following the same idea, one could also support querying for the Top-k nodes used to end a trip. 

\end{itemize}

\subsection{Implementation details} In our implementation of $\csa$, we used the 
$\icsa$\footnote{\url{http://vios.dc.fi.udc.es/indexing}} from \cite{FBNCPR12} briefly discussed 
in Section~\ref{sec:csa}. Yet, we introduced some small modifications:
 
\begin{itemize}
	\item The construction of the Suffix Array $A$ is done with 
	{\em SA-IS} algorithm~\cite{nong2011two}.\footnote{\url{ https://sites.google.com/site/yuta256/sais}} 
	In comparison with the  {\em qsufsort} algorithm\footnote{
		http://www.larsson.dogma.net/research.html}
	%\footnote{https://github.com/y-256/libdivsufsort/}
	\cite{Larsson:2007:FSS:1314704.1314853} used in the original $\icsa$, it achieves a linear time construction 
	and a lower extra working space. 
	
	%\item The construction of the Suffix Array is done with 
	%{\em SA-IS} algorithm\footnote{\url{ https://sites.google.com/site/yuta256/sais}} ~\cite{nong2011two}, 
	% which achieves a linear time construction 
	%and a low memory footprint during construction, when compared with algorithm {\em qsufsort}\footnote{
	%http://www.larsson.dogma.net/research.html} \cite{Larsson:2007:FSS:1314704.1314853} used in \cite{FBNCPR12}.  \marginpar{\tiny Daniil cando mais rapido e canta menos memoria?}
    %
	\item In  $\icsa$, they used a plain representation for bitvector $D$ and additional structures to support
	$rank_1$ in constant time using ($0.375\times n$ bits). With that structure, they could solve $select$ in $O(\log n)$ time (yet 
	they did not actually needed solving $select$ in $\icsa$).
	In our $\csa$, we have used the {\em SDArray} from \cite{okanohara2007practical} to represent $D$. It provides a very 
	good compression for sparse bitvectors, as well as constant-time $select$ operation.
	\item In \cite{FBNCPR12}, $bsearch$ operation was implemented with a simple binary search over $\Psi$ rather than
	using the  backward-search optimization proposed in the original $\csa$ \cite{Sad03}. In our experiments, we used
	backward search since it led to a much lower performance degradation at query time when a sparse sampling of $\Psi$ 
	was used.
	
	%We implemented a backward search for the $bsearch$ operation in the $\csa$. When the pattern $P[1,p]$ is searched, 
	%we start by locating the ranges for $P[p]$ and $P[p-1]$ in $D$. After that, we binary search over $\Psi$ for the symbols
	%in $P[p-1]$ that point to $P[p]$, getting a narrower range. We repeat it recursively until reaching $P[1]$. The main
	%advantage of the backward search in this situation is that we have more control on the accesses on the compressed 
	%$\Psi$, making it possible to search over the samples first, and narrowing that search down to a single compressed 
	%block in each cell, instead of having to rely on random access for long patterns.

\end{itemize}

\section{Temporal component of $\repres$}
\label{sec:time_repr}

In this section we focus on the temporal component associated with each node of the enlarged trips $\mathcal{T}'_i$ 
in our dataset. Recall that in Figure~\ref{fig:tcsa}, sequence $I$ contains the time
associated with each node in a trip, and $Icode$ a possible encoding of times. 
In $\repres$ we focus on the values in $Icode$, yet, since $S$ is not kept anymore in $\repres$, we 
reorganize the values in $Icode$ to keep them aligned with $\Psi$ rather than with $S$. Those
values are represented within array $Icode^{\Psi}$ in Figure~\ref{fig:tcsa}. 
For example, we can see that $Icode^{\Psi}[4]$ corresponds with $Icode[A[4]]=10$, 
$Icode^{\Psi}[15]$ corresponds with $Icode[A[15]]=8$, and so on.
% are aligned with $\Psi$ rather than with $S$.%, and represented with a $\wm$.

Aiming at having a compact representation of $Icode^{\Psi}$ while permitting fast 
access and resolution of range-based queries (that we could use to search for trips within 
a given time interval), we have considered two $\wt$-based alternatives from the ones 
presented in Section~\ref{sec:wt}:

\begin{figure}[htb]
	\vspace{-0.2cm}
	\begin{center}
		%{\includegraphics[width=1.0\textwidth]{figures/WM.eps}}
		{\includegraphics[width=1.0\textwidth]{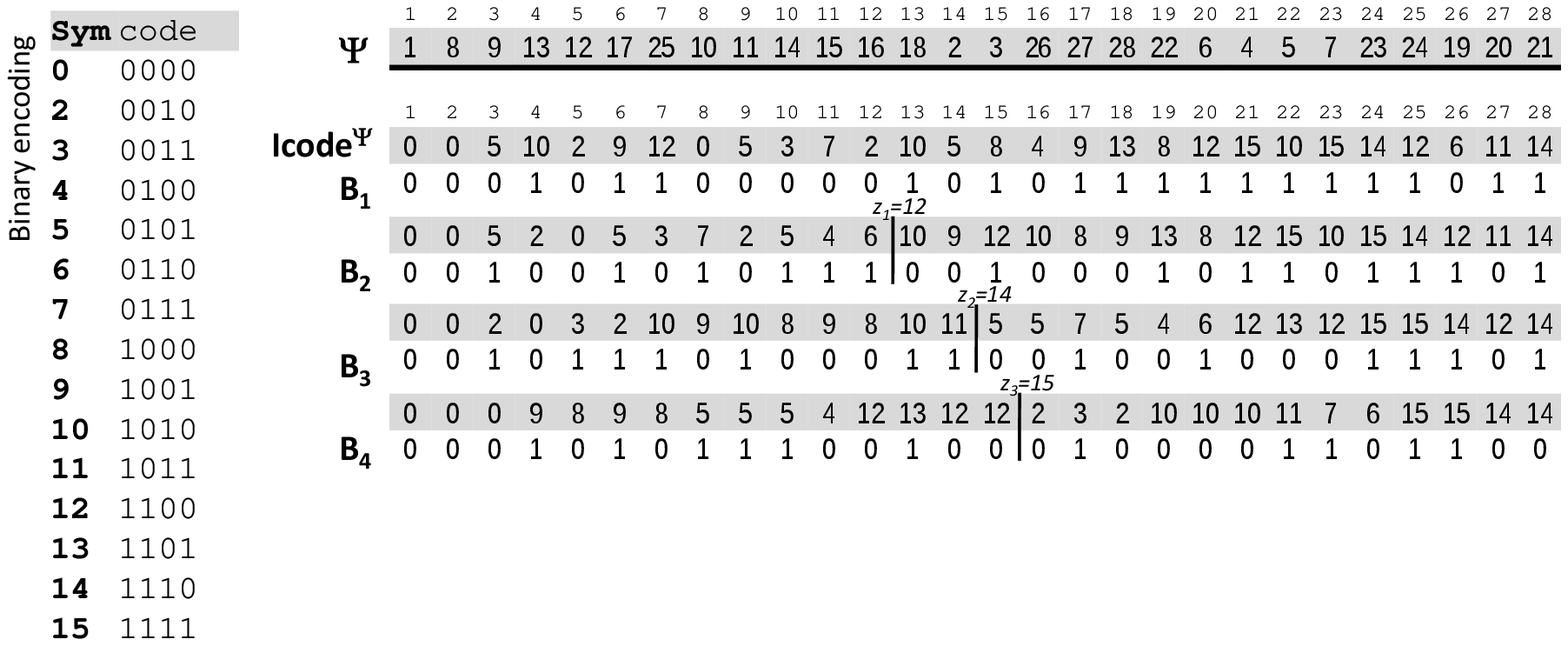}}
		{\includegraphics[width=1.0\textwidth]{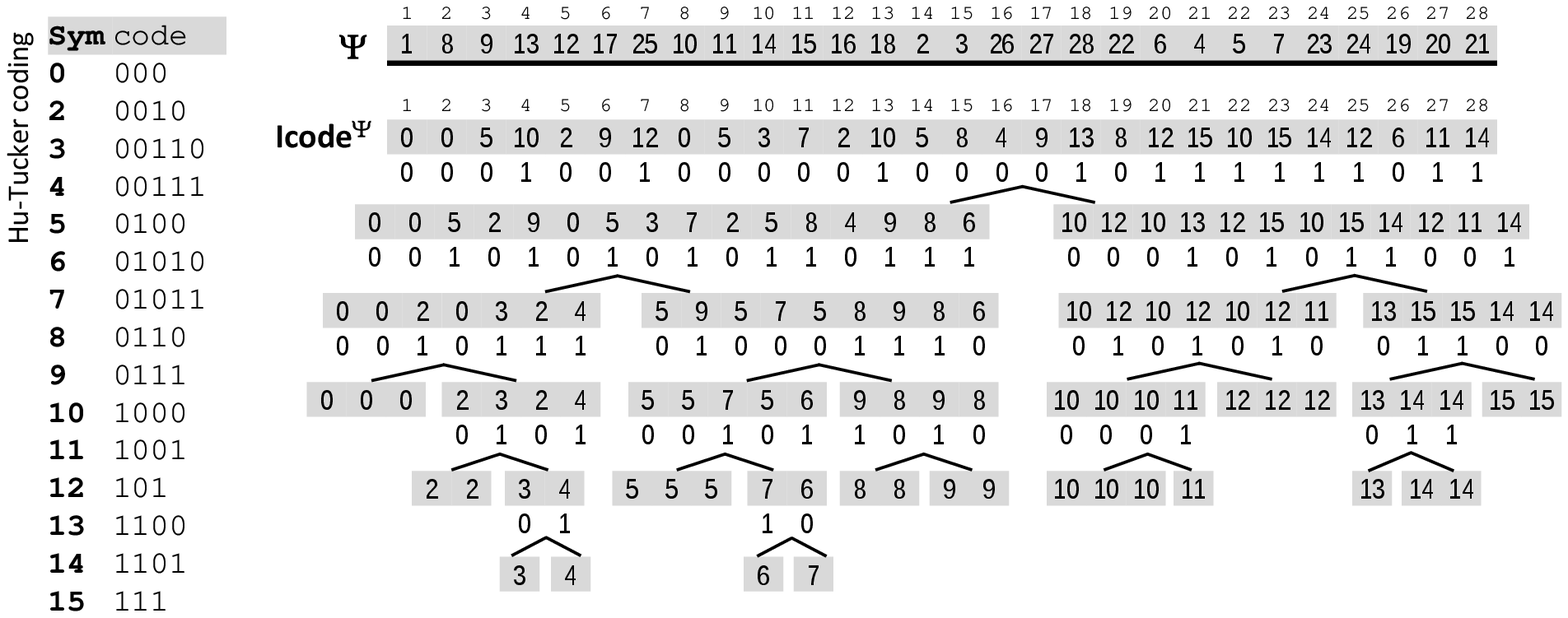}}
	\end{center}
	\vspace{-0.4cm}
	\caption{Balanced $\wm$ (top) and Hu-Tucker-shaped $\wt$ (bottom) for the times within $Icode^{\Psi}$ in Figure~\ref{fig:tcsa}.}
	\label{fig:wm}
	\vspace{-0.1cm}
\end{figure}

\begin{itemize}
  \item A Wavelet Tree ~\cite{WT03} using variable-length Hu-Tucker codes~\cite{hu1971optimal} ($\wtht$).
 % which is an optimal prefix code that preserves the order of the input vocabulary.
  %That means that the lexicographic order of the output variable-length binary codes is the same as the order of the input symbols.
  %Using this codification in a $\wt$ allows us to have a variable number of levels in the tree for each symbol,
  %while preserving some interesting properties that will be explained further in this section.
  Recall this is the $\wt$ variant that permits to compress the original symbols with variable-length codes and 
  still supports $count$ operation in $O(log(\sigma_t))$ time. Since Hu-Tucker coding assigns shorter codes
  to the most frequent symbols, the compression of our $\wtht$ is highly dependent of the distribution
  of frequencies of the values in  $Icode^{\Psi}$. Yet, if our 
  trips represent movements of single users in a transportation network, we could expect to observe two or more periods 
  corresponding to rush hours within a single day. This would lead to obtaining a skewed distribution of the frequencies 
  for the symbols in $Icode^{\Psi}$, and consequently, we could expect to have better compression than if we used a 
  balanced $\wt$. The expected number of bits of our $\wtht$ is $nH_0(Icode^{\Psi})$.

  \item A balanced Wavelet Matrix ($\wm$)~\cite{CNO15}. As we showed in Section~\ref{sec:wt} the $\wm$ is typically the most
  compact uncompressed variant of $\wt$ and it is faster than a pointerless $\wt$. This is the reason why we chose
  a balanced $\wm$ instead of a balanced $\wt$ as this second alternative. Recall that, $Icode^{\Psi}$ contains $n$ symbols, 
  and each symbol can be encoded with $\log  \sigma_t$ bits, hence the 
  balanced $\wm$ will be a matrix of $n  \log\sigma_t$ bits.
  
  %In this case, the symbols
  %in $Icode^{\Psi}$ are given a fixed-length binary code.
  %Recall that a $\wm$ is a grid of $n \times m$ bits. In  our
  %case $n$ is the number of entries in the $\csa$ and $m = \log  \sigma_t$ are the bits
  %needed to represent the different $\sigma_t$ codes for the time instants of interest.
  %This approach grants better compression for large alphabets when compared to the $\wt$, as the $\wm$ does not need
  %to store any kind of pointers.
\end{itemize}

In Figure~\ref{fig:wm}, we show both the $\wm$ and the $\wtht$ built on top of $Icode^{\Psi}$ from Figure~\ref{fig:tcsa}.
The binary code-assignment to the source symbols in $[1,\sigma_t]$ and that obtained after applying Hu-Tucker encoding
algorithm~\cite{hu1971optimal} are also included in the figure.

\subsection{Dealing with Temporal queries}
\label{sec:tq}
With either one of the described alternatives ($\wtht$ or $\wm$) to represent time intervals
we can answer the following pure temporal queries:

\begin{itemize}[leftmargin=3mm]
\setlength{\itemindent}{0mm}
%\item {\em Number of nodes used at instant $t$}. This is computed as $rank_t(n) -rank_t(z+1)$. 
%That is, to discard the occurrences of $t$ in the $\$$ zone, we count the occurrences of instant $t$ in $\wt$, and
%then we subtract $rank_t(z+1)$, where $|\mathcal{T}|$ is the number of trips.

\item {\em Number of trips starting during the time interval $[t_1,t_2]$ (\Tst).} Since we keep the
starting time of each trip within $Icode^{\Psi}[1,z]$, we can efficiently solve this query 
by simply computing $count(1,z,t_1,t_2)$.

\item  {\em Total usage of network stops during the time interval $[t_1,t_2]$ (\Tut).} This query
can be seem as the sum of the number of trips that traversed each network node during $[t_1,t_2]$.
We can solve this query by computing $count(z+1,n,t_1,t_2)$. 

\item {\em Number of trips performed during the time interval $[t_1,t_2]$ (\Ttt).} This is also an
interesting query that permits to know the actual network usage during a time interval. 
To solve this query
we could compute {\em \Ttt} by subtracting the number of trips that started after $t_2$ 
(\textit{\Tst}$(t_2+1,\sigma^t-1) $) and the number of trips that ended 
before $t_1$ (\textit{\Tet}$(0,t_1-1) $) from the total number of trips ($z$). 
However, recall that
$Icode^{\Psi}[1,z]$ has the starting time of each trip, but we do not keep their ending time.
We could solve \textit{\Tet}$(0,t_1-1)$ by taking the first node ($X$) of each trip starting
before $t_1$, then applying $\Psi$ until reaching the ending node ($Y$), and finally getting the ending
time of that trip associated to node $Y$. However, this would be rather inefficient.
A possible solution to efficiently solve \textit{\Tet}$(0,t_1-1)$, would require to increment our temporal
component, in parallel with $Icode^{\Psi}[1,z]$,  with another $\wt$-based representation of the 
ending times for our $z$ trips. This would permit to report the number of trips
ending before $t_1$ as $count'(1,z,0,t_1-1)$, but would increase the overall size of $\repres$.
Yet, note that even without keeping ending-times, we could 
provide rather accurate estimations of \Ttt\ for a system administrator. For example, using \Tut\
to compute the number of times each trip went through any node during the time interval $[t_1,t_2]$, 
and dividing that value by the average nodes per trip. Another good estimation can also be obtained with 
{\em\Tst$(t_1,t_2)$}.

% 
% \item {\em Top-k most used times (\Ttk)}. We can follow a similar procedure to that used to solve the {\em Top-k most used
% 	nodes} discussed in Section~\ref{sec:sq}. We can use both the simpler sequential or a binary-partition approach.
% For the binary-partition approach we proceed similarly as in Figure~\ref{fig:topk_nieves}. In this case,
% we start with the root node in the priority queue, and recursively split the
% extracted node $v$ (whose bitvector is $B_v$) with $n_0 \leftarrow rank_0(B_v,|B_v|)$, 
% obtaining two new nodes $v_0$ and $v_1$: one for the zeros and another for the ones. The priority of $v_0$ and $v_1$ 
% depends on the size of the bitmaps in the nodes, for $v_0$ it is $n_0$, whereas for 
% $v_1$ it is $|B_v| - n_0$. The process stops when the first $k$ leaves are extracted from the queue.
% 
% As we will show in Section~\ref{sec:exp:temp}, the binary-partition approach is preferred when the distribution of times 
% is skewed and the queried $k$ is not large.  Otherwise, the sequential approach is typically preferred. 

\end{itemize}

\subsection{Implementation details} We include here details regarding how we tune our $\wtht$ and $\wm$.
As we discussed in Section~\ref{sec:wt}, both $\wtht$ and $\wm$ are built over bitvectors that require
support for $rank$ and $select$ operations. In our implementations we included two alternative bitvector representations avaliable
at {\em libcds} library:{\footnote{\url{https://github.com/fclaude/libcds}}}

%A plain bitvector \cite{Mun96} with additional structures (extra $o(n)$ bits) to support $rank$ and $select$, and a compressed
%bitvector using RRR technique~\cite{Raman:2002:SID:545381.545411}.

\begin{itemize}
	\item A plain bitvector based on \cite{Mun96} named {\em RG} with 
	additional structures to support $rank$ in constant time ($select$ in logaritmic time).
	{\em RG} includes a sampling parameter ($factor$) that we set to value $32$. In this case,
	our  bitvector {\em RG} uses $n (1+1/32)$ bits. That is, we tune {\em RG} to use a sparse sampling. 
		
	\item A compressed RRR bitvector~\cite{Raman:2002:SID:545381.545411}. The {\em RRR} implementation includes
	a sampling parameter that we tune to values $32$, $64$, and $128$. Higher sampling values achieve better compression.
	
\end{itemize}
%The configuration using {\em RG} always used more space than those using {\em RRR}, and as indicated above,
%$size(RRR_{32}) > size(RRR_{64}) > size(RRR_{128})$.

In advance, when presenting results for $\wtht$ and $\wm$ we will consider the four bitvector configurations
above. Regarding our implementations of $\wm$ and $\wtht$, note that we reused the same implementation of $\wm$ from \cite{CNO15}, 
and we created our custom $\wtht$ implementation, paying special focus at solving $count$ efficiently. 
% %
%\OJOFARI{ 
%Furthermore, in the implementation of \Ttk, we used the simpler linear approach for $\wm$.\footnote{Since linear approach requires 
%keeping track of the position in which each symbol starts and ends in the last level, 
%we explicitly keep pointers that were originally used to speed-up $select$ in \cite{CNO15}.}  However, in our $\wtht$ 
%we also implemented the \Ttk\  using a binary partition approach.
%This allows us to analyze the advantage of linear approach when the distribution is uniform.  
%}
%\marginpar{\tiny ahora que tenemos top-k-bin y top-k-seq para WM Y WTHT, ya creo que mejor no comentar esto, ni lo de la footnote ... no??}

\subsection{Comparing the space/time trade-off of $\wm$ and $\wtht$}
\label{sec:time_exp}
In order to compare the efficiency of our $\wtht$ (that uses variable-length codes and supports $count$ efficiently)
 with a balanced $\wm$ alternative under 
different time distributions (recall that this $\wm$ is time distribution invariant), 
we run some experiments that evaluate the average time to execute $count$ operation on
both representations.

\begin{figure}[tbp]
	\begin{center}
		\includegraphics[angle=-90,width=0.75\textwidth]{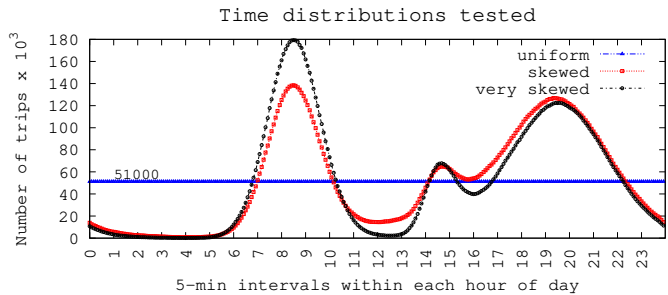}		
		\caption{Time distributions used. The y-axis indicates the number of passengers per each 5-min interval.}
		\label{fig:distrib}
	\end{center}
	\vspace{-3mm}
\end{figure}

We used a dataset of generated trips  (Refer to Section~\ref{sec:ed} for the details about Madrid dataset) and we
generated three kinds of time distributions for our evaluation. We refer to them as: uniform, skewed and very skewed. They are shown
in Figure~\ref{fig:distrib}. 
According to the total number of passengers in a day, in the uniform distribution $51,\!000$ passengers 
use the network for each 5-minute interval. 
We also generated a skewed distribution for the time interval frequencies in an effort to
model the usage of a public transportation network in a regular working day, where the starting time of a trip
is generated according to the following rules:
\begin{itemize}
	\item With 30\% of probability, a trip occurs during a morning rush hour.
	\item With 45\% of probability, a trip occurs in an evening rush hour.
	\item With 5\% of probability, a trip occurs during lunch rush hour.
	\item The remaining 20\% of probability is associated to unclassified trips, starting at a random hour of the day, which may
	also fall into one of the three previous periods discussed.
\end{itemize}
In the very skewed distribution we increase the rush-hour probabilities with
40\% for the morning rush hour, 50\% for the evening rush hour, 8\% for lunch period and only
2\% of random movements.
\medskip

Then we built the $\wtht$ and the $\wm$ considering two different granularities for the discretization of times: 
five-minute and thirty-minute intervals. Then, we generated $10,\!000$ random intervals of times $[t_1,t_2]$ over the whole 
time sequence of the dataset considering interval widths of five minutes, one hour, and six hours.  
Finally, we run $10,\!000$  $count(z+2,n,t_1,t_2)$ queries (we show average times) from each query set over 
the six configurations of $\wtht$ and $\wm$  
(2 different granularities for the time discretization and 3 datasets).

%We made a custom $\wt$ implementation with Hu-Tucker coding, where we focused in optimizing the $count$ operation. We also implemented the $top-k$ mentioned in Section~\ref{sec:tq} using a binary partition approach.

%For the $\wm$ we reuse the implementation from \cite{CNO15}, with $top-k$ implemented with a simple linear approach, using leaf pointers that were already stored in the structure. That allows us to analyze the advantage of the linear when the distribution is uniform.

%F%our different bitvectors were used to evaluate our representations, sorted by the compression achieved in all experiments:
%\begin{itemize}
%  \item Plain bitvector described in \cite{Mun96} with constant time rank
%  and a logarithmic time binary search for select with a sampling parameter that
%  was set to the typical value of 32.
%  % a block size of 32 integers of 32 bits each (1024 bits blocks),
%  % that does not use any kind of compression.
%
%  \item A compressed RRR bitvector~\cite{Raman:2002:SID:545381.545411} with %a block size of 15 bits and
%  three sampling rates: every 32, 64 and 128 blocks. Higher sampling achieves better compression.
%\end{itemize}

\begin{figure}[htb]
	\begin{center}
		\includegraphics[angle=-90,width=0.45\textwidth]{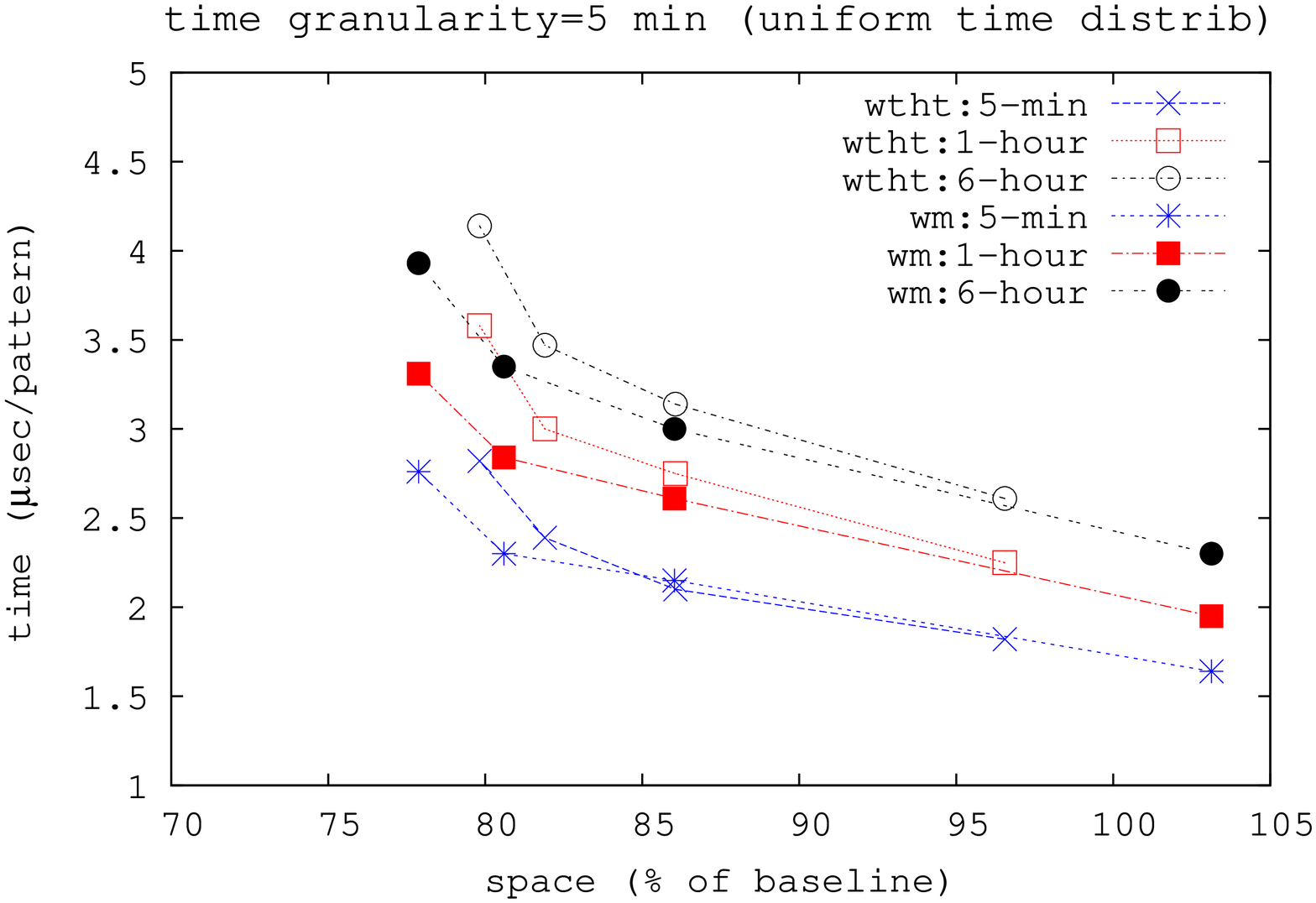}
		\includegraphics[angle=-90,width=0.45\textwidth]{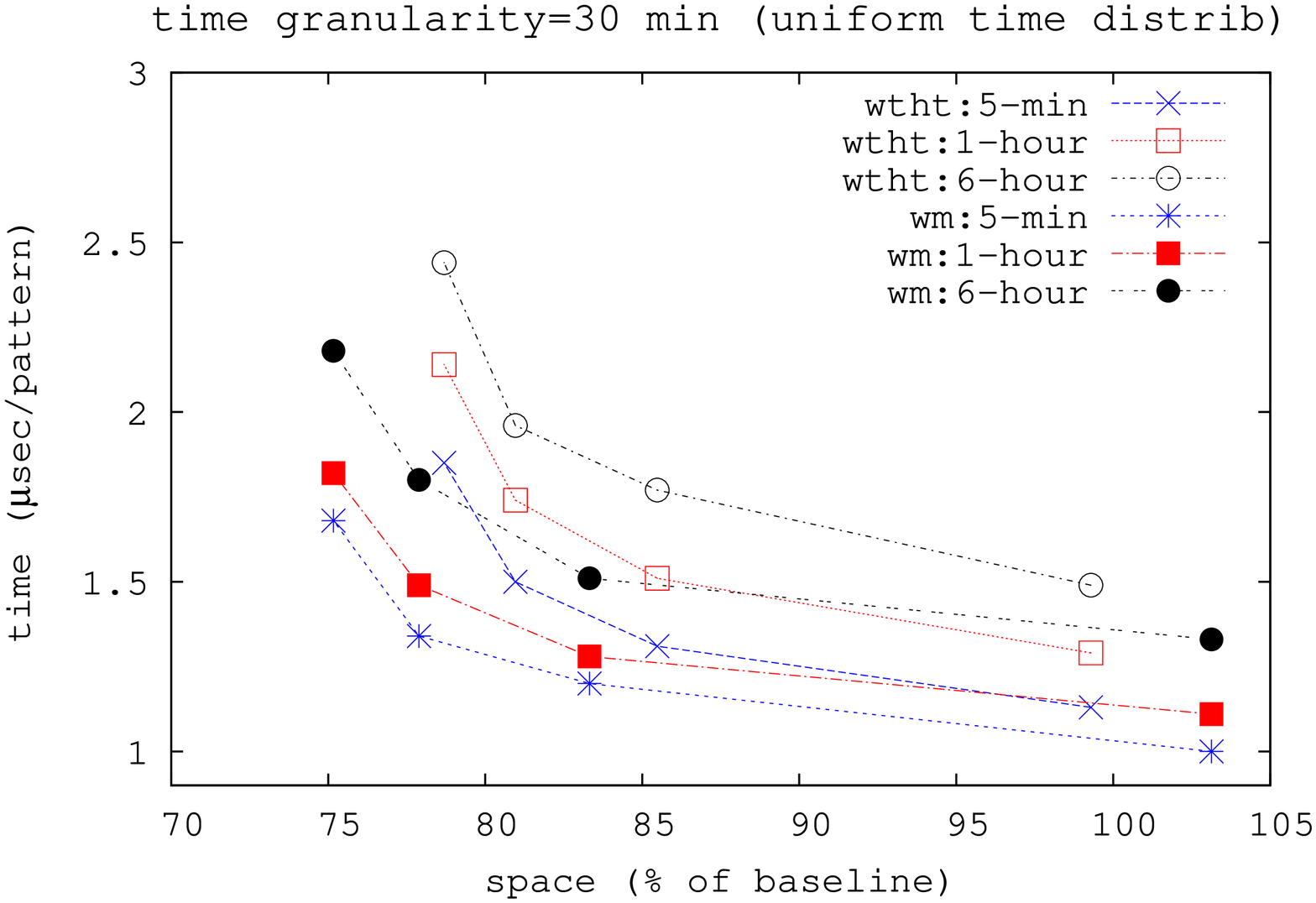}
		\includegraphics[angle=-90,width=0.45\textwidth]{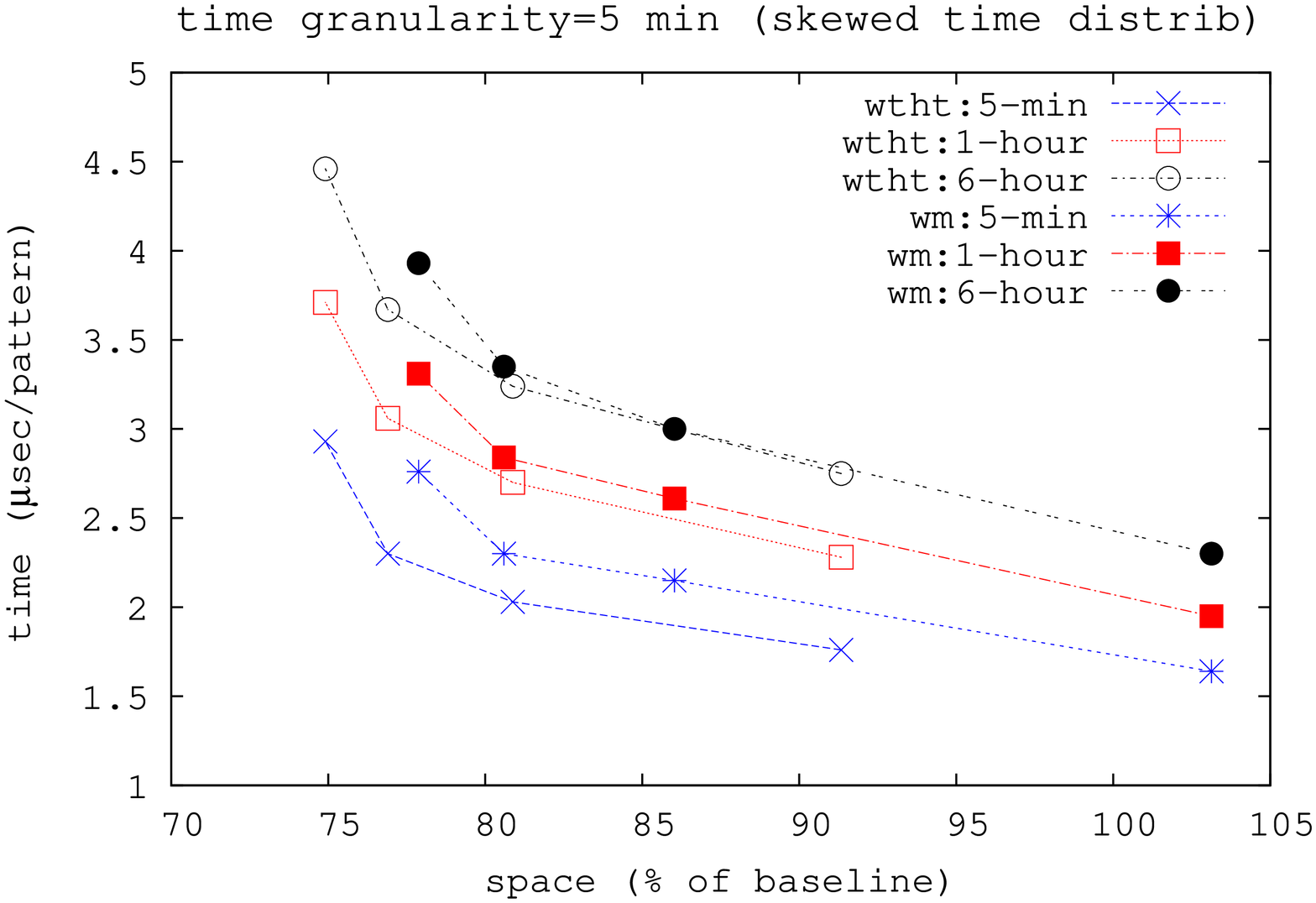}
		\includegraphics[angle=-90,width=0.45\textwidth]{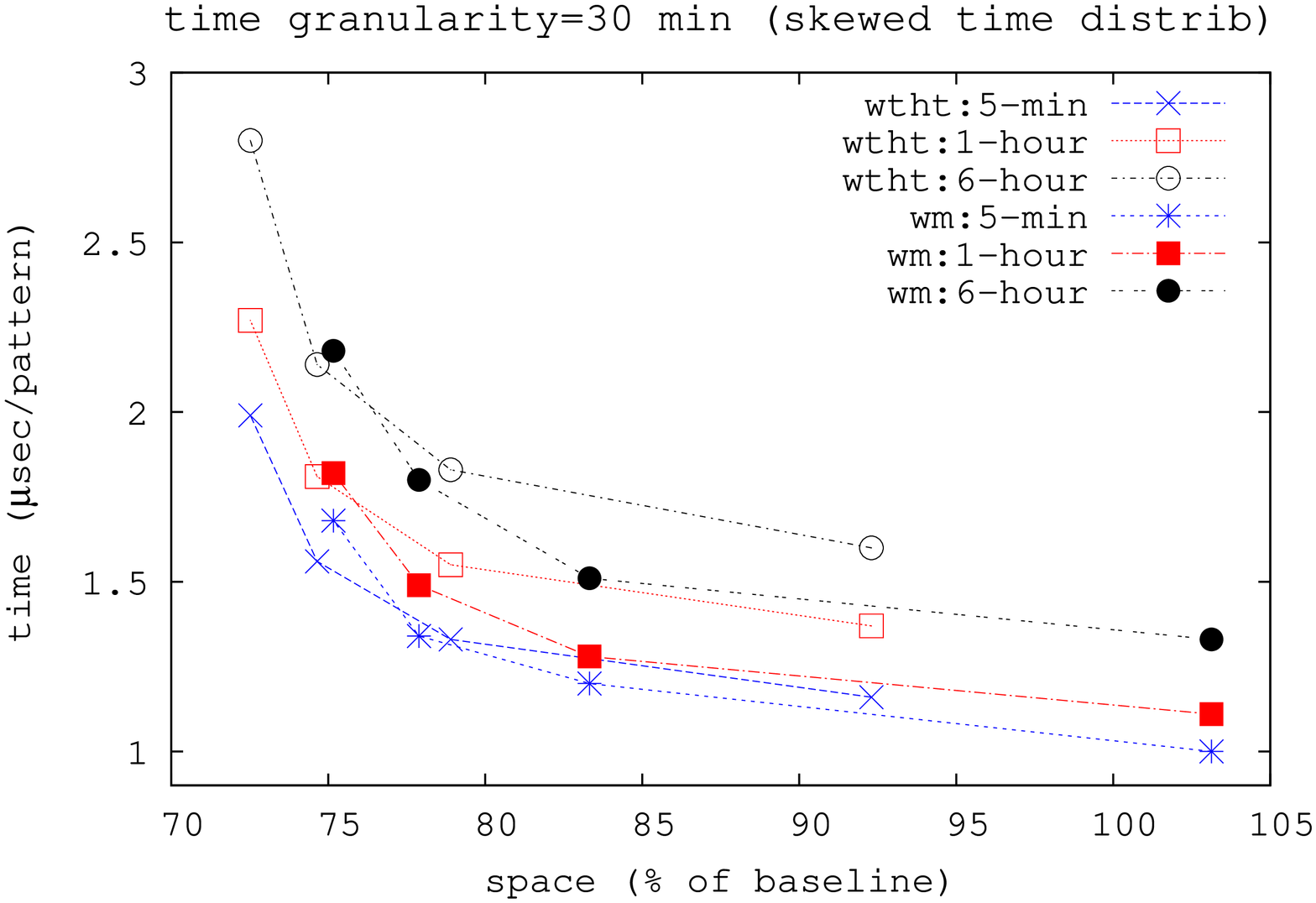}
		\includegraphics[angle=-90,width=0.45\textwidth]{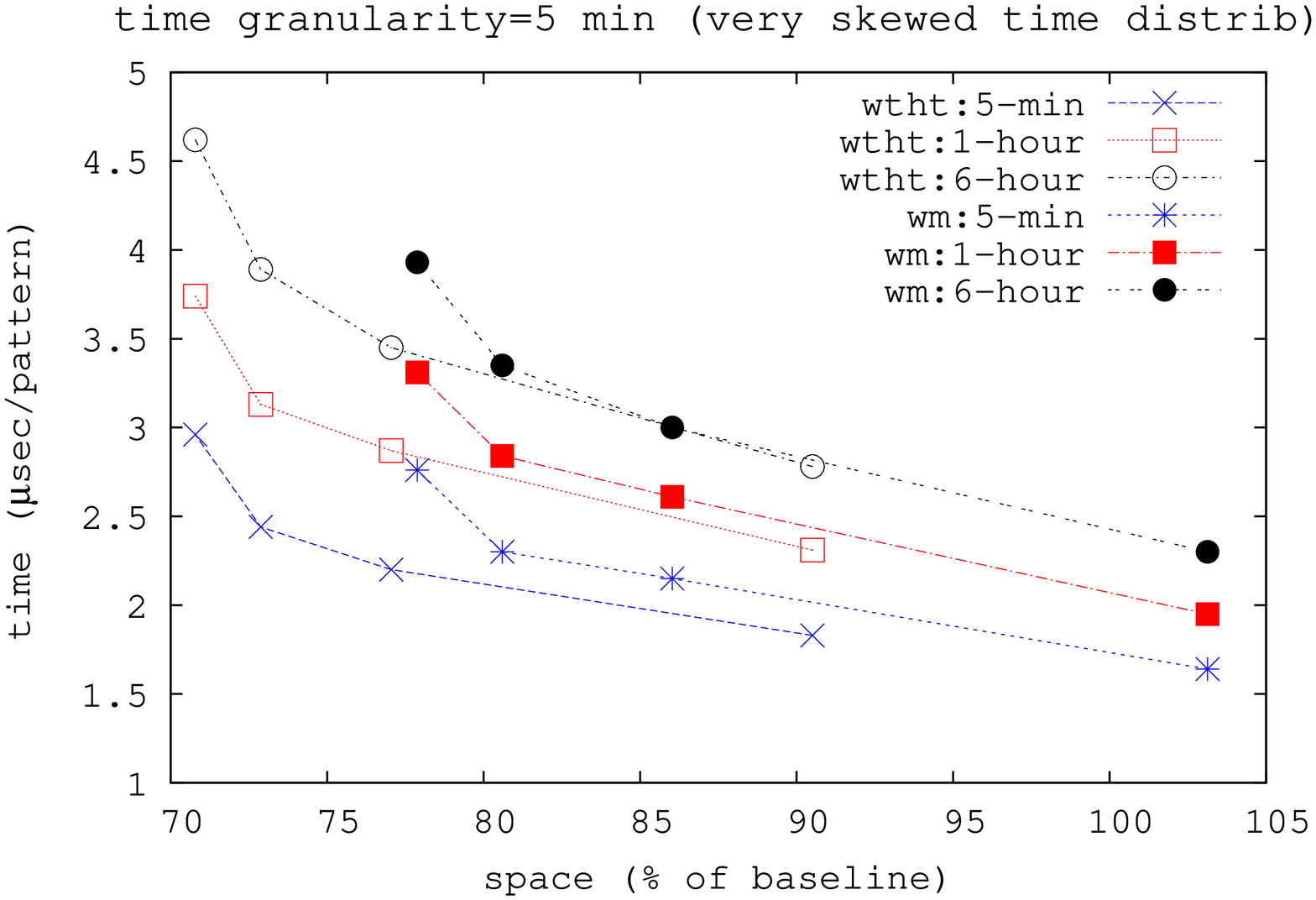}
		\includegraphics[angle=-90,width=0.45\textwidth]{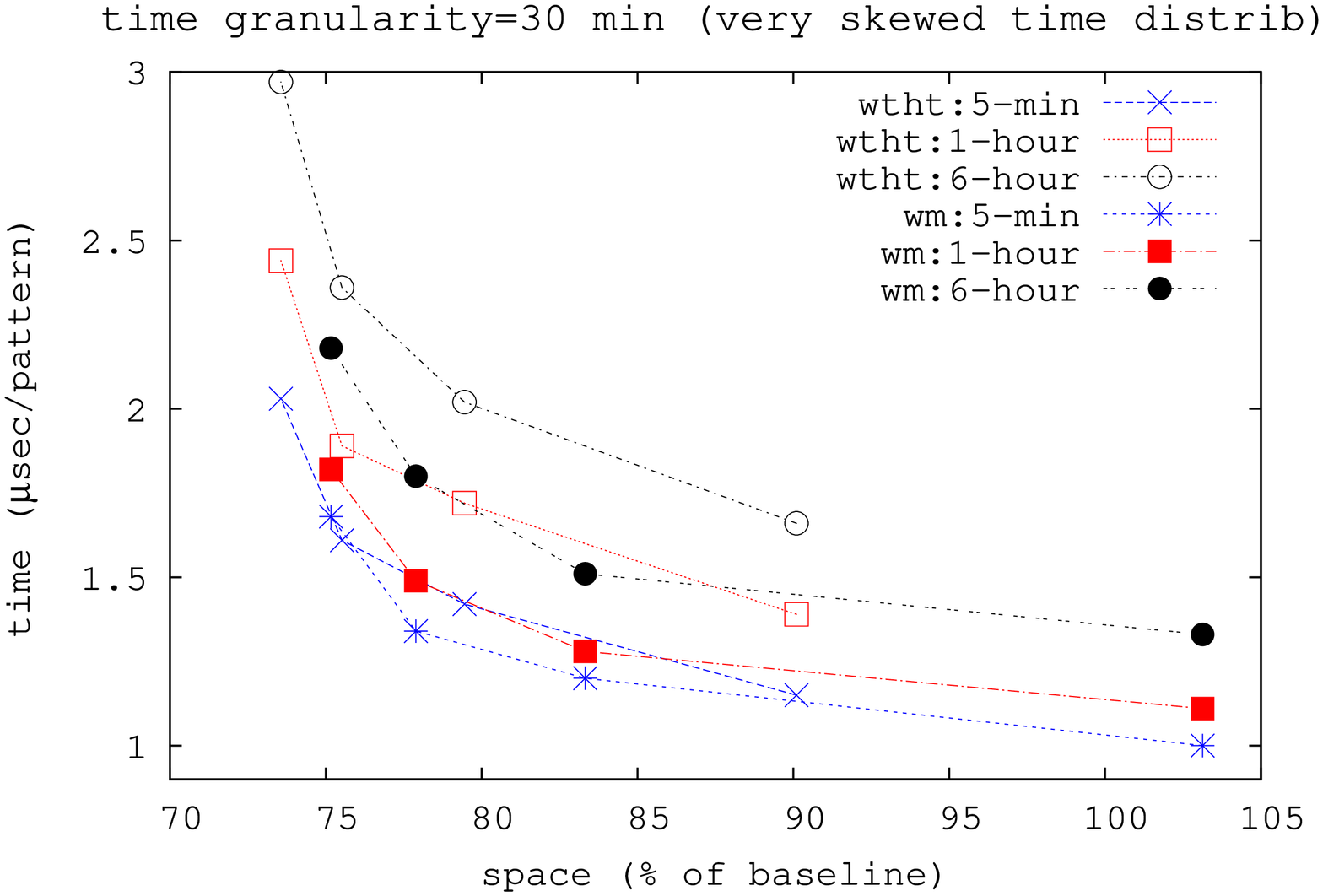}
		\caption{Space/time trade-offs for {\em count} queries depending on the time distribution: 
			uniform (top), skewed (middle), and very skewed (bottom).
			Time granularity for the time index is 5 minutes (left) or 30 minutes (right).
		}
		\vspace{-3mm}
		\label{fig:study}
	\end{center}
\end{figure}

In Figure~\ref{fig:study}, we show the results of our experiments. In the upper part of the figure, we 
include the results for $\wtht$ and $\wm$ built over the times assuming uniform frequency distribution.
In the middle part we assume times follow a the skewed distribution, and in the bottom of the figure we show
results when considering a very skewed distribution. Moreover, figures in the left column show results
for our structures considering that a 5-minute granularity is chosen for the discretization of times, whereas
figures on the right column assume time granularity is 30 minutes. For each scenario we include plots
\texttt{wtht:5-min}, \texttt{wtht:1-hour}, and \texttt{wtht:6-hour} for $\wtht$ (range width for $count$ is respectively
5-minutes, 1-hour, and 6-hours). We also present those plots for $\wm$ (\texttt{wm:5-min}, \texttt{wm:1-hour}, and \texttt{wm:6-hour}).

The baseline used for the space usage (x-axis) is the size of an array of
fixed-length time-interval IDs represented with the least number of bits needed (12 bits and 9 bits
respectively for 5-minute and 30-minute granularity, see Section~\ref{sec:ed}).
\medskip

When times are uniformly distributed, our $\wtht$ can only exploit the redundancy introduced by
the $\$$ symbols. This fact permits $\wtht$ to obtain only a minimal compression (around $96-98\%$ of the baseline)
when using a $RG$ (plain) bitvector, whereas $\wm$ uses more space than the baseline (around $104\%$).
Recall that for each plot we present four points corresponding (left-to-right) 
to $RRR_{128}$, $RRR_{64}$, $RRR_{32}$, and $RG$ bitvectors.
When using compressed  bitvectors ($RRR$), $\wm$ becomes the best choice. It is both more compact 
(bitvectors in $\wm$ are more compressible) and faster than $\wtht$. 
In any case, using $RRR$ clearly slows down queries.

A skewed distribution favors the compression for a statistical coder like Hu-Tucker,
which explains the higher compression obtained. However,
it also slightly increases the query times, especially in the wider
one-hour and six-hours query sets. This happens because the probability of having a query 
that forces to descend completely up to the leaves of the $\wtht$ increases.

For a very skewed distribution, the gap in compression between $\wtht$ and $\wm$ increases clearly (around 
$5$ percentage points), whereas query times remain similar to those in the previous scenario.

%In case of a $\wm$ representation, the time and space efficiency is the same
%regardless of the statistical distribution of the times in the dataset, as no
%statistical coder has been used. As expected, in Figure~\ref{fig:study}
%the version with the plain bitvector {\em RG} uses even more space than the original unindexed
%representation, as the bitvectors need additional structures for
%$rank$ and $select$.
%
%Regarding query times, there is a slight advantage of the $\wm$ that is mostly
%because the Hu-Tucker codes are longer than the original binary codes in some cases,
%increasing the height of the $\wt$. As we query random time intervals, a slightly
%deeper level was reached for the $\wt$ than for the $\wm$ for some of the queries.

As a conclusion of the experiments discussed in this section, we have shown that
the distribution of the sequence of times can be
exploited by our $\wtht$ to achieve a better compression and even improved
query times than the balanced $\wm$ counterpart.

%%%%%%%%%%%%%%%%%%%%%%%%%%%%%%%%%%%%%%%%%%%%%%%%%%%%%%%%%%%%%%%%%%%%%%%%%%%%%%%%%%

\section{Dealing with Spatio-temporal queries}
\label{sec:stq}
Apart from the pure spatial and temporal queries discussed in the previous sections, % in Sections~\ref{sec:sq}~and~\ref{sec:tq}, 
we can combine both the self-indexed spatial and temporal components from $\repres$ to answer spatio-temporal queries. 
The idea is to restrict spatial queries to a time interval $[t_1, t_2]$.  An example of this type of query is to return  the
{\em number of trips starting at node $X$ that occurred between $t_1$ and $t_2$}, which we can solve by first finding the
range in the $\csa$ of the trips starting in $X$ and then relying on the  ${count}$
operation in the $\wtht$ (or $\wm$). The following spatio-temporal queries can be solved by $\repres$:
%\marginpar{\tiny ojo, comente operacion ``report''}

\begin{itemize}[leftmargin=3mm]

	\setlength{\itemindent}{0mm}
	\item {\em Number of trips starting at node $X$ during time interval $[t_1,t_2]$ (\Tswx).}
	Recall that in the time sequence we also included timestamps associated with the area of $\$$-symbols in $\Psi$.
	Particularly, for each $\$$, we keep the time of the first node of its trip. Therefore, 
	we can perform $[l,r]~\leftarrow~bsearch(\$X)$ as in a regular spatial query to find the
	range $\Psi[l,r]$ ($[l,r]\subseteq [2,z+1]$) that corresponds to $\$$ symbols that end a trip
	which started at node $X$. Then, since the time sequence $Icode^{\Psi}$ 
	(represented with either a $\wtht$ or $\wm$) is
	 aligned with $\Psi$, we can filter out those trips that started within $[t_1,t_2]$ performing
	operation $count(l,r,t_1,t_2)$. In Figure~\ref{fig:search2} (steps \textcircled{1} and \textcircled{2}) we can see the steps involved.

\begin{figure}[thb]
	%\vspace{-0.4cm}
	\begin{center}
		{\includegraphics[width=0.90\textwidth]{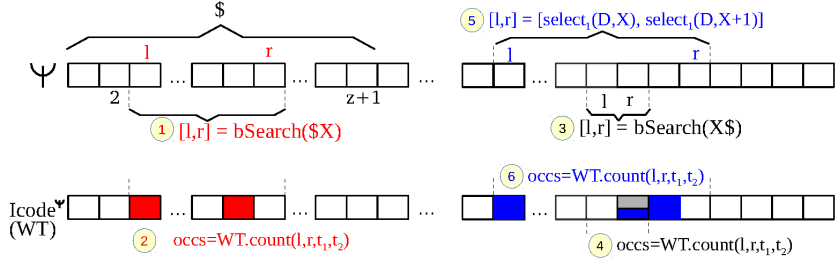}}
	\end{center}
	\vspace{-0.3cm}
	\caption{Trips staring at, ending at, or using node $X$  during time interval $[t_1,t_2]$.}
	\label{fig:search2}
	%\vspace{-0.6cm}
\end{figure}
	
	\item {\em Number of trips ending at node $X$ during the time interval $[t_1,t_2]$ (\Tewx). }
	As above, we initially perform the spatial query $[l,r]~\leftarrow~bsearch(X\$)$ to 
	obtain the range in $\Psi$ that corresponds to the pattern $X\$$ (trips ending at node $X$).
	Then, we use  $count(l,r,t_1,t_2)$ operation to count how many of those trips match the temporal constraint.
	See steps \textcircled{3} and \textcircled{4} in Figure~\ref{fig:search2}.
	
	\item {\em Number of trips using node $X$ during the time interval $[t_1,t_2]$ (\Tux).}
	As in the corresponding spatial query, the range $[l,r]$ in $\Psi$  is obtained with two $select_1$ operations on $D$.
	Finally, $count(l,r,t_1,t_2)$ finds the occurrences within the time interval $[t_1,t_2]$.
	See steps \textcircled{5} and \textcircled{6}  in Figure~\ref{fig:search2}.

\begin{figure}[thb]
	%\vspace{-0.4cm}
	\begin{center}
		{\includegraphics[width=0.90\textwidth]{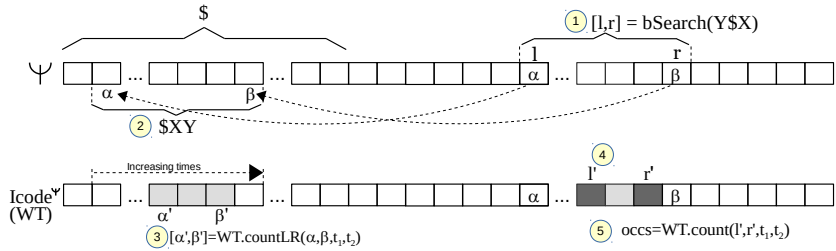}}
	\end{center}
	\vspace{-0.3cm}
	\caption{Trips staring at $X$ and ending at $Y$ during time interval $[t_1,t_2]$.}
	\label{fig:search}
	%\vspace{-0.6cm}
\end{figure}	
	\item {\em Number of trips starting at $X$ and ending at $Y$ occurring during  time interval $[t_1,t_2]$ (\Tfxty).}
	We consider two different semantics. A query with  {\em strong semantics} will obtain trips
	that start and end within  $[t_1,t_2]$. Whereas, a query with  {\em weak semantics} will obtain trips
	whose time intervals overlap  $[t_1,t_2]$ and, therefore, they could actually start before $t_1$ or end after $t_2$.
	
	In Figure~\ref{fig:search}, we show the step-by-step process to solve this type of queries.
	As in a spatial query, we start by searching for the range $[l,r] \leftarrow bsearch(Y\$X)$ in $\Psi$ corresponding 
	to trips starting at $Y$ and ending at $X$ ({\em step-\textcircled{1}}). Next, due to our sorting of trips,  the range for $Y\$X$ in $\Psi[l,r]$
	can be mapped to a continuous range $[\alpha,\beta]$ of the same size in the $\$XY$ region of $\Psi$. We compute $\alpha \leftarrow \Psi[l], 
	\beta\leftarrow\alpha+r-l$ ({\em step-\textcircled{2}}). Furthermore, note that the range for $\$XY$ preserves the same order as that for $Y\$X$.
	
	At this point, since $Icode^{\Psi}$  was aligned with $\Psi$, 
	we could check ending-time constraints within $Icode^{\Psi}[l,r]$  and starting-time constraints 
	within $Icode^{\Psi}[\alpha,\beta]$ (recall we keep starting times associated with the corresponding $\$$ of each trip).
	Note also that, due to our sorting (by starting-node, ending-node, starting-time,$\dots$) the times in $Icode^{\Psi}[\alpha,\beta]$ are 
	increasing (all of them correspond to trips with the same starting-node $X$ and ending-node $Y$). Therefore,
	we can find the continuous subrange $[\alpha',\beta'] \subseteq [\alpha,\beta] $ corresponding to trips
	that start within $[t_1, t_2]$ ({\em step-\textcircled{3}}). We refer to this operation as $countLR$ in Figure~\ref{fig:search}.
	Thus, assuming that $Icode^{\Psi}[\alpha,\beta]$ are increasing,  
	$[\alpha',\beta'] \leftarrow countLR(\alpha,\beta,t_1,t_2)$ would report the 
	positions $[\alpha',\beta'] \subseteq [\alpha,\beta]$ such that $\alpha' = argmin_{x} (Icode^{\Psi}[x] \geq t_1)$ and
	$\beta' = argmax_{x} (Icode^{\Psi}[x] \leq t_2)$.
	
	Using a $\wt$, a simple way to implement $countLR$  consists in performing  two binary searches within 
	$[\alpha,\beta]$ to find $[\alpha',\beta']$, where at each step we would use $access$ operation. This would cost $O(\log n \log \sigma)$. 
	Yet, we could also regard on $count$ operation to obtain a more efficient and also rather straightforward 
	implementation of $countLR$ so that we set $\alpha' \leftarrow count(\alpha,\beta,-1,t_1-1)$ and
	$\beta' \leftarrow \alpha'+ count(\alpha,\beta,t_1,t_2)$. It costs $O(\log\sigma)$.

	\begin{itemize}
		\item {\em Strong semantics (\Tfxtys).} Note that the subrange $[\alpha',\beta']$ (containing trips starting within $[t_1,t_2]$) 
		has a matching subrange $[l',r'] \subseteq [l,r]$ ({\em step-\textcircled{4}}), where some of the ending times of these trips will fall inside 
		$[t_1, t_2]$ (this allows us to check the ending time constraint). By performing  $count(l',r',t_1,t_2]$  we get the final result 
		({\em step-\textcircled{5}}). 
		To sum up, answering this query  requires: one $bsearch$ over $\Psi$ (to find $[l,r]$), one $access$ to $\Psi$ to obtain
		 $\alpha$ ($\beta\leftarrow \alpha+r-l$), one $countLR$ to find $[\alpha',\beta']$, and one $count$ 
		 (to count the valid ending times in $[l',r']$).

		\item {\em Weak semantics (\Tfxtyw).}
		The size of $[\alpha',\beta']$ is already a partial answer. To get the final result, we need to add 
		also the occurrences of those trips starting before $t_1$ that end at $t_1$ or later. 
		To do so, if $l<l'$, we need to compute $count(l,l'-1, t_1, {\sigma_t})$. This gives us the number of time 
		instants in the range $[l,l')$ of  $Icode^{\Psi}$ that fall inside $[t_1, {\sigma_t}]$. 
		That is, ending times equal or after $t_1$.
	\end{itemize}

\item {\em Top-k most used nodes during  time interval $[t_1,t_2]$ (\STtk).}
Both the sequential and binary-partition approaches discussed in Section~\ref{sec:sq} can easily
be extended to support this query. The idea is that, when we add a node either to the min-heap or priority-queue
respectively, we compute its frequency within time interval $[t_1,t_2]$ (using $count$ operation) 
rather than using its overall frequency.

\begin{itemize}
	\item In the {\em sequential approach (\Stkseq)}, given a node whose corresponding range
	in $\Psi$ is $D[l,r]$, we compute its frequency using $count(l,r,t_1,t_2)$ instead of simply using  $r-l+1$. 
	The rest of the process is exactly as discussed for the pure spatial \Stkseq\ query.
	
	\item In the {\em binary-partition approach (\Stkbin)}, we have to consider the priority of a
	given segment as the number of trips covered by that segment that occurred during $[t_1,t_2]$. Again, given
	a segment $[l,r]$ in $\Psi$ we compute that priority as $p_l^r \leftarrow count(l,r,t_1,t_2)$ instead of 
	$p_l^r \leftarrow  r-l+1$. Apart from
	that, the only modifications that we must consider over the pure spatial \Stkbin\ algorithm in 
	Figure~\ref{fig:topk_nieves} are:
	we replace (l.2) by $p_l^r \leftarrow count(select_1(D,2),n,t_1,t_2)$; $Q.push(\langle2, |V|\rangle, \langle select_1(D,2), n\rangle, \underline{p_l^r})$,
	and we replace  (l.12) and (l.13) respectively by 
	   $ Q.push(\langle i, m-1 \rangle, \langle l, q-1 \rangle, \underline{count(l,q-1,t_1,t_2)})$ and
	   $ Q.push(\langle m,   j \rangle, \langle q,   r \rangle, \underline{count(q,r,t_1,t_2)})$.

\end{itemize}

\item {\em Top-k most used nodes to start a trip during time interval $[t_1,t_2]$(\STtks).}
Following the same guidelines discussed above for \STtk, adapting the  sequential and 
binary-partition solutions for the spatial \Stks\ to include temporal constraints is straightforward.

\end{itemize}

%%%%%%%%%%%%%%%%%%%%%%%%%%%%%%%%%%%%%%%%%%%%%%%%%%%%%%%%%%%%%%%%
\section{Experimental evaluation} \label{sec:experiments}
%%%%%%%%%%%%%%%%%%%%%%%%%%%%%%%%%%%%%%%%%%%%%%%%%%%%%%%%%%%%%%%%
We have run experiments to evaluate both the space requirement and performance at query time of $\repres$
when dealing with spatial, temporal and spatio-temporal queries over two different datasets 
(Porto and Madrid) that are described in Section~\ref{sec:ed}. 

We have used several configurations of $\repres$ by tuning both its
spatial and temporal components. In the spatial part, %(Section~\ref{sec:transnet_repr}), 
we set the  $\Psi$ sampling parameter ($t_{\Psi}$) to the values $t_{\Psi} = \{32, 128, 512\}$. 
For the temporal component, % (Section~\ref{sec:time_repr}), 
we have tested both the 
balanced $\wm$, and the Hu-Tucker-shaped $\wt$ ($\wtht$) using the same bitvector configurations
discussed in Section~\ref{sec:time_exp}. That is, using either a plain bitvector $RG$ with a 
sparse sampling ($RG_{32}$), or a $RRR$ bitvector with sampling parameter $\in \{32,64,128\}$ 
($RRR_{32}, RRR_{64}$, and $ RRR_{128})$.

\subsection{Experimental datasets}
\label{sec:ed}
\begin{figure}[tbp]
	\begin{center}
		\includegraphics[angle=-90,width=0.85\textwidth]{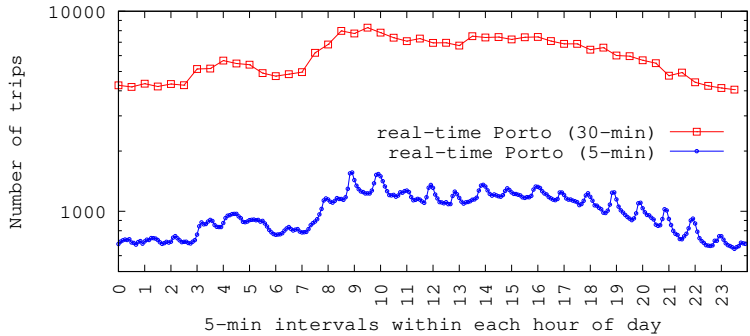}		
		\caption{Real time distribution from Porto. The y-axis (in log-scale) indicates the  average number of passengers per day for each time interval considering both
			5-min and 30-min time intervals.}
		%Para cada uno de los 365 días del año muestreado, mostramos el promedio diario para cada franja de 5 minutos
		\label{fig:distribporto}
	\end{center}
	\vspace{-3mm}
\end{figure}

We used two different datasets of trips in our experiments:
\begin{itemize}
   %%%%%%%%%	
  \item {\bf Madrid dataset}:
  Using GTFS\footnote{GTFS is a well-known specification for representing an urban transportation network. See
  \url{https://developers.google.com/transit/gtfs/reference?hl=en}} data from the  public transportation network of 
  {Madrid},\footnote{Data from
  the EMT corporation at \url{https://www.emtmadrid.es/movilidad20/googlet.html}} we generated a dataset of 
  synthetic trips combining the subway network with the Spanish commuter rail system (called {\em cercanías}).
  In total, there are $313$ different stations/nodes from $23$ lines.
 
We generated $10$ million trips with lengths varying from $2$ to $31$ nodes traversed. Those lengths follow a binomial 
distribution. The average length of the trips is $11.81$ nodes. 
%
%	\marginpar{\tiny \OJOFARI{Daniil, se hace asi realmente? porfa check}}
%	In the generation of a trip of length $l$, we randomly choose a starting node from a line and the starting direction. Then, we follow that line until 
%	we reach a switching node. At this node, we choose the line and direction to follow with uniform distribution. We avoid repeated nodes in the trip.
%	Generation ends when $l$ nodes have been added to the trip.

In the generation of a trip of length $l$, we randomly choose a starting node from a line, and the starting direction. 
Then, we follow that line until we reach a switching node. At this node, we decide whether to follow the current 
line or to switch to a new line. We allow only up to four line switches for a given trip, and use fixed probability 
values to decide whether to switch line or not. Such probability is $0.5$, $0.1$, $0.05$, and $0.02$ respectively 
for the first, second, third, and fourth line switch in a trip. 
We also avoid revisiting nodes in the same trip. 
The generation process ends when $l$ nodes have been added to the trip, or a dead end is reached.

As a baseline, the plain representation of the generated trips using a $9$-bit integer ($\lceil\log_2 314\rceil= 9$) 
for every node-ID (and $\$$ separator) would require $137.47$ MiB.

%We also generated synthetic times for those trips that can be seen in Figure~\ref{fig:distrib} with the label 
%\textit{skewed}, so most of the trip timestamps belong to rush hours. 
We also generated synthetic times for those trips following the same rules used to create the time distribution
named \textit{skewed} in Figure~\ref{fig:distrib}, so most of the trip timestamps belong to rush hours. 
Yet, instead of using only regular working days, we distinguished four kinds of days in a week: 
regular working days; Fridays and holiday eves; Saturdays; and Sundays and holidays. We also 
assume that there are two kinds of weeks related to high and low season periods. 
Therefore, a time interval may belong to eight types of day. 
When discretized at five-minute intervals we obtain $2,\!304$ distinct time intervals, 
while when we use thirty-minute intervals we obtain $384$. In the former case, our  baseline
 for the generated times using  $12$ bits per time-ID would occupy $183.30$ MiB. In the latter one, each
time-ID requires  $9$ bits and the  temporal baseline requires $137.47$ MiB.

%
%\marginpar{\tiny OJO Tal y como se generan las probabilidades de tiempos para cada tiempo, sólo estamos considerando un tipo de tiempo}
%

   %%%%%%%%%
   \item \textbf{Porto dataset}:
    We downloaded a collection of $1,\!710,\!671 $ trajectories from the city of {Porto} corresponding to taxi trips during 
    a full year (from July 1, 2013 to June 30, 2014), provided by \cite{moreira2013predicting}.\footnote{Description at \url{http://www.geolink.pt/ecmlpkdd2015-challenge/dataset.html}. Download at \url{https://archive.ics.uci.edu/ml/machine-learning-databases/00339/train.csv.zip}} Among other
    fields those data include, for each taxi ride, a list of GPS coordinates and times gathered every $15$ seconds of
    the trip. We adapted such data to our needs by using a map matching algorithm provided by the Graphhopper library,\footnote{\url{https://github.com/graphhopper/map-matching}}
     and OpenStreetMap cartography.\footnote{\url{http://www.openstreetmap.org/}} This permitted us to
  figure out the streets that trips were passing through. Finally, trips were encoded as a sequence of identifiers
  corresponding to adjacent stretches of street (that is, basic street segments with no intersections) the trip traversed, 
  each one of them tagged with a timestamp.

  %[1] https://archive.ics.uci.edu/ml/machine-learning-databases/00339/train.csv.zip
  %[2] https://archive.ics.uci.edu/ml/datasets/Taxi+Service+Trajectory+-+Prediction+Challenge,+ECML+PKDD+2015
  
  After filtering incomplete matches, $1,\!617,\!774$ trips, built  over $59,\!618$ distinct street segments, were used for the dataset. 
  Due to the nature of the network and the trips, 
  %the average length per trip is longer than in Madrid, of $64.74$ street segments. 
  the average number of street segments per trip is $64.74$; that is, the length of the trips is longer than in Madrid dataset.
  Since we needed $16=\lceil\log_2 59,\!618 \rceil$ bits to represent each segment in a trip, the total size of our
  plain spatial baseline is $202.85$ MiB.

  For the temporal part, we considered only one kind of day. Therefore, when we sample those $24$ hours into five-minute intervals,
  we obtain $288$ distinct time intervals that are given a $9$-bit time-ID.  Consequently the overall size of the temporal
  baseline becomes $114.10$ MiB. 
  However, if we split those $24$ hours into thirty-minute intervals, only $48$ time intervals 
  arise. In this case, each time-ID needs only $6$ bits and the total size of the temporal baseline is $76.07$ MiB.
  The average number of daily passengers for each time interval is shown in Figure~\ref{fig:distribporto}.
\end{itemize}

\subsection{Space Requirements of $\repres$}\label{sec:exp.space}
We show the compression obtained by $\repres$ when built on our two test datasets. Compression is shown
as the percentage of the size of the plain baselines discussed above.
Using different configurations of $\repres$, we will show the compression of the spatial component ($\csa$), 
that of the temporal component ($\wtht$ and $\wm$), and finally the overall compression of $\repres$.

\begin{table}[htb]
%\vspace{-6mm}
\begin{center}
\scriptsize
  \begin{tabular}{|l|*{3}{r}|}
  \cline{2-4}
  \multicolumn{1}{c|}{} & \multicolumn{3}{c|}{$t_{\Psi}$ } \\
  \multicolumn{1}{c|}{} & 32 & 128 & 512 \\
  \hline
  Madrid & 41.32\% & 26.80\% & 23.06\% \\
  Porto & 23.66\% & 15.49\% & 13.37\% \\
  \hline
  \end{tabular}
%\vspace{-5mm}
%\caption{Compression of the spatial component with respect to the spatial-only baseline.}
\caption{Compression of $\csa$ with respect to the spatial baseline.}
\label{table:spatial_spaces}
\vspace{-3mm}
\end{center}
\end{table}

Results regarding the compression obtained by $\csa$ are given in Table~\ref{table:spatial_spaces}.
The compression ratio is calculated over a plain spatial-only (stop-IDs or street-segment-IDs in each case) representation.   %\marginpar{\tiny Daniil, incluyes separadores aqui ?}
%Note that an $\icsa$  built on English text~\cite{FBNCPR12} typically
%reached the compression of {\em gzip} (around 35\% in compression ratio).
%As expected, the high compressibility of our sorted datasets of trips permits $\csa$ to improve 
%those numbers with compression ratios under 30\% in the most sparse setup, while also offering indexing features 
%that allow us to perform efficient searches.
In a rather dense 
configuration of $\csa$ with $t_{\Psi}=32$ we obtain compression ratios around $41$\% and $23$\% for Madrid and Porto datasets respectively.
Those results are interesting from the simple point that the baseline representations were only using respectively 
$9$-bits per node (Madrid) and $16$-bits per segment (Porto). 
As expected, compression improves as we increase the $\Psi$ sampling parameter $t_{\Psi}$. We show that by tuning 
$\csa$ in a more sparse setup we can almost halve the space needs of using $t_{\Psi}=32$. Yet, the resulting $\csa$ would become
much slower as we will see in the next section.
In general, we can see that $\csa$ obtains better compression in Porto than in Madrid. This is probably due to the longer and
more predictable trips. Note that is not common to arrive at an intersection having more than two valid 
street links where to navigate to.

\begin{table}[htb]
%\vspace{-6mm}
\begin{center}
\scriptsize
\setlength\tabcolsep{2pt}
  \begin{tabular}{|l|*{4}{r}|}
  \cline{2-5}
  \multicolumn{1}{c|}{} & \multicolumn{4}{c|}{Type of bitvector in $\wm$/$\wtht$}\\

  \multicolumn{1}{c|}{}   & $RG_{32}$& $RRR_{32}$& $RRR_{64}$& $RRR_{128}$\\
  \hline                                             
  Madrid ($\wtht$, 5-min) &  91.33\% &	 80.89\% &	 76.90\% & 	 74.90\% \\
  Madrid ($\wm$, 5-min)   & 103.13\% &	 86.03\% &	 80.61\% & 	 77.88\% \\
  Madrid ($\wtht$, 30-min)&  92.30\% &	 78.90\% &	 74.66\% &	 72.52\% \\
  Madrid ($\wm$, 30-min)  & 103.14\% &	 83.32\% &	 77.90\% &	 75.18\% \\
  \hline                  
  Porto ($\wtht$, 5-min)  &  93.52\% &	102.61\% &	 98.27\% &	 96.11\% \\
  Porto ($\wm$, 5-min)    & 103.13\% &	106.88\% &	101.41\% &	 98.66\% \\
  Porto ($\wtht$, 30-min) &  96.00\% &	103.78\% &	 99.08\% &	 96.74\% \\
  Porto ($\wm$, 30-min)   & 103.12\% &	107.00\% &	101.50\% &	 98.75\% \\

  \hline
  \cline{2-5}
  \end{tabular}
%\vspace{-1mm}
\caption{Compression of $\wm$ and $\wtht$ with respect to the temporal baseline.}
\label{table:ctr_wt_spaces}
\vspace{-3mm}
\end{center}
\end{table}

\medskip
In Table~\ref{table:ctr_wt_spaces}, we focus on the space needed by the temporal component of $\repres$. 
In this case we show the compression ratios obtained by $\wtht$ and $\wm$ %(our two alternative representations) 
considering that time is either discretized into $5$-min or $30$-min intervals. Recall that the size of the 
plain baseline representations differs depending on the discretization period. Both $\wtht$ and $\wm$ were tuned by
using bitvector representations $RG_{32}$, $RRR_{32}$, $RRR_{64}$, and $ RRR_{128}$.

It is interesting to see that in the synthetic dataset from Madrid, $RRR$ bitvectors always lead to a better 
compression than the plain $RG$, while in the real dataset from Porto that is never the case. 
In some cases $RRR$ does not compress the times at all. Consequently, for Porto dataset, the faster plain $RG$ bitvectors are probably the best choice. In Madrid dataset, we can see an actual space/time trade-off: $RRR$ obtains better compression but will be slower (as we will see in the next section).

To understand why $RRR$ is much more effective in Madrid than in Porto, recall that the values in $Icodes$ are aligned to the suffix Array ($A$). Recall also that, within the range in $A$ corresponding to each node $X$, suffixes are sorted by ending node, then by starting time, and finally by the remaining nodes. 
Therefore, at least all the trips that start in node $X$ and end in the same node $Y$ are sorted by
time, and consequently, the corresponding range in $Icodes$ keeps non-decreasing values.
We have measured the number of times that the “starting-time” component was used during the suffix-sort step from the construction of $\repres$. We obtained that, for Madrid, it was used $42,051,591$ times, while for Porto, it was only used $1,313,269$ times (note that this is not the number of repeated trips from $X$ to $Y$, but the number of times the ``starting-times" were actually compared during suffix-sort). Since the number of entries in $Icodes$ is rather similar in both datasets (around $128\!\times\!10^6$ in Madrid, and $106\!\times\!10^6$ in Porto)  we could expect that the sequence of $Icodes$ is much more regular in Madrid than in Porto. Consequently, this could explain why $RRR$ performs much better in Madrid than in Porto.

\begin{table}[ht!]
%\vspace{-6mm}
\begin{center}
\scriptsize
\setlength\tabcolsep{2pt}
  \begin{tabular}{|l|*{4}{c}|*{4}{c}|}
  \cline{2-9}
  \multicolumn{1}{c|}{} & \multicolumn{4}{c|}{Type of bitvector in $\wm$/$\wtht$}& \multicolumn{4}{c|}{Type of bitvector in $\wm$/$\wtht$} \\

  \multicolumn{1}{c|}{}     &$RG_{32}$& $RRR_{32}$& $RRR_{64}$&$RRR_{128}$&$RG_{32}$& $RRR_{32}$& $RRR_{64}$&$RRR_{128}$ \\
  \hline                                             
  Madrid ($\wtht$, 5-min)  & 69.90\% &   63.93\% &   61.65\% &   60.51\% & 62.07\% &	56.10\% &	53.82\% &	 52.68\% \\
  Madrid ($\wm$, 5-min)     & 76.64\% &   66.87\% &   63.77\% &   62.21\% & 68.81\% &	59.04\% &	55.94\% &	 54.38\% \\
  Madrid ($\wtht$, 30-min) & 66.81\% &   60.11\% &   57.99\% &   56.92\% & 57.68\% &	50.98\% &	48.86\% &	 47.79\% \\
  Madrid ($\wm$, 30-min)    & 72.23\% &   62.32\% &   59.61\% &   58.25\% & 63.10\% &	53.19\% &	50.48\% &	 49.12\% \\
  \hline
  Porto ($\wtht$, 5-min)   & 48.81\% &   52.08\% &   50.52\% &   49.74\% & 42.22\% &	45.49\% &	43.93\% &	 43.15\% \\
  Porto ($\wm$, 5-min)      & 52.27\% &   53.62\% &   51.65\% &   50.66\% & 45.68\% &	47.03\% &	45.06\% &	 44.07\% \\
  Porto ($\wtht$, 30-min)  & 43.39\% &   45.51\% &   44.23\% &   43.59\% & 35.91\% &	38.03\% &	36.75\% &	 36.11\% \\
  Porto ($\wm$, 30-min)     & 45.33\% &   46.39\% &   44.89\% &   44.14\% & 37.85\% &	38.91\% &	37.41\% &	 36.66\% \\
  \hline
  \cline{2-9}
  \multicolumn{1}{c|}{} & \multicolumn{4}{c|}{$t_{\Psi}=32$}& \multicolumn{4}{c|}{$t_{\Psi}=512$} \\
    \cline{2-9}
  \end{tabular}
%\vspace{-5mm}
\caption{Overall Compression of $\repres$ including different configurations for both the spatial and temporal components.  }
\label{table:ctr_spaces}
\vspace{-4mm}
\end{center}
\end{table}

\medskip
Finally, in Table~\ref{table:ctr_spaces}, we show the overall compression ratios of  $\repres$.
We use the same configurations for $\wtht$ and $\wm$  as in Table~\ref{table:ctr_wt_spaces}, and both the
most dense and sparse tuning of $\csa$ ($t_{\Psi}= 32$ and $t_{\Psi}= 512$ respectively).
For Madrid dataset, the pair (node,timestamp) is represented with $9+9=18$ bits in our baseline representation 
when time is discretized into $30$-minute intervals, and with $9+12=21$ when we use $5$-minute intervals.
In the case of Porto dataset, when using $30$-minute intervals, each pair (node,timestamp) from the baseline requires $16+9=25$ bits. 
If discretization considers $5$-minute intervals, the baseline requires $16+6=22$ bits. We can see that the overall
compression of $\repres$ in Madrid dataset ranges between $76\%$ and $50\%$. Also we show that Porto dataset is much
more compressible, obtaining compression ratios from around $50\%$ to $35\%$.

\subsection{Performance at query time}
Through this section, we evaluate the time performance of $\repres$ when solving spatial, temporal, and spatio-temporal queries.
We have randomly generated $10,\!000$ query patterns from our two datasets for each type of query.
Each time measurement presented below is the average execution time of $10,\!000$ runs using the corresponding query patterns, 
except for the \Stk\ queries where we perform $100$ runs of the {\em top-k} algorithms with $k=\{10,100\}$.

Our test machine has an Intel(R) Core(tm) i5-4690@3.50GHz CPU (4 cores/4 siblings) and 8GB of DDR3 RAM. 
It runs Ubuntu Linux 16.04 (Kernel 4.4.0-21-generic). The compiler used was g++ version 5.4.0 and we set compiler optimization flags to $-O9$. All our experiments run in a single core and time measures refer to CPU user-time.

During the generation of query patterns, for those queries involving only one node $X$ from the network, 
we have randomly chosen $X$ $10,\!000$ times from the available network nodes. 
This is the case of the query patters used both for
the spatial queries \Sswx, \Sewx, and \Sux\ or the spatio-temporal \Tswx, \Tewx, and \Tux.
In the case of the spatial \Sfxty\ and the spatio-temporal \Tfxtys, and \Tfxtyw\ the pair of
network nodes $\langle X,Y \rangle$ that compose our
query patterns were generated by randomly choosing
$10,\!000$ trips and then extracting the initial $X$ and ending $Y$ nodes of those trips.

Moreover, we also generated the time intervals $[t_1,t_2]$ required for 
the spatio-temporal queries. Considering the different available time-IDs, we chose a random starting
instant $t_1$ and then randomly generated the width of that interval from five minutes to two hours.
Note that if we discretized time into $5$-minute intervals and $interval$-$width=59$ minutes, our time
interval $[t_1,t_2]$ would contain exactly $12$ time IDs ($t_2\leftarrow t_1+11$). However, if time was
discretized into $30$-minute intervals, $[t_1,t_2]$ would contain only $2$ time IDs ($t_2\leftarrow t_1+1$).
We followed the same procedure to gather the query patterns used for the pure temporal queries \Tut\ and \Tst.
%queries \Tuti\ and \Tutr. 
%Yet, since \Tuti\ only needs a unique time-ID, we only randomly extracted $t_1$.

%Each time result was the average of the execution of 10000 random queries of that type, except for the top-k queries. 
%The time intervals for the spatio-temporal queries were generated with a random size from five minutes to two hours.

%\OJOFARIDONE{posiblemente no tabla con todas las queries, pq -las probamos todas-}

\subsubsection{Space/time trade-off when dealing with spatial queries} \label{sec:exp:temp}

\begin{figure}[htb]
	%\vspace{-0.4cm}
	\begin{center}
		{\includegraphics[angle=-90,width=0.45\textwidth]{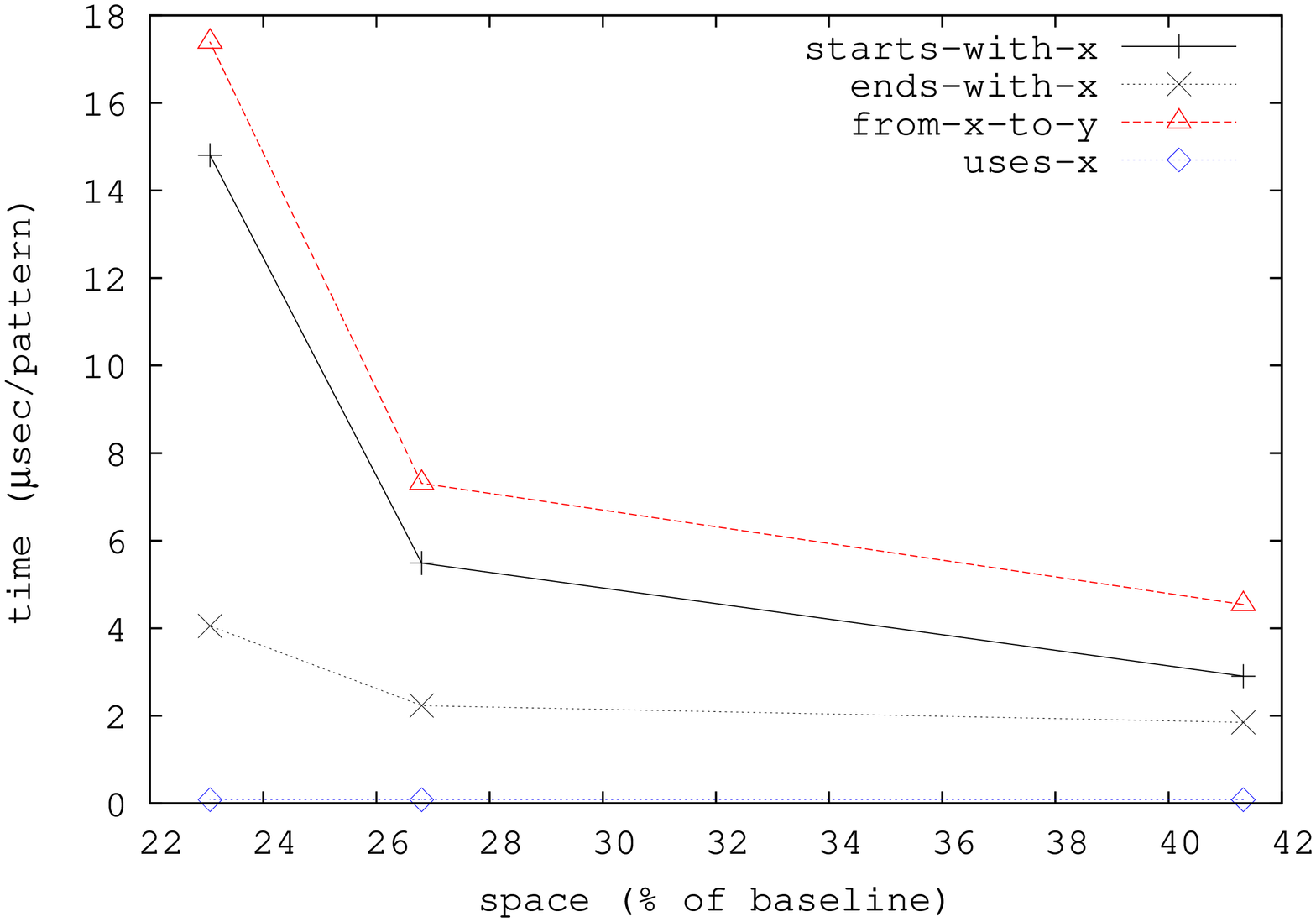}}
		{\includegraphics[angle=-90,width=0.45\textwidth]{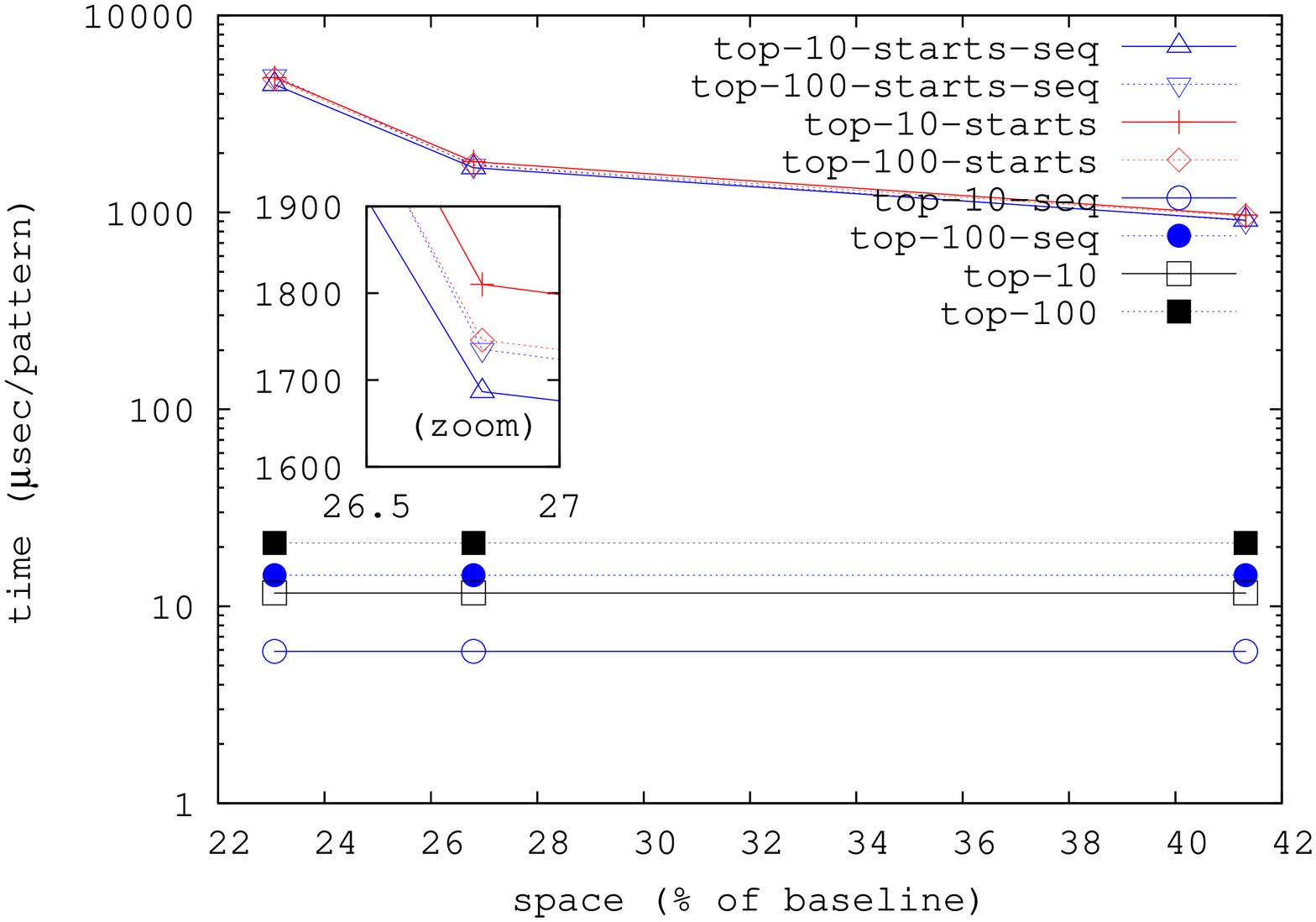}}
	\end{center}
	\vspace{-0.3cm}
	\caption{Spatial queries (left) and spatial {\em top-k} queries (right) for Madrid.}
	\label{fig:madridsp}
	\vspace{-0.3cm}
%\end{figure}

%%%%%%%%%%% PORTO - SPATIAL %%%%%%%%%%%%%
%\begin{figure}[htb]
	%\vspace{-0.4cm}
	\begin{center}
		{\includegraphics[angle=-90,width=0.45\textwidth]{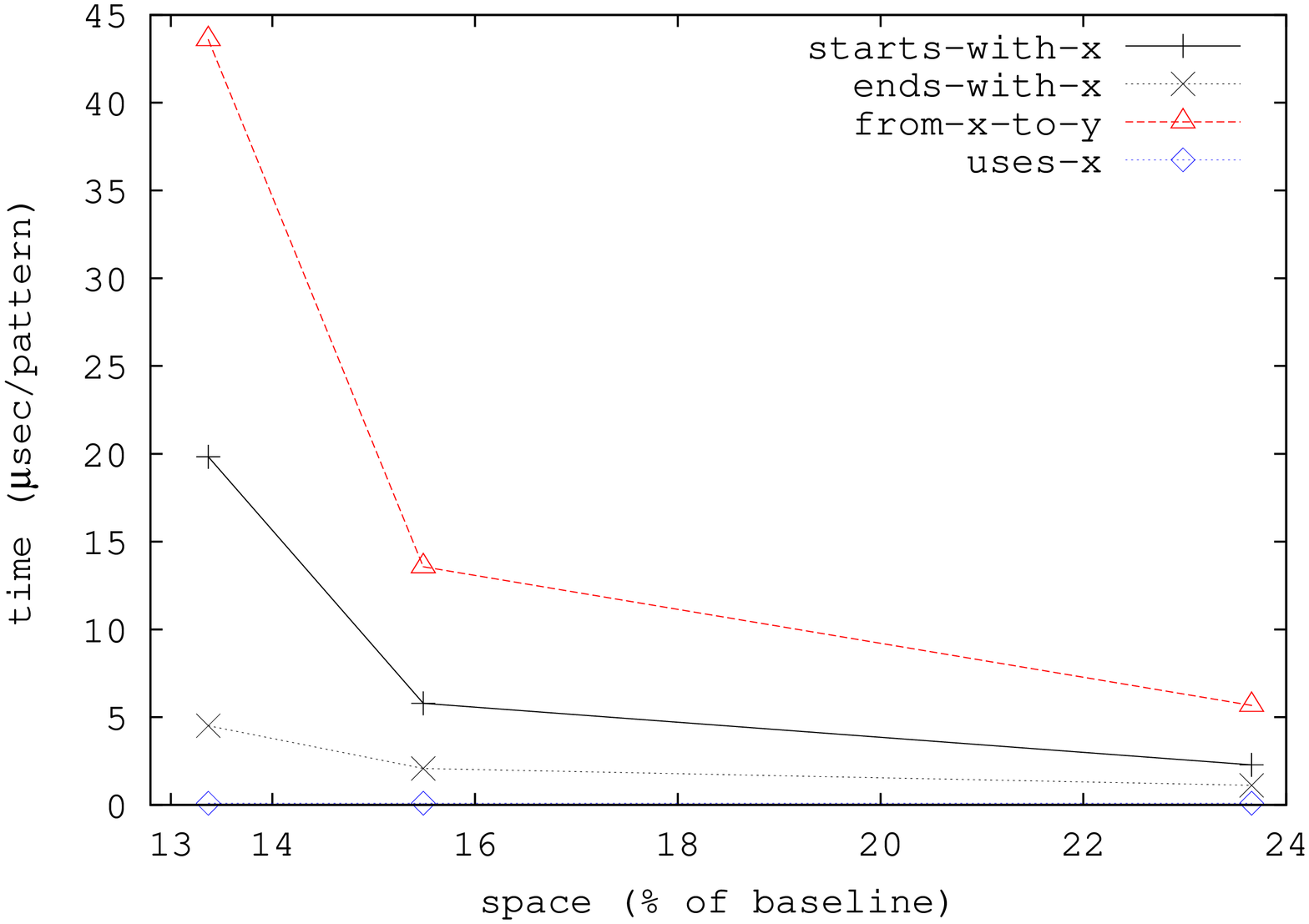}}
		{\includegraphics[angle=-90,width=0.45\textwidth]{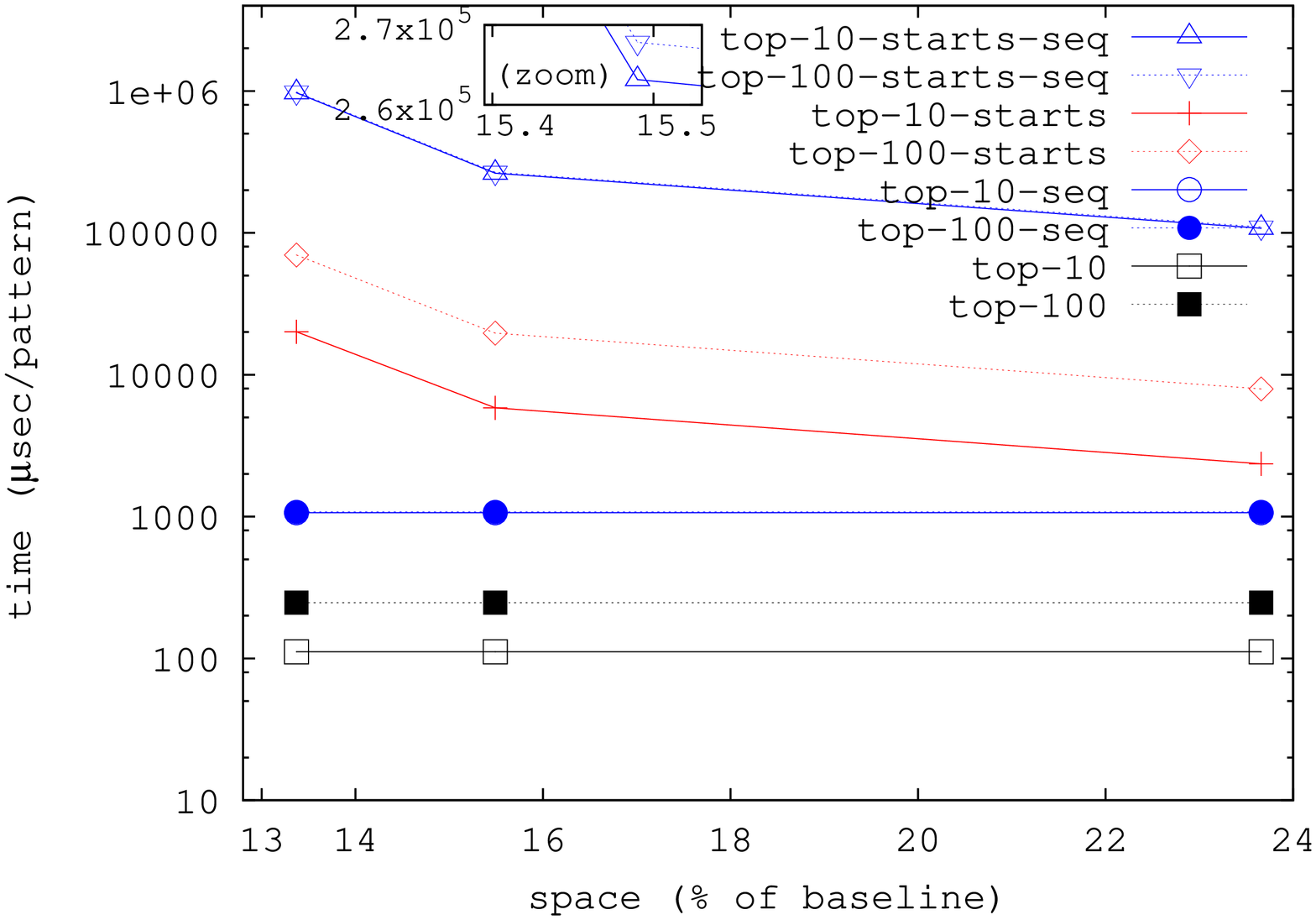}}
	\end{center}
	\vspace{-0.3cm}
	\caption{Spatial queries (left) and spatial {\em top-k} queries (right) for Porto.}
	\label{fig:portosp}
	%\vspace{-0.6cm}
\end{figure}

In Figures~\ref{fig:madridsp} and \ref{fig:portosp}, we show the performance of $\repres$ at
solving spatial queries for Madrid and Porto datasets respectively. 
Note that all these queries can be answered using only the $\csa$ component
of $\repres$. Therefore, the size of the temporal component 
is not considered here and compression values (x-axis) refer only to the size of $\csa$ with
respect to the spatial baseline as in Table~\ref{table:spatial_spaces}. 
We show the average query time (in $\mu$s) depending on the
space used by $\csa$ with three different 
sampling configurations ($t_{\Psi} =\{512, 128, 32\}$).

%It is worth noting 
Results show that the queries that involve searching in the $\$$ region of 
$\Psi$, such as \Sswx\ or \Sfxty\ are considerably slower than queries \Sewx\ and \Sux\ 
due to the large size of that region. Recall there is one $\$$ for every trip. 
%
%This difference is much more accentuated if a forward search was used in the $\csa$.

%\marginpar{\tiny Daniil, como se implementa backward search? supongo que usas select para localizar zona para X y luego bsearch dentro de esa %zona? Ya me comentaras...}  Ok, correcto! ;)

%%%%%%%%%%% MADRID -SPATIAL %%%%%%%%%%%%%

In both datasets, we can see that \Sux\ (solved using {\em select} on $D$ rather than
$bsearch$ on $\Psi$) is the fastest query. On average, it takes only around $10$ns
per query. Except in the most sparse configuration of $\csa$, queries \Sewx, \Sswx, and 
\Sfxty\ require typically less than $10\mu$s. This basically shows the cost of performing
$bsearch$ on a compressed $\Psi$. In the most sparse setup, times for \Sswx\ and \Sfxty\ are always better 
in Madrid than in Porto dataset, and \Sewx\ draws rather identical times.
With the densest configuration  ($t_{\Psi}=32$), \Sewx\ and \Sfxty\ are respectively 
around $10$-$20$\% fastest in Madrid dataset (\Sewx\ takes $4.05\mu$s and $4.51\mu$s respectively, and
\Sfxty\ takes $4.54\mu$s and $5.66\mu$s). However, \Sswx\ performs around $20$\% faster in Porto dataset 
($2.28\mu$s vs $2.90\mu$s).
\medskip

%Ends-with-X   Madrid/Porto (4.05/4.51)
%F-X-to-Y     Madrid/Porto (4.54/5.66)
%Starts-with-X   Madrid/Porto (2.90/2.28)

%Regarding \Stk\ queries we can see that the fact of having less trips, but a larger vocabulary, make
%\Stk\ queries to perform slower at Porto than at Madrid dataset.

Focusing on \Stk\ queries, we can see huge differences between \Stks\  
and the rest of the \Stk\ queries, as the former needs to perform {$ bsearch$} over the compressed $\Psi$
instead of a $select$ on $D$. 

We can also see that due to the small number of stops in Madrid dataset, it is always more efficient 
to use the sequential version of \Stks\ and \Stk\ algorithms. This is also because a rather uniform frequency among nodes 
increases the number of insertions in the priority queue ($i$) of the binary algorithm
needed for retrieving the first $k$ nodes ($i \approx |V|$). Moreover, note that for the sequential algorithm 
$i$ is  at most $|V|$, whereas for the binary-partition counterpart it could become up to $2|V|-1$.

However, in Porto dataset, where nodes follow a biased distribution (some streets are far more used than others
by taxis), and whose vocabulary 
is $190$ times larger than that of Madrid's, the binary-partition version of \Stks\ and \Stk\ algorithms is clearly
faster than the sequential counterpart (\Stkseq\ and \Stksseq).
Note that in Madrid dataset, \Stcien\ returns 32\% of the nodes (hence sequential processing worths it) 
whereas in Porto dataset less than 0.2\% of the nodes are returned. 

The gap between \Stdiezseq\ and \Stcienseq\ that we can clearly appreciate in Madrid dataset 
is due to the cost of the insertion of nodes in the min-heap. However, the gap between 
the binary \Stdiez\ and \Stcien\ 
is mainly related to the number of iterations performed until the binary-partition algorithm gathers the first
$10$ and $100$ {nodes}  returned respectively. The same discussion applies for \Stks\ queries.

%\begin{itemize}
%	\item top-100-starts en Madrid: seq y bin son similares en tiempo: solo hay 314 nodos en la red y 
%	top-100starts devuelve el 33\% de los nodos. En Porto, top-100-starts devuelve el 0.2\% de los nodos 
%	de la red por ello bin es mucho mejor opcion que seq.
%	\item top-10-starts
%\end{itemize}

%We evaluated $\repres$ over the Porto dataset, with the same index configurations and type of queries as for Madrid. 
%Figure~\ref{fig:portosp} shows the performance of the spatial component. It is significantly slower than with Madrid 
%in the sparse configuration for the \texttt{starts-with-x} or \texttt{from-x-to-y}.

%\OJOFARI{DANIIL dice aqui que la siguiente frase no es valida: en concreto me dice que: esa frae la puse para explicar
% starts-with-x y from-x-to-y, pero estoy viendo que no es valida para explicar form-x-to-y. Asi que no se porque  
% from-x-to-y es mas lenta en un lado que en el otro (Madrid vs Porto)} 
%This is because the generated queries are with random street segment identifiers, while most of them have never been 
%used as a starting stop for a taxi, thus triggering the worst case of a binary search.

%We can also see that because of the increased size of the vocabulary, the sequential versions of top-k perform much 
%slower than their binary counterparts.

% % % % % % % % % % % % % % % % % % % % % % % % % % % % % % % % % % % % % % % % % % % % % % % % % % % % % % %
% % % % % % % % % % % % % % % % % % % % % % % % % % % % % % % % % % % % % % % % % % % % % % % % % % % % % % %
\subsubsection{Space/time trade-off when performing temporal  queries}
% % % % % % % % % % % % % % % % % % % % % % % % % % % % % % % % % % % % % % % % % % % % % % % % % % % % % % %
% % % % % % % % % % % % % % % % % % % % % % % % % % % % % % % % % % % % % % % % % % % % % % % % % % % % % % %

\begin{figure}[!ht]
	%\vspace{-0.4cm}
	\begin{center}
		\begin{center}
			{\includegraphics[angle=-90,width=0.45\textwidth]{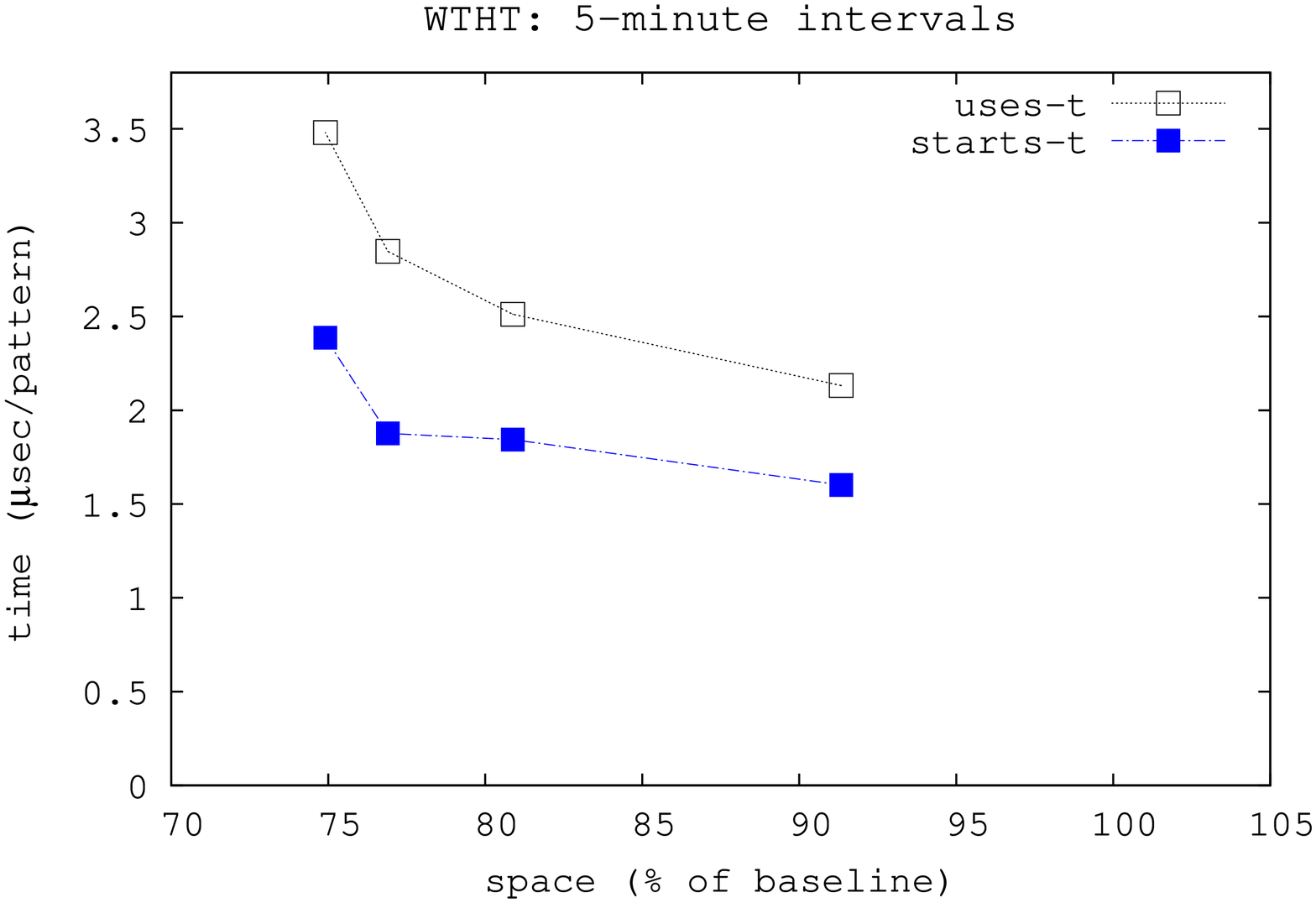}}
			{\includegraphics[angle=-90,width=0.45\textwidth]{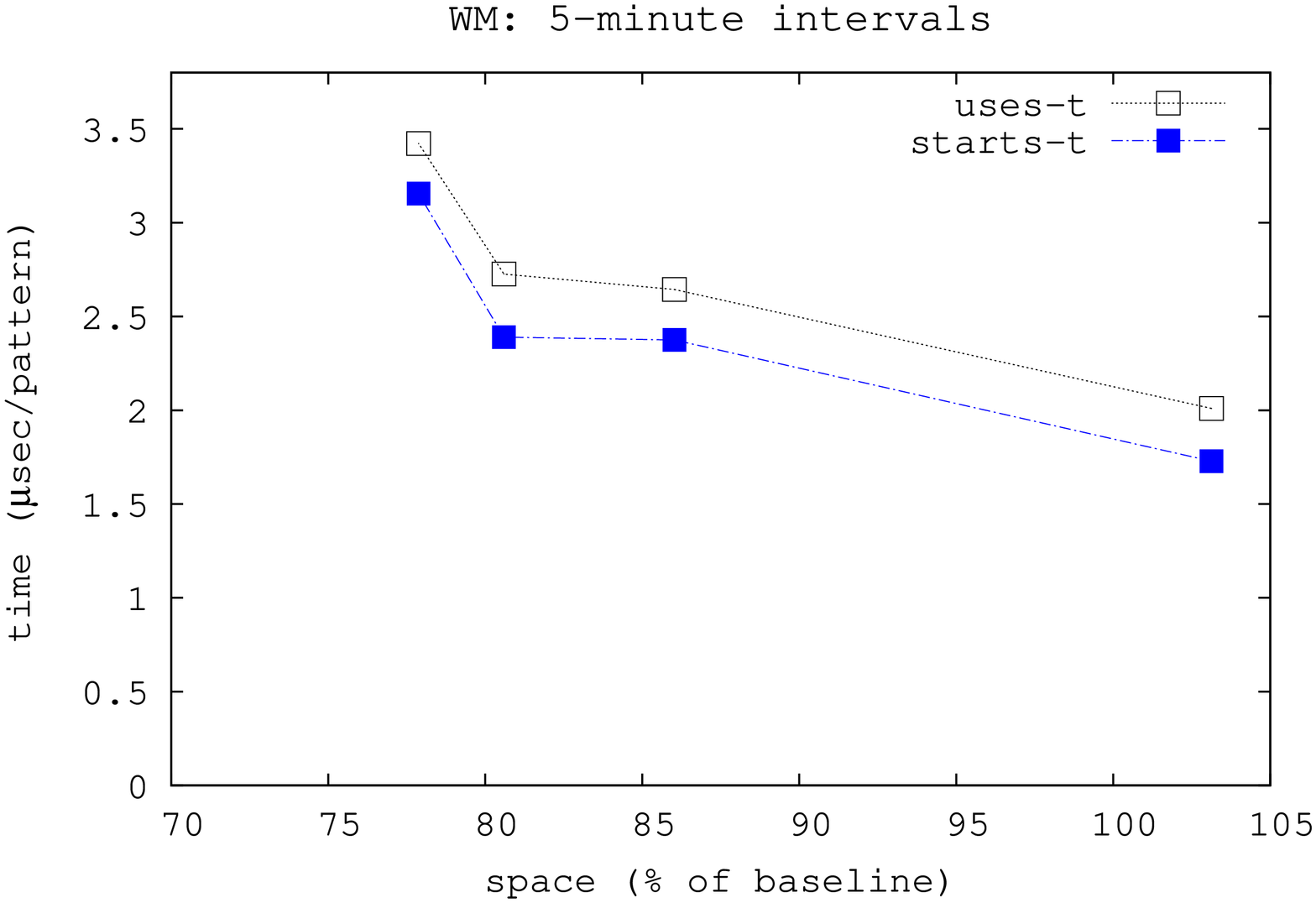}}
			{\includegraphics[angle=-90,width=0.45\textwidth]{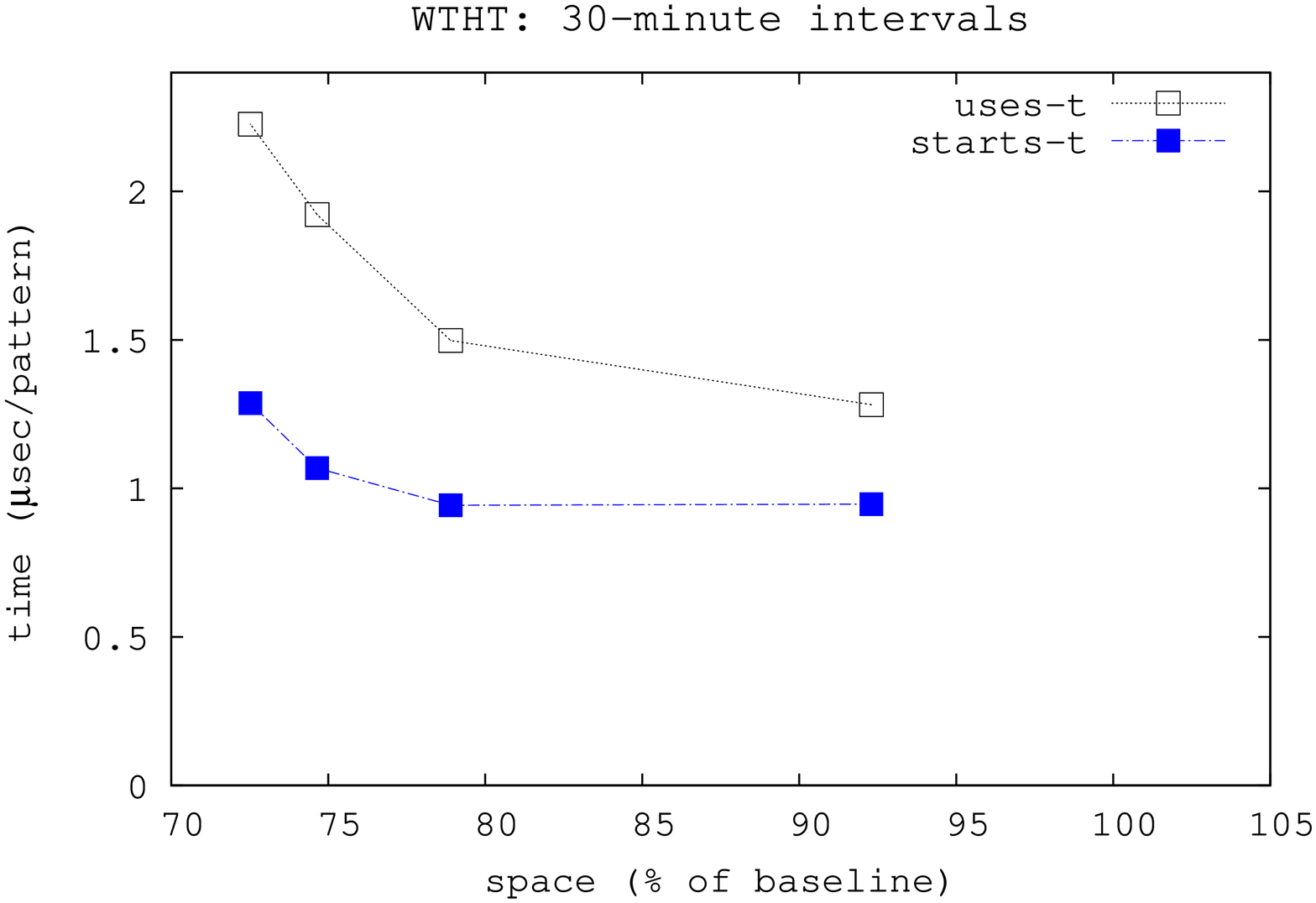}}
			{\includegraphics[angle=-90,width=0.45\textwidth]{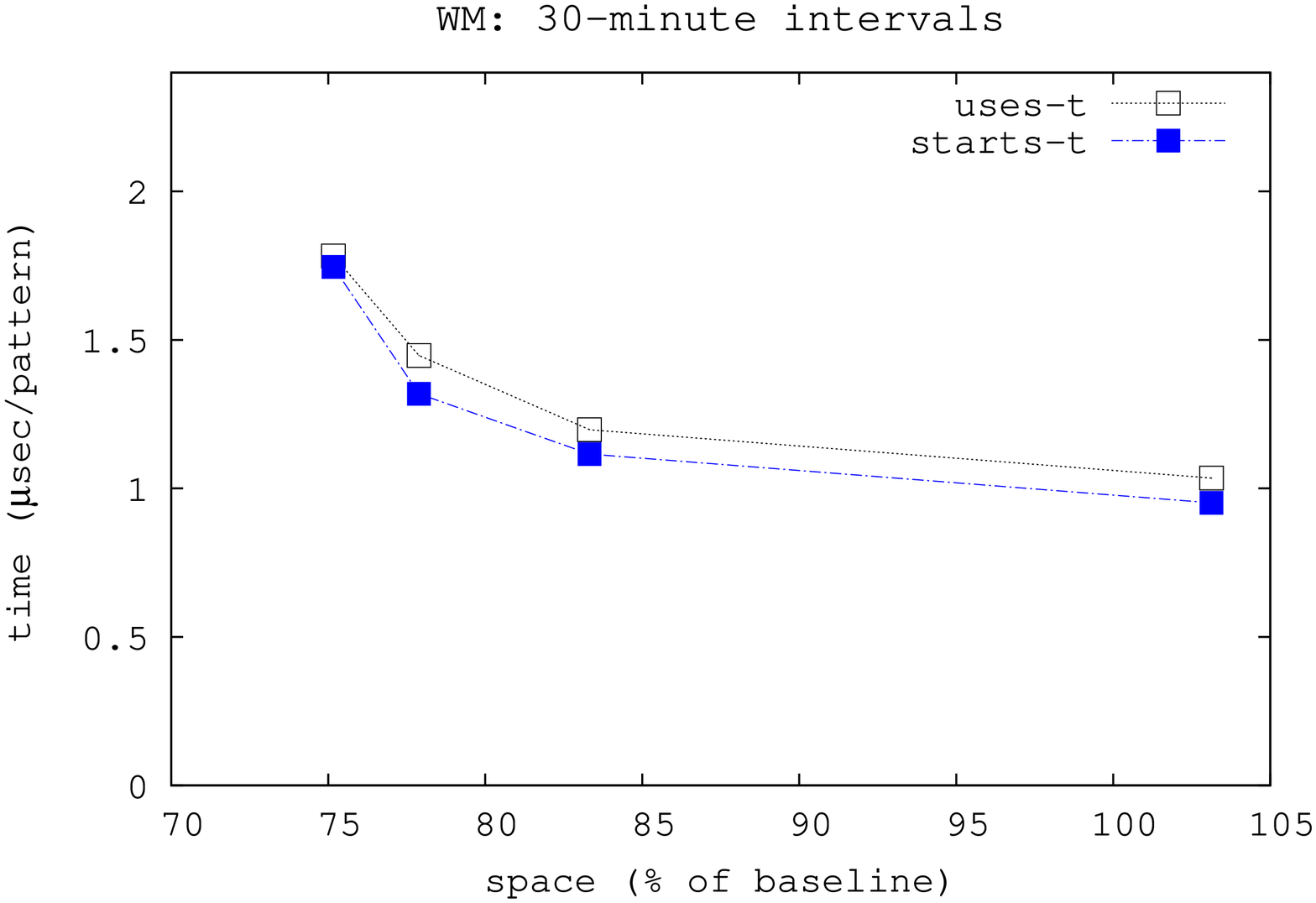}}
		\end{center}
	\end{center}
	\vspace{-0.5cm}
	\caption{Pure temporal queries for Madrid. $\repres$ uses either a $\wtht$ (left) or a $\wm$ (right). 
		Time granularity is $5$ minutes (top) or $30$ minutes (bottom).}
	\label{fig:madridst_topk}
	\vspace{-0.3cm}
%\end{figure}

%%%%%%%%%%% PORTO - PURE TEMPORAL %%%%%%%%%%%%%

%\begin{figure}[htb]
	%\vspace{-0.4cm}
	\begin{center}
			{\includegraphics[angle=-90,width=0.45\textwidth]{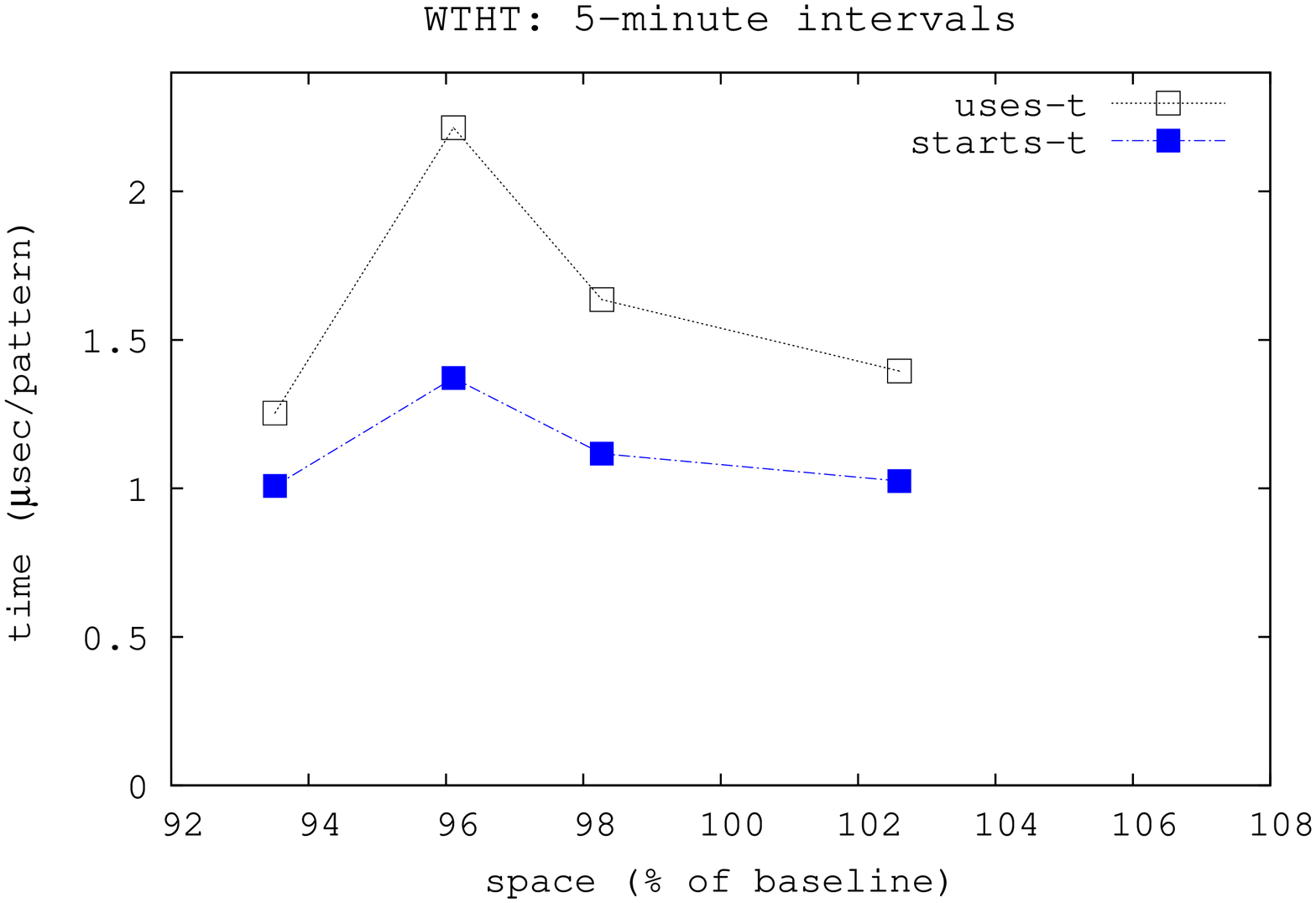}}
			{\includegraphics[angle=-90,width=0.45\textwidth]{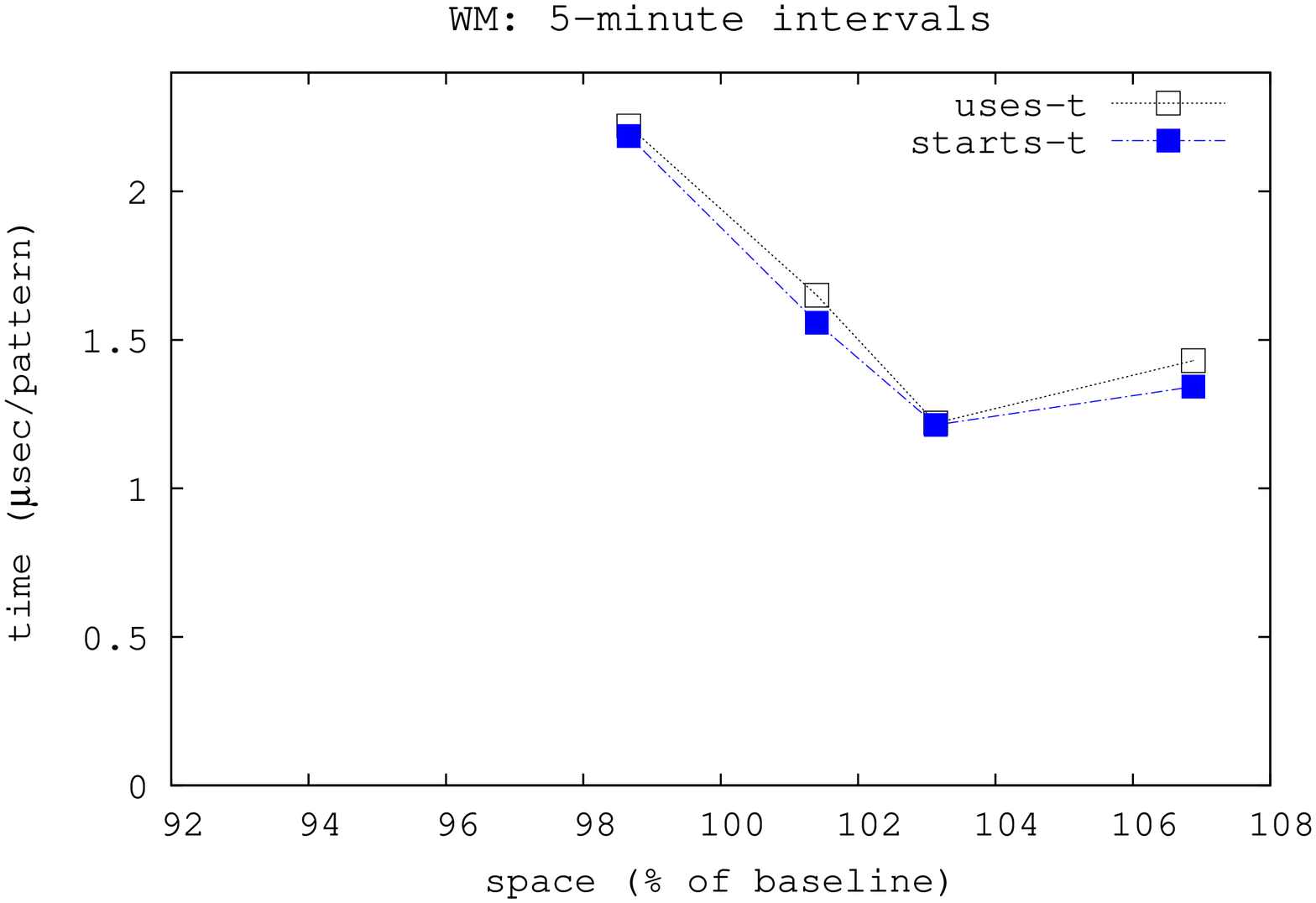}}
			{\includegraphics[angle=-90,width=0.45\textwidth]{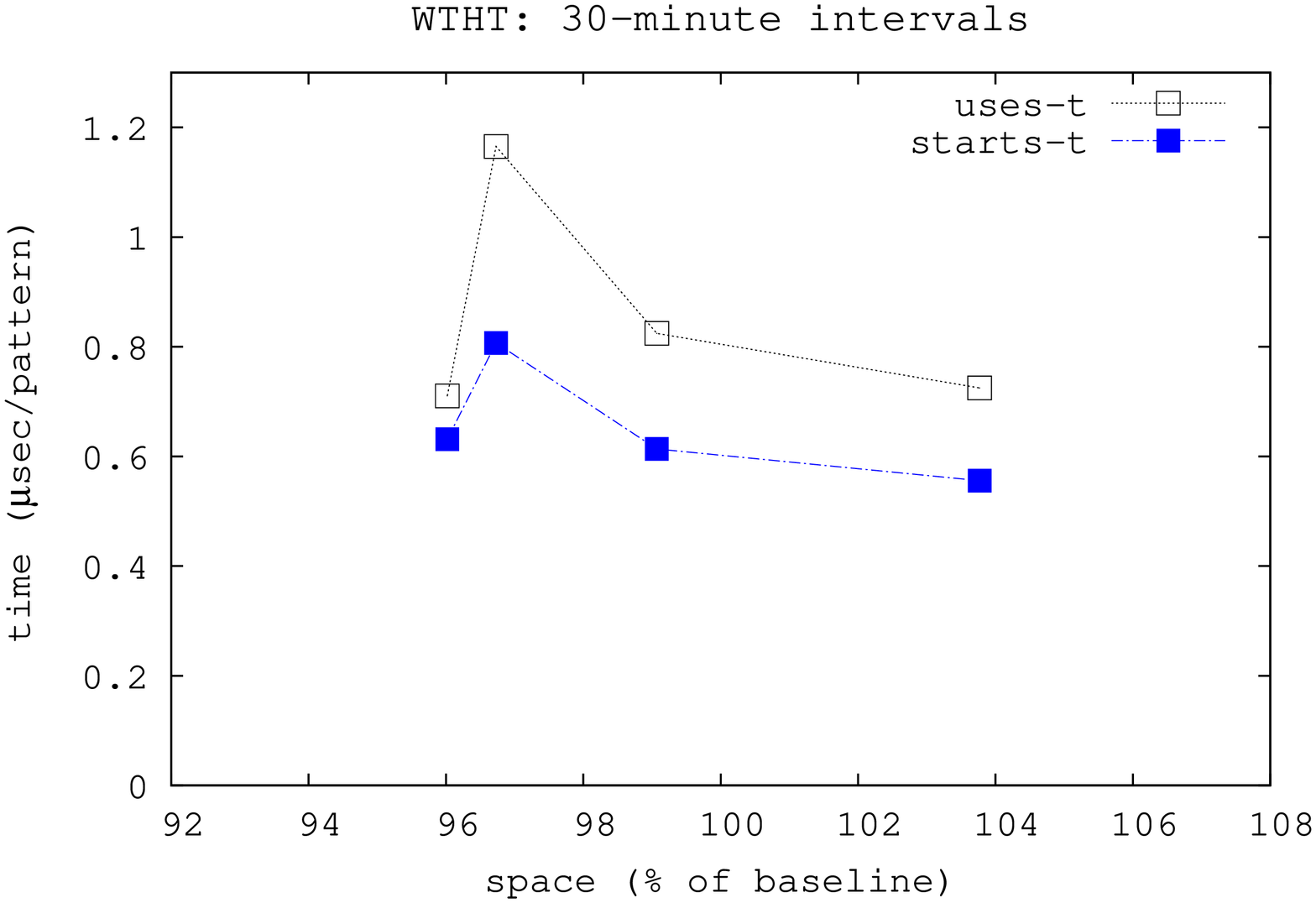}}
			{\includegraphics[angle=-90,width=0.45\textwidth]{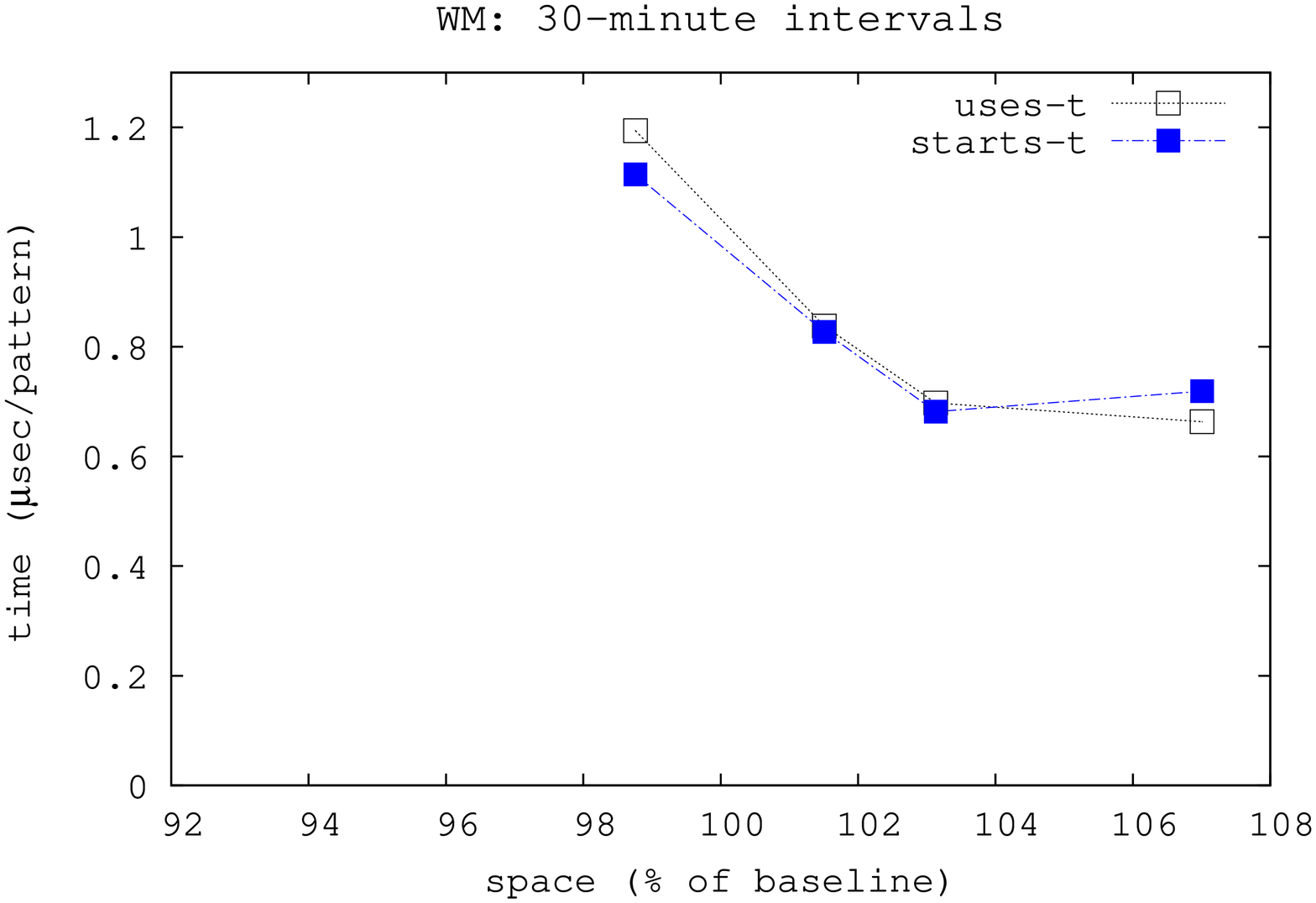}}
	\end{center}
	\vspace{-0.5cm}
	\caption{Pure temporal queries for Porto. $\repres$ uses either a $\wtht$ (left) or a $\wm$ (right). 
		Time granularity is $5$ minutes (top) or $30$ minutes (bottom).}
	\label{fig:portost_topk}
	%\vspace{-0.6cm}
\end{figure}

In this section we focus on the performance of the temporal component of $\repres$. We use the same 
configurations as in Table~\ref{table:ctr_wt_spaces} for $\wm$ and $\wtht$, and 
show the space/time trade-offs obtained when solving pure temporal queries. Figures~\ref{fig:madridst_topk}
and \ref{fig:portost_topk} present the results obtained at  \Tut\ and \Tst\ queries for Madrid and Porto datasets respectively. 
Note that, in this case, since the $\csa$ is not actually needed to solve temporal queries, we do not include its size
within the compression values (x-axis).

We can see that when running \Tut\ queries, both $\wtht$ and $\wm$ obtain rather similar times (requiring less than 4$\mu$s 
to perform a $count$ operation in all cases) and that those times improve as the height of the structure decreases. We can see 
that in the highest $\wtht$ and $\wm$, corresponding to using $5$-min intervals in Madrid dataset, \Tut\ requires less
than $3.5\mu$s. Then, when using $30$-min intervals, the time required to solve \Tut\ is always below $2.3\mu$s (yet
$\wm$ performs faster than $\wtht$ here), and those times are similar to the ones obtained for Porto dataset when
using $5$-min intervals. And finally, the best query times (below $1.2\mu$s) are obtained for Porto dataset with 
$30$-min intervals.

Regarding $\Tst$, recall that it also performs a $count$ operation, but within a smaller range ($[1,z]$) in comparison with the range
$[z+1,n]$ where  $count$ is performed for \Tut. We can see that, whereas $\wm$ obtains similar times to those of 
$\Tut$ query,  \Tst\ performs clearly faster than \Tut\ over $\wtht$.

%Due to the smaller vocabulary of times, in Porto dataset, we can see that all the indexes performs better for the top-10 times 
%query in Figure~\ref{fig:portost_topk}, although the binary version of the $\wt$ is still faster because the time 
%distribution is not uniform either.

As a final note, recall that in Madrid dataset, bitvector $RG$ always needs more space than $RRR$ counterparts 
whereas in Porto dataset (as discussed in Section~\ref{sec:exp.space})
$RG$ obtains the best space values when using $5$-min intervals and still requires less space than $RRR_{32}$ when using
$30$-min intervals. 
This is the reason why while plots for Madrid dataset are decreasing from left to right, in Porto
dataset the first point ($RG$) in the left figures ($5$-min intervals), and the third point ($RG$) 
in the right figures ($30$-min intervals) require less space than the others ($RRR$) and are also typically  faster. 
%This is mainly noticeable for queries \Tfxtys\ and \Tfxtyw.

% % % % % % % % % % % % % % % % % % % % % % % % % % % % % % % % % % % % % % % % % % % % % % % % % % % % % % %
% % % % % % % % % % % % % % % % % % % % % % % % % % % % % % % % % % % % % % % % % % % % % % % % % % % % % % %
\subsubsection{Space/time trade-off when performing  spatio-temporal queries}
% % % % % % % % % % % % % % % % % % % % % % % % % % % % % % % % % % % % % % % % % % % % % % % % % % % % % % %
% % % % % % % % % % % % % % % % % % % % % % % % % % % % % % % % % % % % % % % % % % % % % % % % % % % % % % %

\begin{figure}[!ht]
	%\vspace{-0.4cm}
	\begin{center}
		{\includegraphics[angle=-90,width=0.45\textwidth]{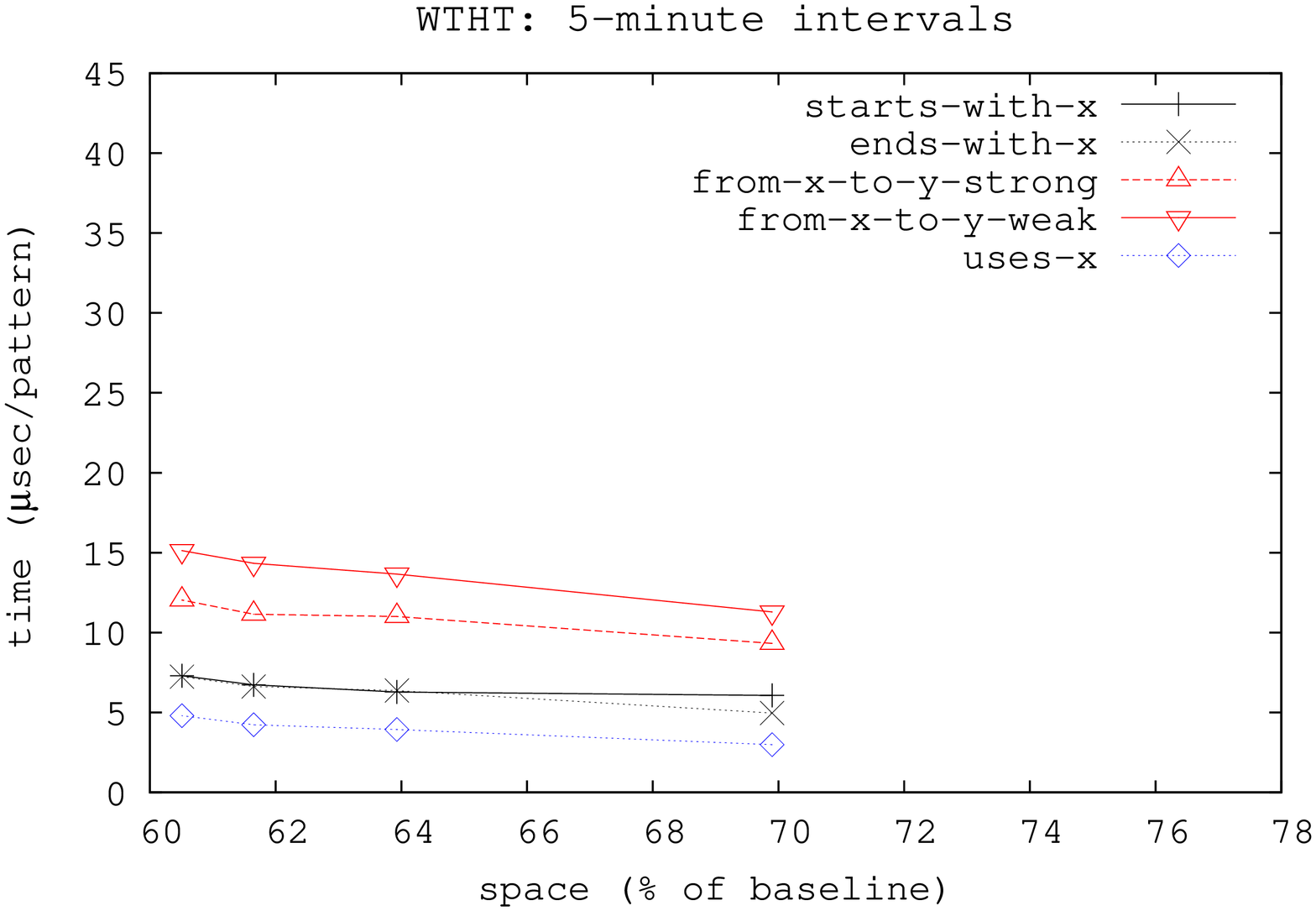}}
		{\includegraphics[angle=-90,width=0.45\textwidth]{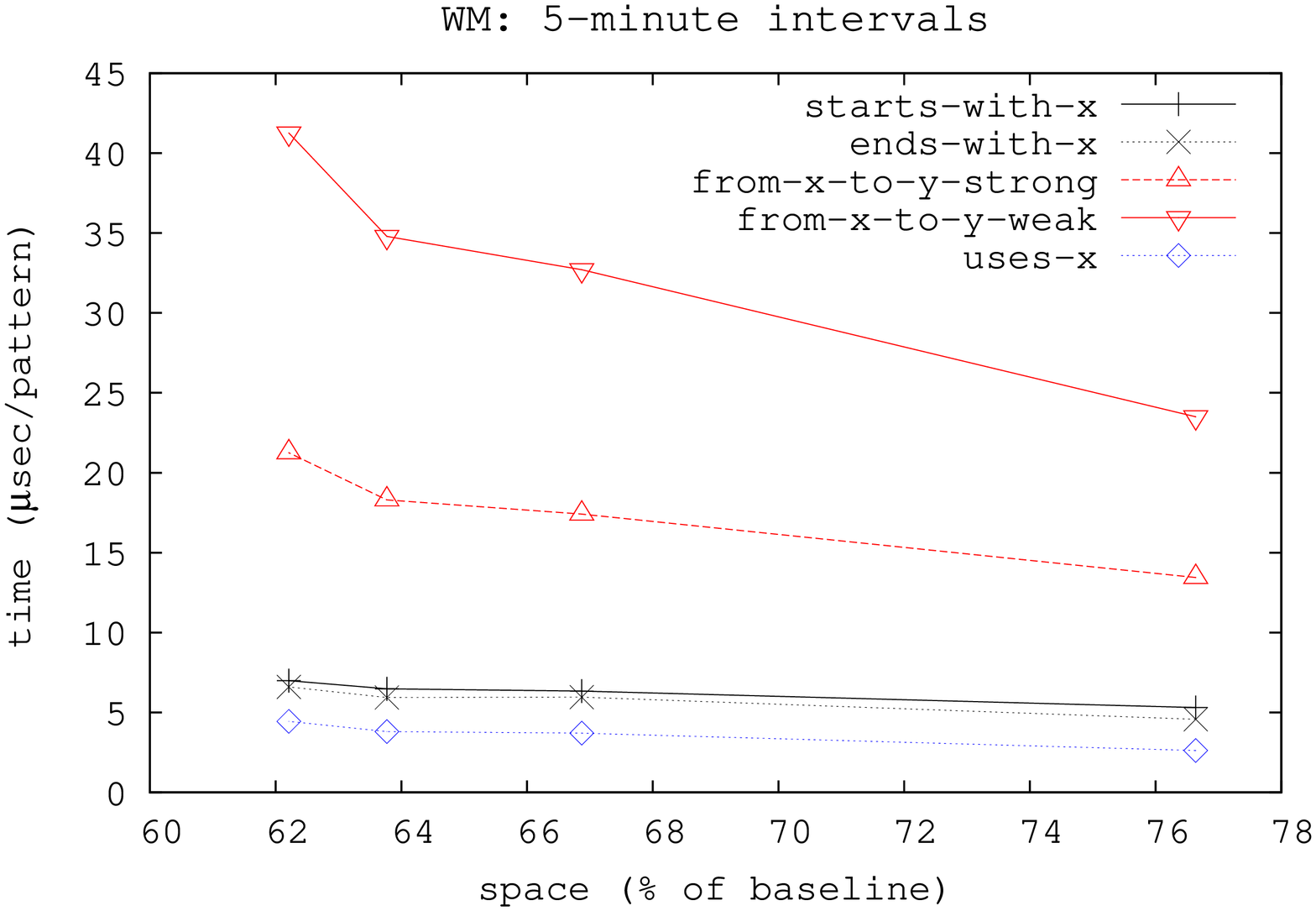}}
		{\includegraphics[angle=-90,width=0.45\textwidth]{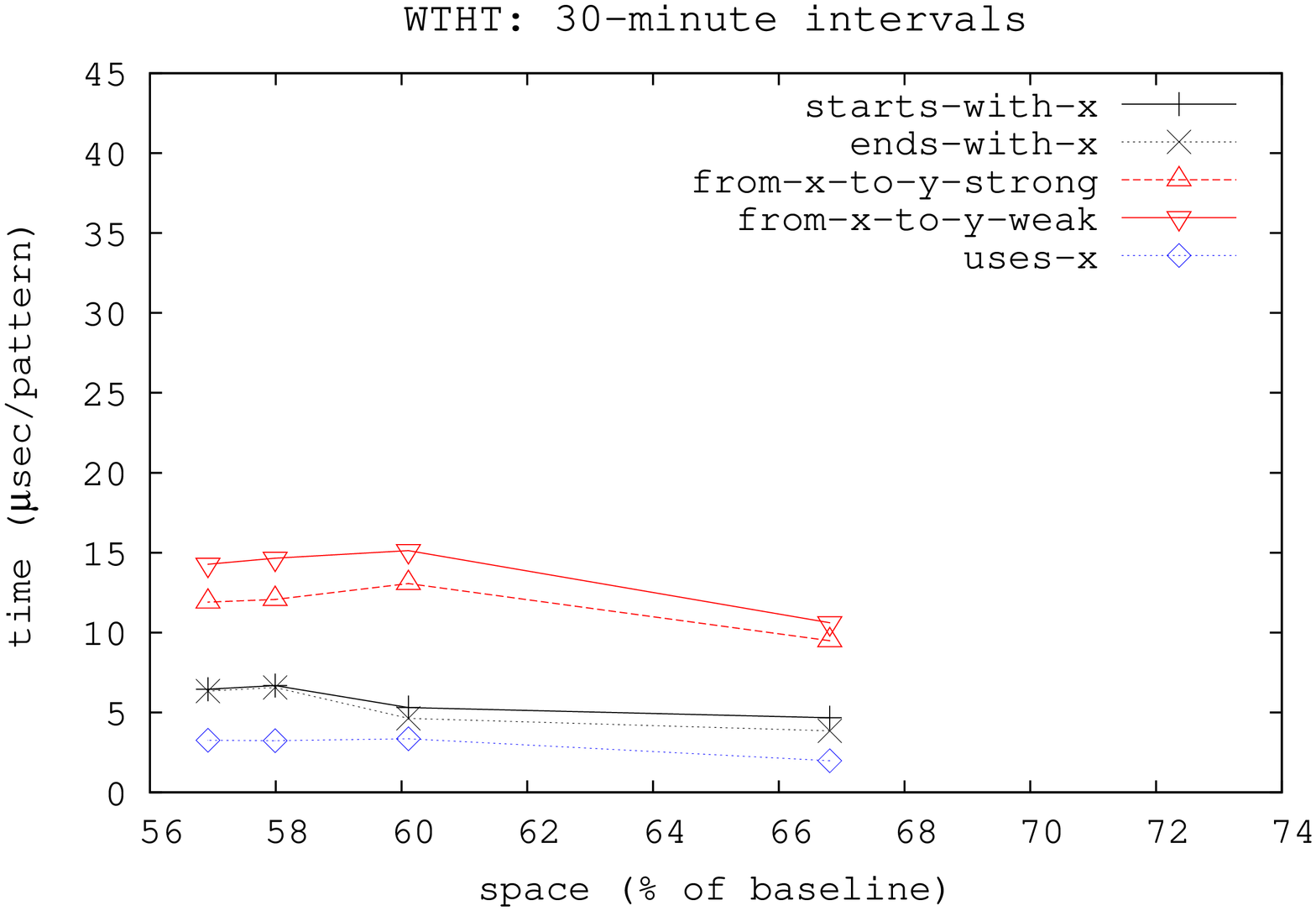}}
		{\includegraphics[angle=-90,width=0.45\textwidth]{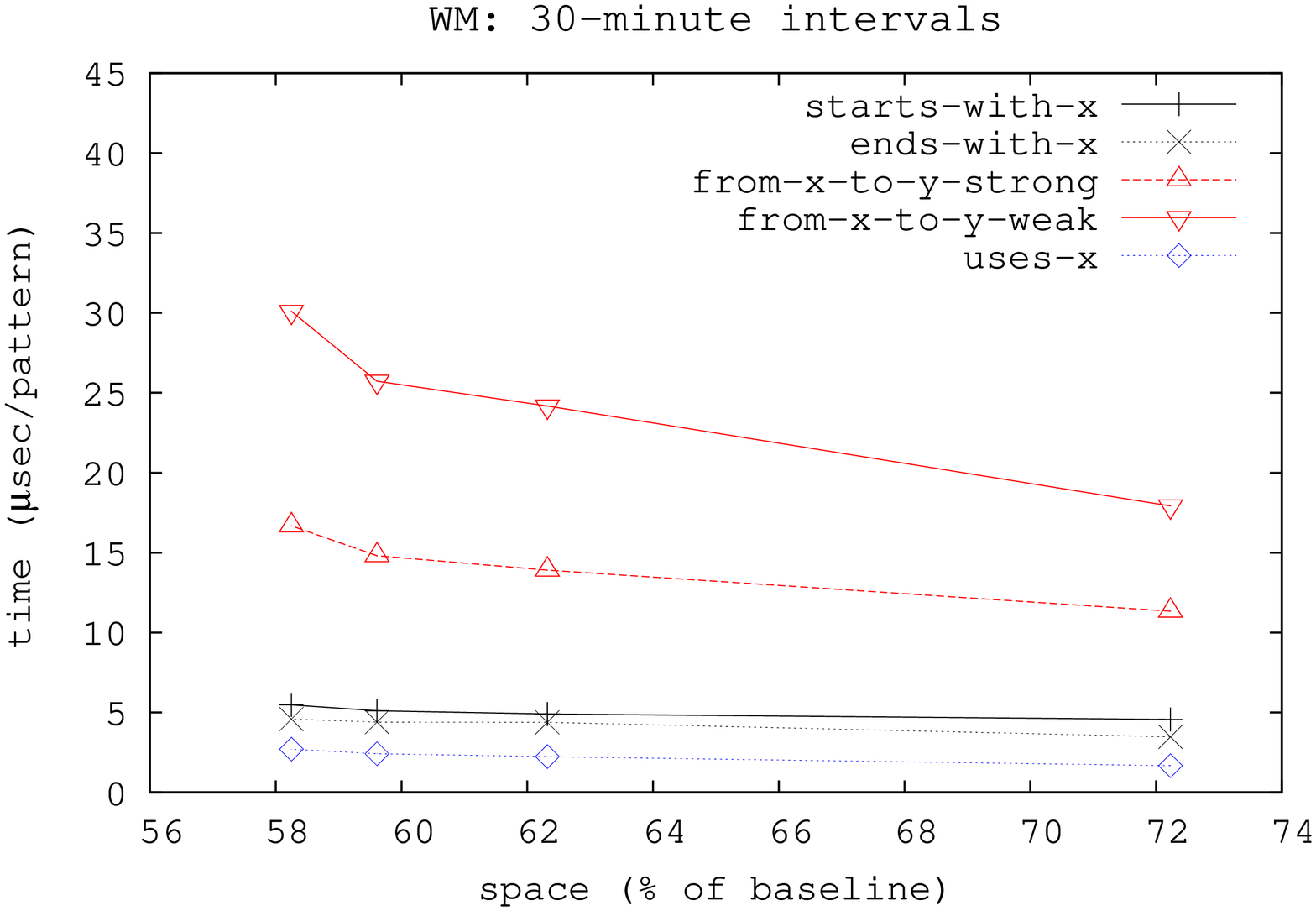}}
		
	\end{center}
	\vspace{-0.3cm}
	\caption{Spatio-temporal queries for Madrid. $\repres$ uses either a $\wtht$ (left) or a $\wm$ (right). 
		Time granularity is $5$ minutes (top) or $30$ minutes (bottom). 
	}
	\label{fig:madridst}
	\vspace{-0.3cm}
%\end{figure}

%%%%%%%%%%% PORTO - SPATIO-TEMPORAL %%%%%%%%%%%%%

%\begin{figure}[!ht]
	%\vspace{-0.4cm}
	\begin{center}
		{\includegraphics[angle=-90,width=0.45\textwidth]{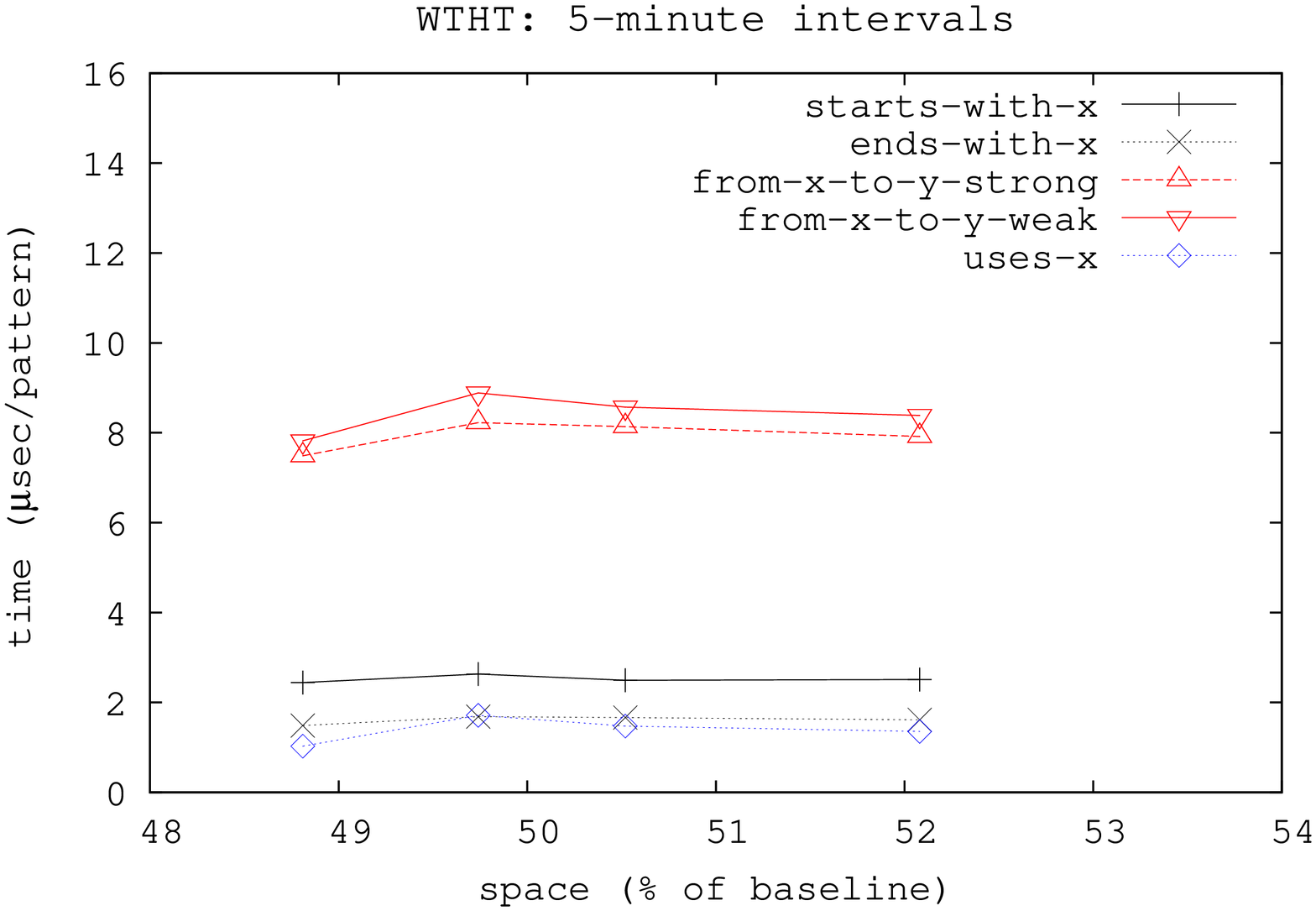}}
		{\includegraphics[angle=-90,width=0.45\textwidth]{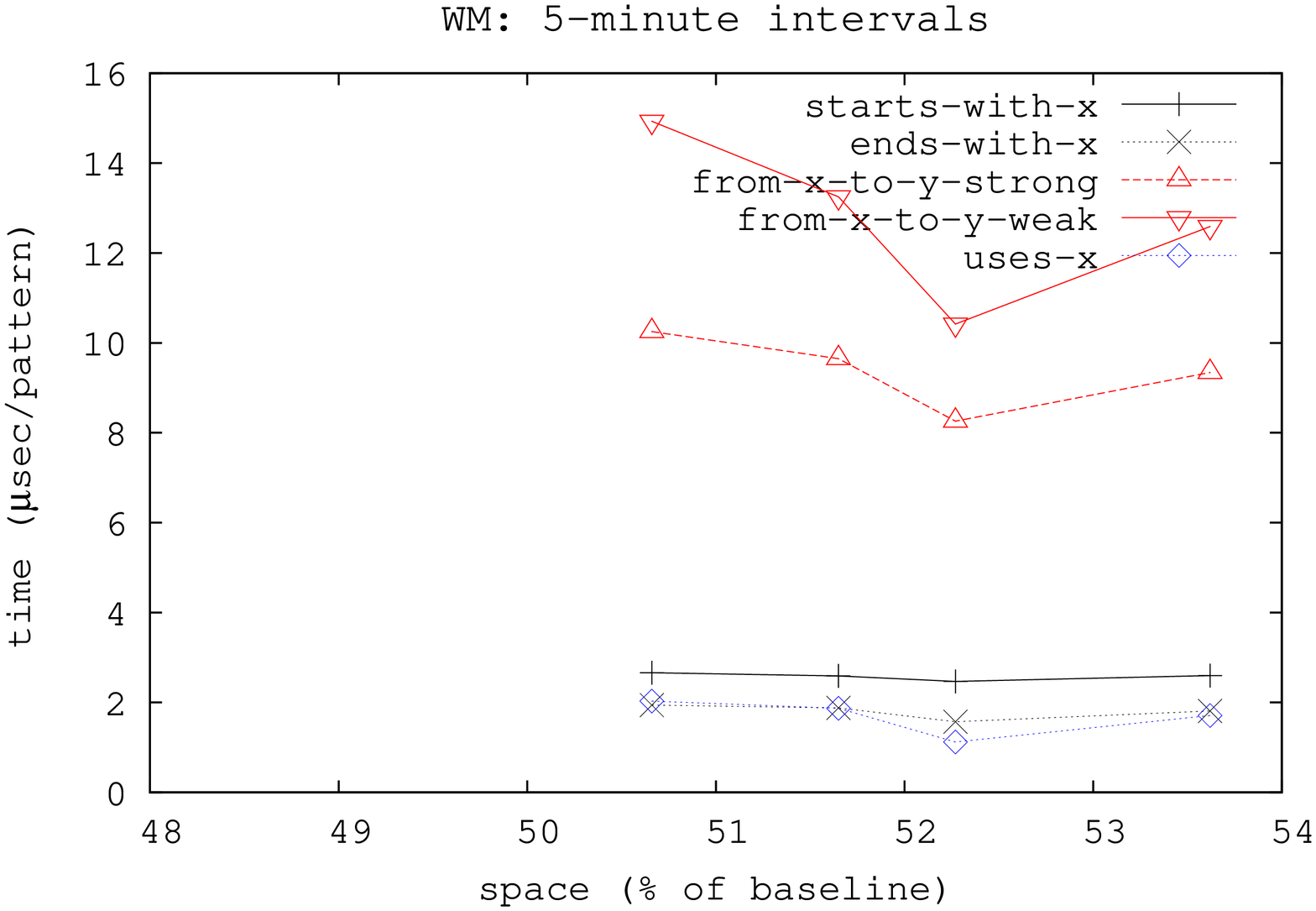}}
		{\includegraphics[angle=-90,width=0.45\textwidth]{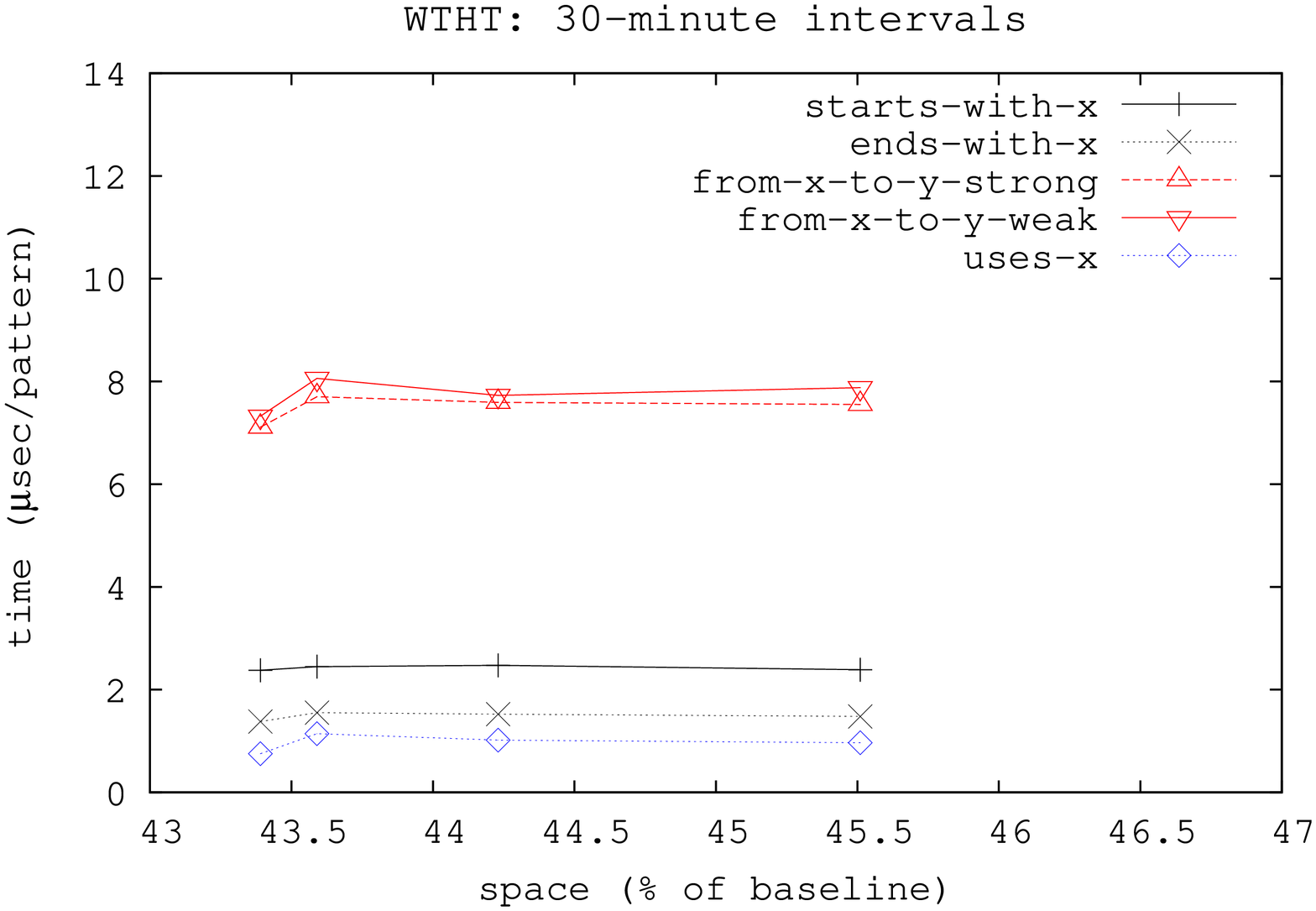}}
		{\includegraphics[angle=-90,width=0.45\textwidth]{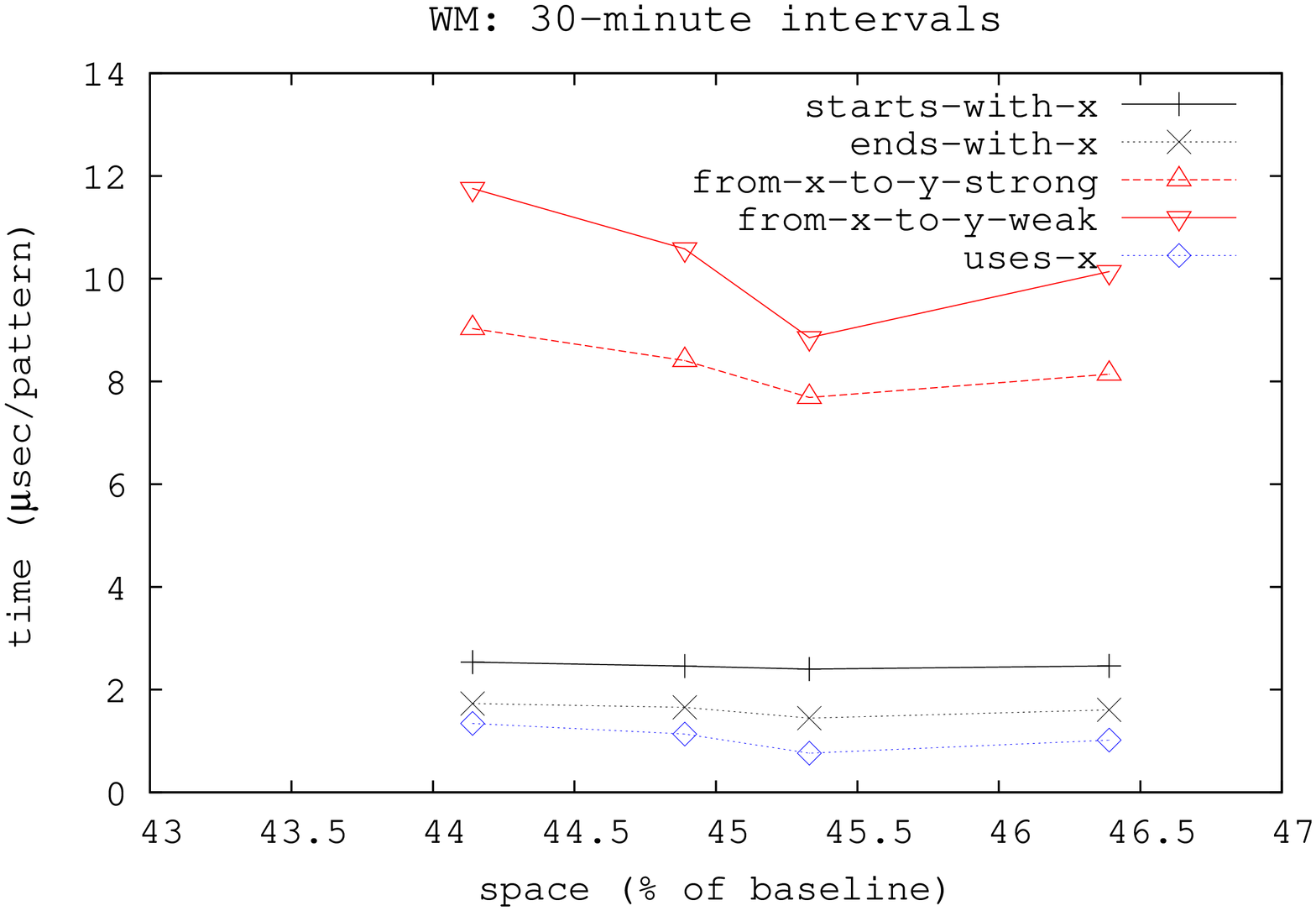}}

	\end{center}
	\vspace{-0.3cm}
	\caption{Spatio-temporal queries for Porto. $\repres$ uses a fixed $t_{\Psi}=32$ for $\csa$, 
		and either a $\wtht$ (left) or a $\wm$ (right). 
		Time granularity is $5$ minutes (top) or $30$ minutes (bottom). 
	}
	\label{fig:portost}
	%\vspace{-0.6cm}
\end{figure}

In Figures~\ref{fig:madridst} and \ref{fig:portost} we show the space/time tradeoff obtained by $\repres$ when 
dealing with spatio-temporal queries. Recall that this type of queries require both using the $\csa$, to 
exploit indexed access to the nodes in the trips, and the 
temporal component of $\repres$ to handle temporal constraints. In this case, the space values showed in
the figures include both the size of $\csa$ and that of either $\wm$ or $\wtht$. Therefore, we also show the
overall space needs of $\repres$. In the case of $\csa$ we
have set $t_{\Psi}=32$ (a fixed dense sampling), and for $\wm$ and $\wtht$ we used again the same 
configurations as in the previous sections obtained by varying the bitvectors and the temporal discretization. 

%%%%%%%%%%% MADRID -SPATIO-TEMPORAL %%%%%%%%%%%%%

For queries \Tswx, \Tewx, and \Tux\ we can see typically small differences between using $\wm$ or $\wtht$. In 
Madrid dataset, $\wm$ overcomes $\wtht$ being $2$-$30$\% faster in these types of queries. 
However, in Porto dataset $\wtht$ is slightly 
faster (from $1$ to $25$\%) than its $\wm$ counterpart.

For queries \Tfxtys\ and \Tfxtyw\ we can see a big gap between the times reported by $\wtht$ and $\wm$.
This gap arises because in $\wm$ we have used exactly the $countLR$ operation discussed in Section~\ref{sec:stq}
that is implemented with two calls to the $count$ operation from the $\wm$.\footnote{For $\wm$ we used exactly the same 
	implementation in \cite{CNO15} and simply added the new operation $countLR$ that calls the underlying $count$ from the
	$\wm$. }  
However, in our implementation of
$\wtht$ we have engineered an improved version of $countLR$ where, during the execution of $count$, we also report
$\alpha'$ and  $\beta'$, hence avoiding two calls to $count$.
 
%
%\OJOFARI{Reescribir para  decir que WM usa 2 counts para implementar countLR, mientras que WTHT usa una 
%versión modificada del count (en la que se hace 1 sola bajada). Modifcar tambien en sec 5, y quitar la busqueda binaria}

%As a final note, recall that in Madrid dataset, bitvector $RG$ always needs more space than $RRR$ counterparts 
%whereas in Porto dataset (as discussed in Section~\ref{sec:exp.space})
%$RG$ obtains the best space values when using $5$-min intervals and still requires less space than $RRR_{32}$ when using
%$30$-min intervals. 
%This is the reason why while plots for Madrid dataset are decreasing from left to right, in Porto
%dataset the first point ($RG$) in the left figures ($5$-min intervals), and the third point ($RG$) 
%in the right figures ($30$-min intervals) requires less space than the others ($RRR$) and is also typically  faster. 
%This is mainly
%noticeable for queries \Tfxtys\ and \Tfxtyw.

%%%%%%%%%%% MADRID -SPATIO-TEMPORAL - TOP-K %%%%%%%%%%%%%
\begin{figure}[!ht]
	%\vspace{-0.4cm}
	\begin{center}

		{\includegraphics[angle=-90,width=0.45\textwidth]{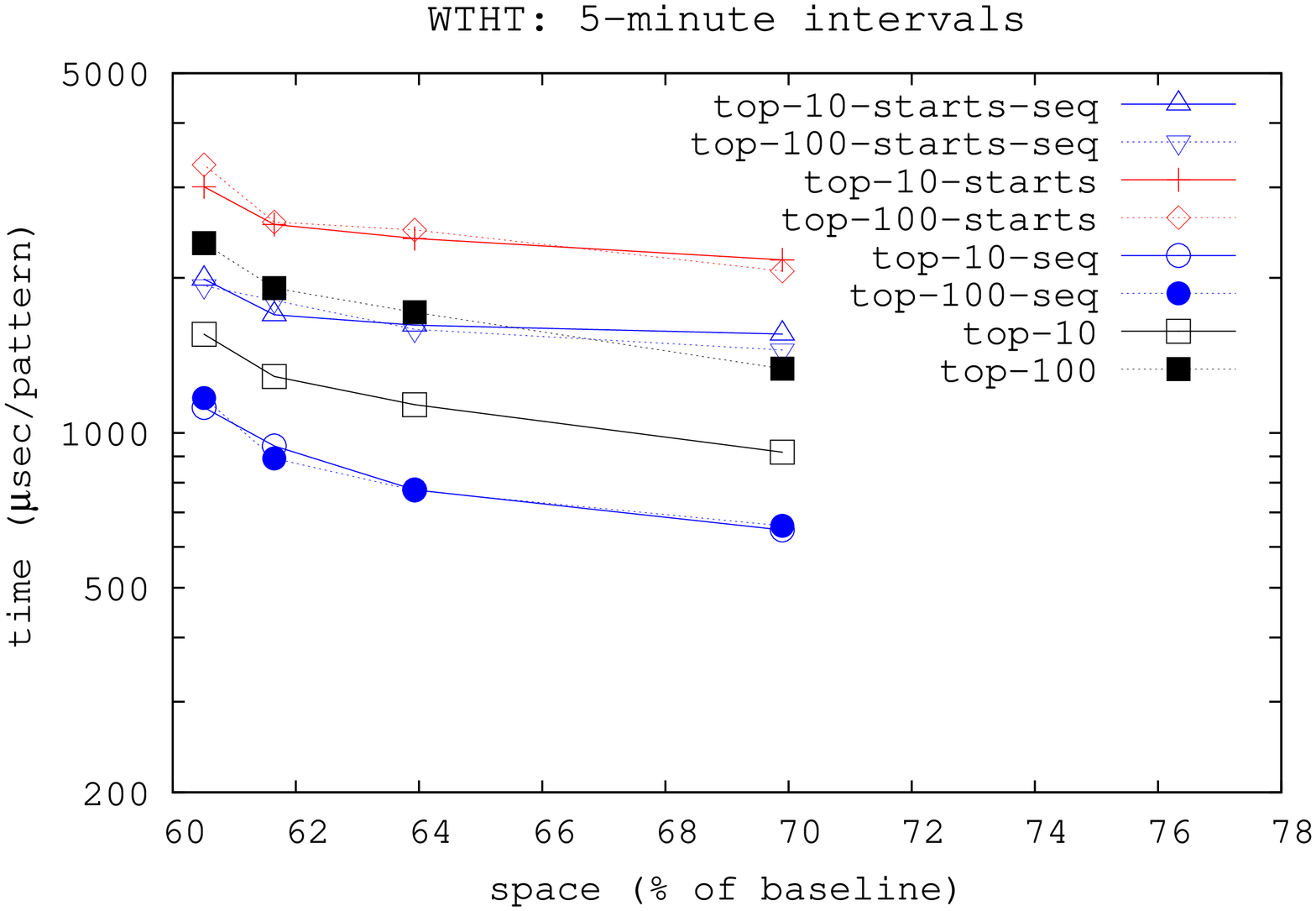}}
		{\includegraphics[angle=-90,width=0.45\textwidth]{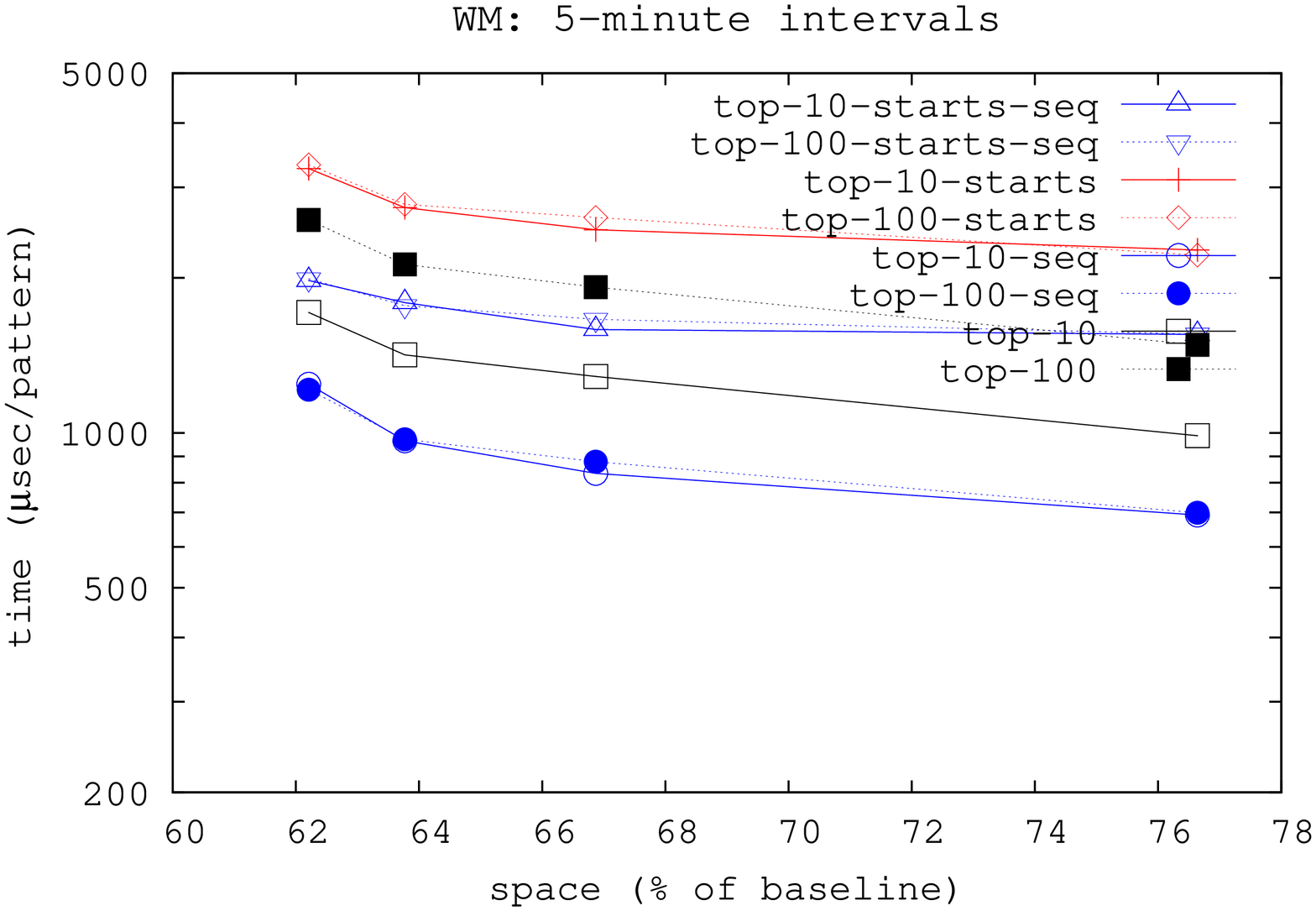}}
		{\includegraphics[angle=-90,width=0.45\textwidth]{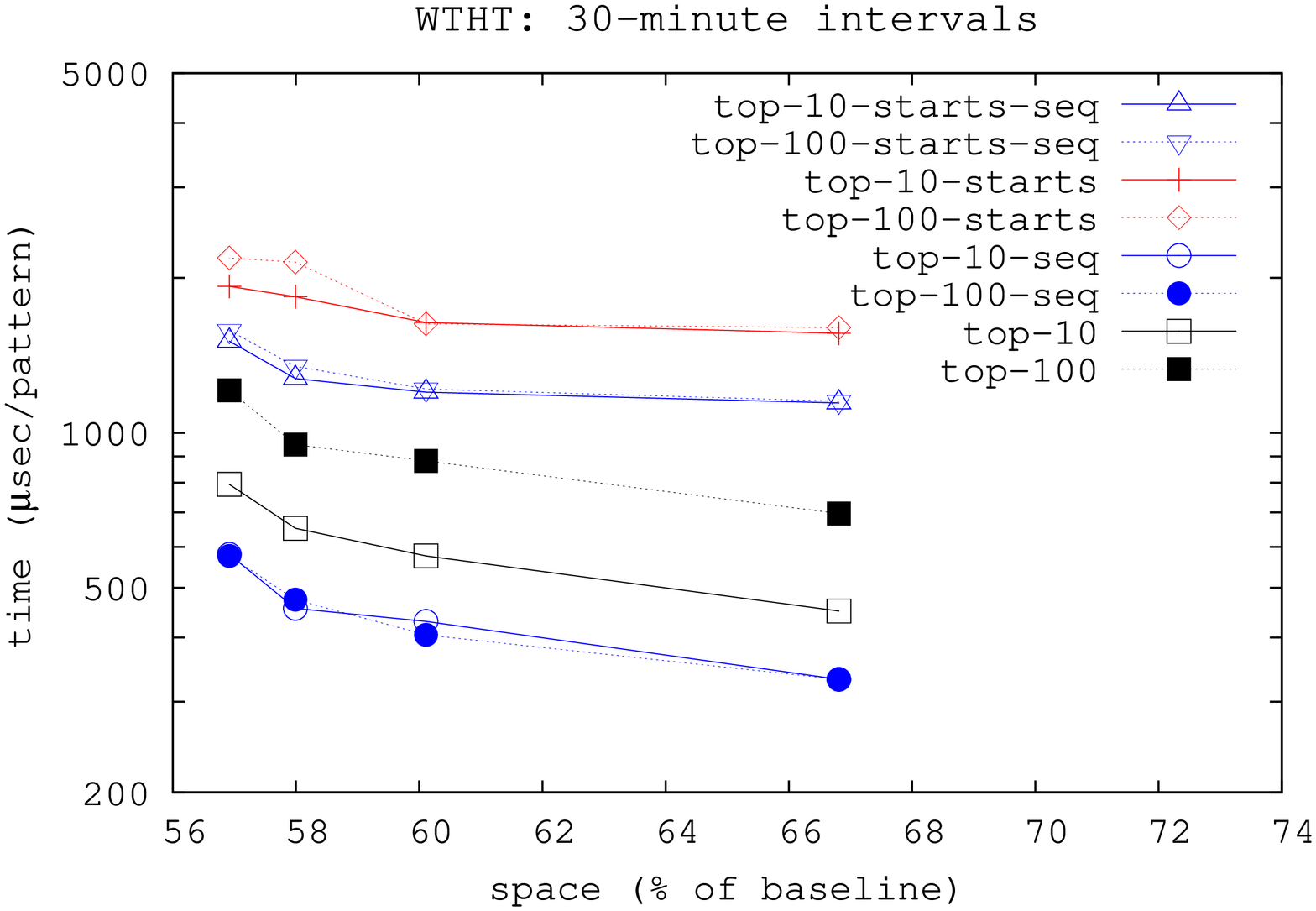}}
		{\includegraphics[angle=-90,width=0.45\textwidth]{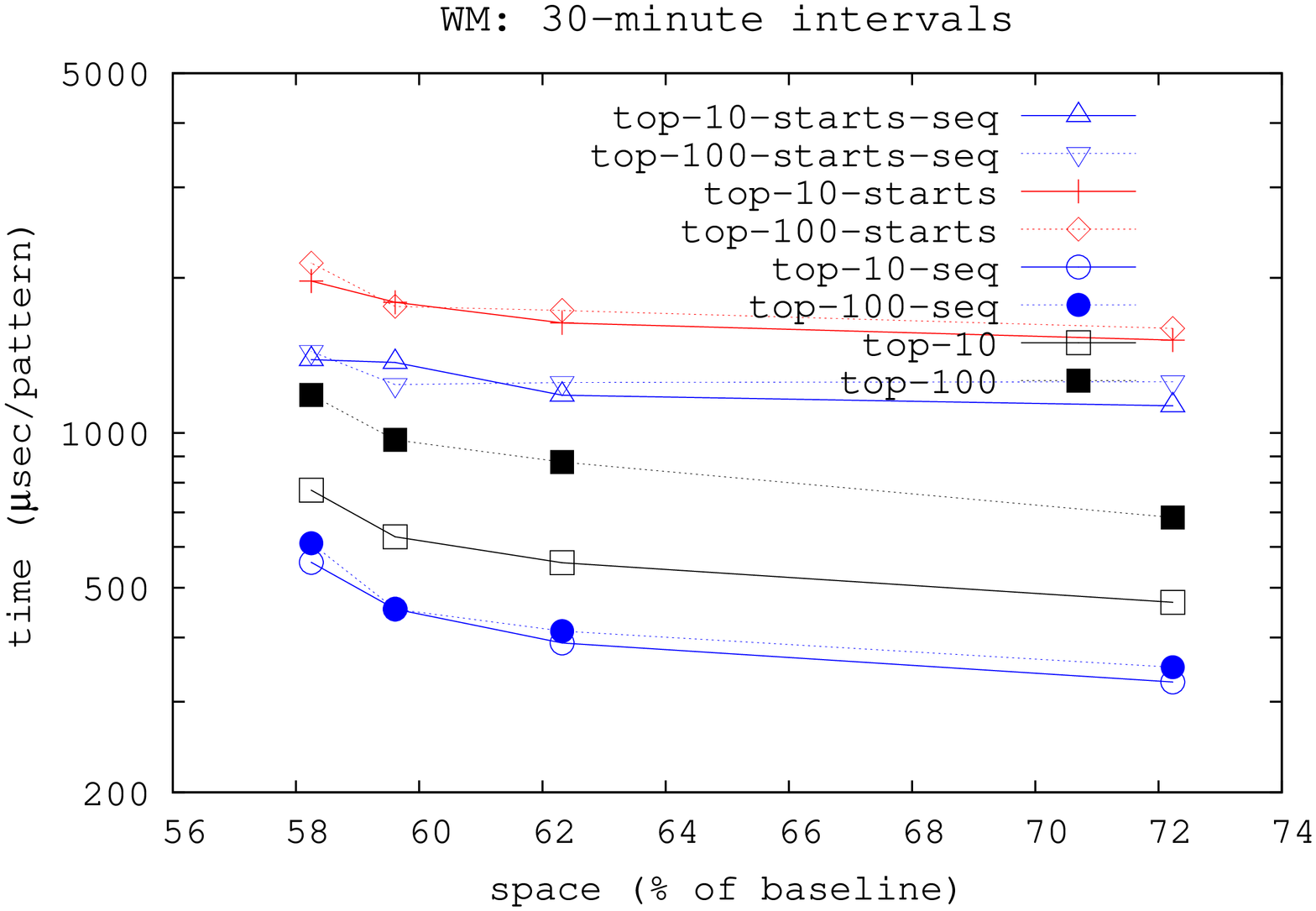}}
		
	\end{center}
	\vspace{-0.3cm}
	\caption{Spatio-temporal {\Stk\ and \Stks} queries for Madrid. $\repres$ uses either a $\wtht$ (left) or a $\wm$ (right). 
		Time granularity is $5$ minutes (top) or $30$ minutes (bottom). 
	}
	\label{fig:madridst.tk}
	%\vspace{-0.3cm}
%\end{figure}

%%%%%%%%%%% PORTO - SPATIO-TEMPORAL - topk %%%%%%%%%%%%%
%\begin{figure}[!ht]
	\vspace{-0.2cm}
	\begin{center}

		{\includegraphics[angle=-90,width=0.45\textwidth]{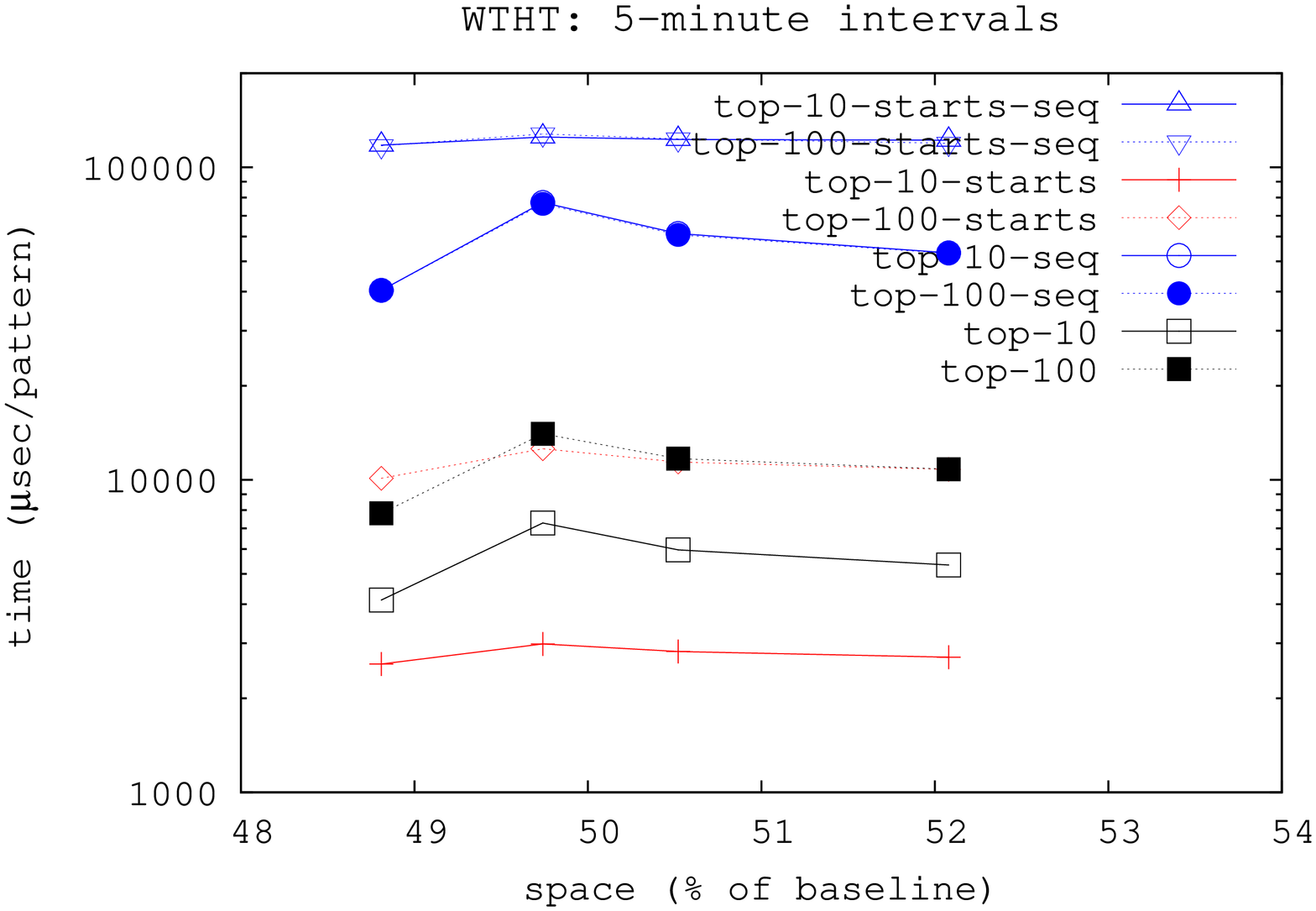}}
		{\includegraphics[angle=-90,width=0.45\textwidth]{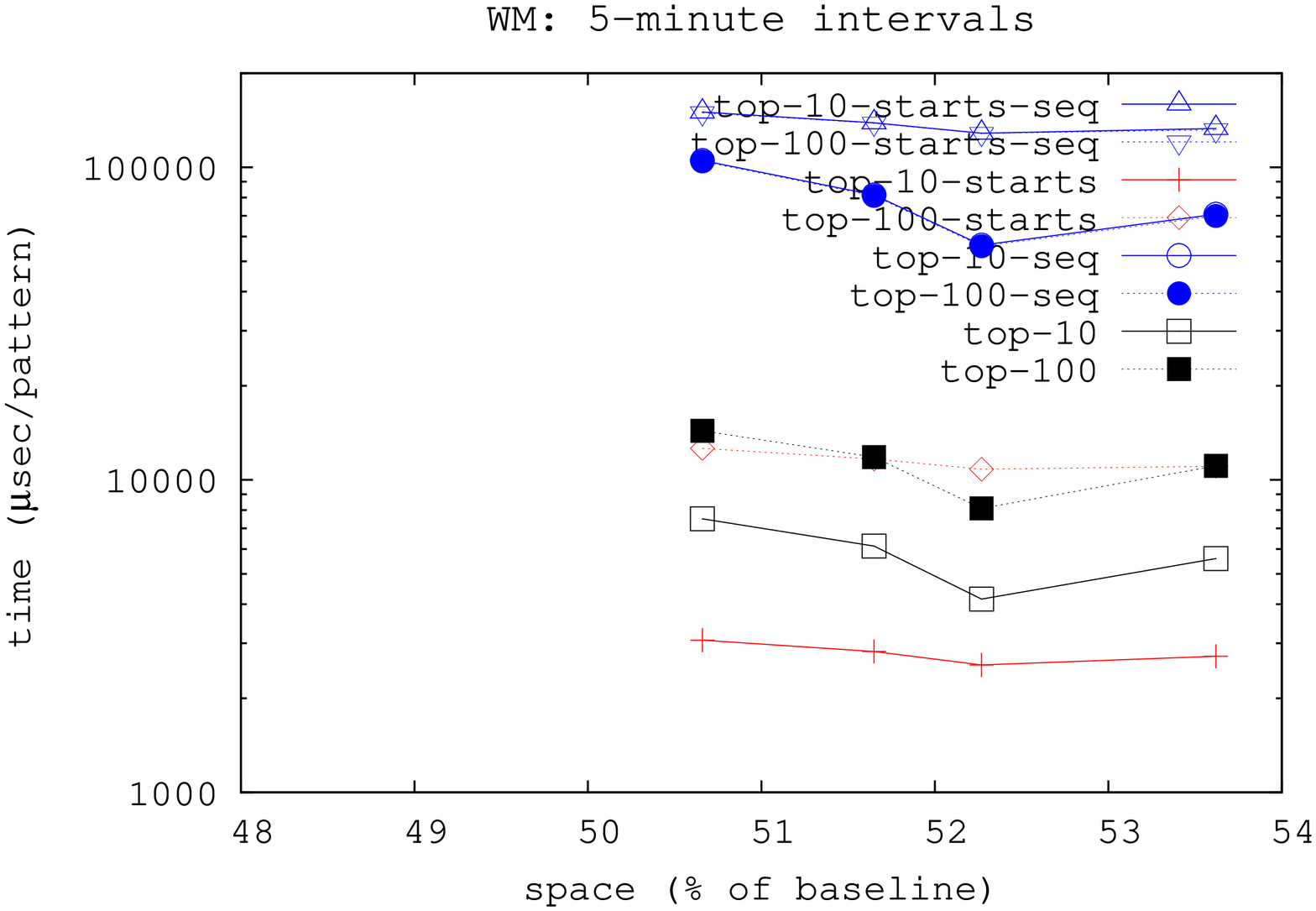}}
		{\includegraphics[angle=-90,width=0.45\textwidth]{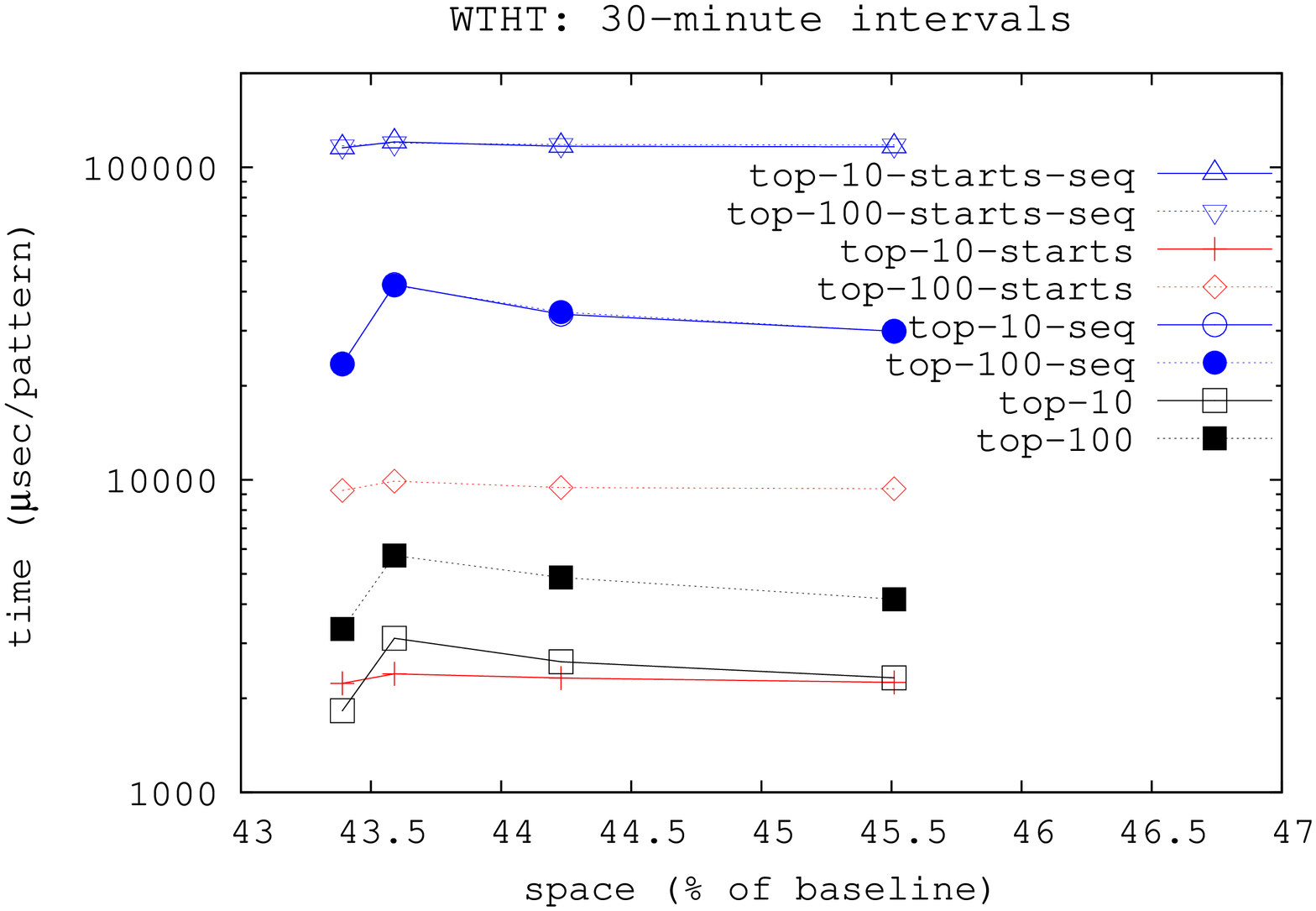}}
		{\includegraphics[angle=-90,width=0.45\textwidth]{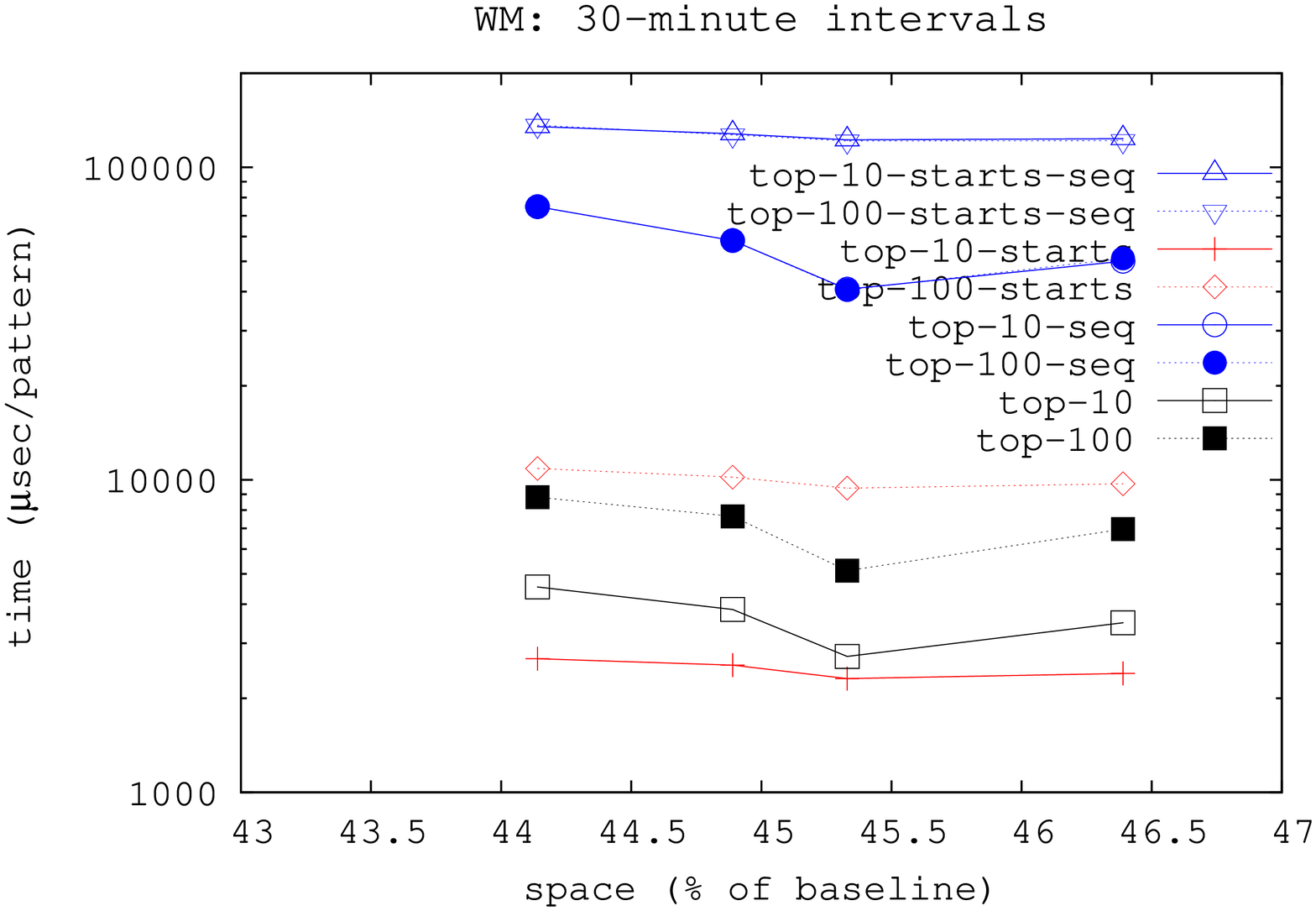}}
		
	\end{center}
	\vspace{-0.3cm}
	\caption{Spatio-temporal {\Stk\ and \Stks} queries for Porto. $\repres$ uses a fixed $t_{\Psi}=32$ for $\csa$, 
		and either a $\wtht$ (left) or a $\wm$ (right). 
		Time granularity is $5$ minutes (top) or $30$ minutes (bottom). 
	}
	\label{fig:portost.tk}
	\vspace{-0.3cm}
\end{figure}

%The experimental results for the spatio-temporal queries (as well as the pure temporal \texttt{uses-t} query) can be found 
%in Figure~\ref{fig:madridst}. This time we calculate the space usage of the whole $\repres$ structure, with a fixed 
%$\Psi_{sampling}$ of 32 and varying the bitvectors for the temporal structures.

%\DAGAL{Oculto la implementación para obtener $[\alpha',\beta']$ de forma óptima a propósito: la implementación es un cristo
% y sólo se usa para una consulta...}
%There is a big difference in the \texttt{from-x-to-y} queries between the Hu-Tucker $\wt$ and the $\wm$ structures as the 
%former has a more efficient implementation for reporting the limits of $[\alpha',\beta']$ (mentioned in Section~\ref{sec:stq}). 
%If needed, the $\wm$ could be improved with a similar optimization in the future. For the rest of the queries, the 
%difference in time efficiency is barely noticeable for both structures and even both time intervals.$\Psi_{sampling}$ 
%of 32 and varying the bitvectors for the temporal structures.

Finally, we also include results for \STtk\ and \STtks\ queries in Figures~\ref{fig:madridst.tk} and \ref{fig:portost.tk}. 
As explained in Section~\ref{sec:exp:temp}, the sequential approach is preferred when the frequency distribution of nodes is
rather uniform (Madrid dataset). Otherwise, the binary-partition counterpart outperforms it. The need for applying a temporal
constraint simply accentuates this effect in comparison with the corresponding pure spatial queries.

%\ART{Si yo hiciera la consulta espacio-tiempo de los top-k stops during a time  interval, pero considerara todo el %interval de tiempo del sistema. Que tan diferente es a solo hacerla espacial? }

\subsection{Discussion: solving queries on CTR Vs Pre-computing counters}
Along this section we have focused our experiments on using $\repres$ to answer our  set of queries. However, all those queries could somehow be pre-computed in such a way that the could be solved faster at the cost of dealing with additional supporting structures. For example, all the spatial queries could be pre-computed with tables that store for each node, or pair of nodes, the corresponding counters. In any case, all those tables would occupy less than 2MB. 
In the case of spatio-temporal queries, we could create a straightforward structure where, for each node, we used a sparse array to keep the counters for each time instant. These would permit us to solve time interval queries by summing up the counters matching the temporal constraint of the query. We have implemented those  structures for queries \Tswx, \Tewx, and \Tux\ for Madrid and Porto datasets when using $5$-minute intervals. In all cases the pre-computed structure for each query occupies around 20MB and permits us to answer those queries from $1$ to $2$ orders of magnitude faster than $\repres$. 
Finally, we also created a simple pre-computed structure to handle the spatio-temporal query \Tfxty. We used a sparse array that, for each pair $\langle X,Y \rangle$, keeps a counter for
all those trajectories that started at $t_s$ and ended at $t_e$.\footnote{Source code at \url{https://github.com/dgalaktionov/compact-trip-representation/blob/master/src/buildFacade.h}} In this case, the resulting structure roughly occupies from $130$ to $160$ MB in memory. That is, around  $40$-$50$\% the size of our plain baseline, which is approximately the same space required by $\repres$. Again times are roughly from $3$ to $50$ times faster than in $\repres$. 

We have seen  above that a simple and straightforward implementation of additional pre-computed structures can handle most of the queries proposed in this work and improve the query times obtained with a solution based on compact data structures such as $\repres$. Yet, $\repres$ still owns some advantages: {\em i)} $\repres$ actually keeps all the trips implicitly in a compressed and self-indexed way. Therefore, it avoids the need of storing them apart for the case in which we had to support further queries. {\em ii)} In some management scenarios, not all the queries can be pre-computed. For example, some indicators in the context of transportation networks require {\em ``counting the number of trips that went through two nodes $X$ and $Y$"}. Using $\repres$ we could relay on the underlying $\csa$ to efficiently locate the ranges corresponding to $X$ and $Y$ and apply $\Psi$ to extract the original trips to check if they contain $Y$.  Other queries such as {\em ``Count how many of the trips from X to Y passed through Z"} could be solved similarly by initially locating the range corresponding to $Y\$Z$ and then applying $\Psi$ to check if those trips between $X$ and $Y$ contain $Z$.

%%%%%%%%%%%%%%%%%%%%%%%%%%%%%%%%%%%%%%%%%%%%%%%%%%%%%%%%%%%%%%%%%%%%%%%%%%%%%%%%
\section{Conclusions and future work} \label{sec:conclusions}
%%%%%%%%%%%%%%%%%%%%%%%%%%%%%%%%%%%%%%%%%%%%%%%%%%%%%%%%%%%%%%%%%%%%%%%%%%%%%%%%

With the installation of better user-tracking mechanisms in public transportation networks, or the 
fact that a simple app installed in a mobile phone permits us to track user movements, the problem 
of storing user trips to finally support network analysis operations has been gaining increasing 
interest in multiple scenarios. For example, we could consider a network management administration, a 
taxi company, services like Uber, Cabify, Car2go, or simply end-user applications.

With enough data of vehicle trips from a
significant amount of drivers over the network composed of the streets
in a city, it would be possible to infer traffic rules by examining
turns that nobody takes, their usual driving speed across the
network, congestion points at a given time, and other useful information. 
Also, a taxi company (or similar services) could benefit from knowing the
city areas where it is more probable that a user would  start a trip, the average time
to go from one area to  another, etc. This also applies for the administrators of
public transportation networks including buses, trains, subway, etc.

We have presented $\repres$ and showed that it is a powerful tool to represent user trips.
Actually, we have used $\repres$ to handle user trips from two different scenarios: the network of subway 
and local trains from Madrid, and taxi trips from Porto. 
$\repres$ uses compact data structures to store both the nodes traversed (spatial component) by an user during a 
trip and the corresponding timestamps (temporal component). This permits us not only to reduce the amount of data to store
but also to efficiently perform spatial, temporal, and spatio-temporal queries  that can help us to analyze the actual usage of the network.

In particular, we used the well-known $\csa$ to represent the spatial component of the trips. For Madrid dataset, 
the size of $\csa$ is around $20$-$40$\% the size of the source data. Porto dataset is still more compressible and
$\csa$ requires only around $13$-$24$\% the space of the original data. This structure is enough to solve typical 
spatial queries within microseconds and \Stk\ queries in milliseconds.
For the temporal component, we used two $\wt$-based structures. We adapted the existing balanced $\wm$ and we created 
a {\em Hu-Tucker-shaped $\wt$} ($\wtht$) that permits to exploit a biased distribution of times to gain compression. 
These structures obtained only a moderate improvement in compression with respect to a plain representation of times 
(compression ratio from $70$ to $105$\%), but they provided indexed access to the temporal data, and consequently 
allowed us to support temporal queries very efficiently.
Finally, we have also shown that the overall $\repres$, including both $\csa$ and either $\wtht$ or $\wm$, permits also  
to efficiently  solve spatio-temporal queries (within microseconds); that is, spatial queries constrained to a time period.
The overall compression obtained by $\repres$ is around $55$-$75$\% in Madrid dataset and around $43$-$54$\% in Porto dataset.

We have presented $\repres$ as a proof of concept development, and we have shown how to solve different types of queries. Yet, based on the underlying data-structures, $\repres$ is flexible enough to allow us to increase its functionality.
 %As future work, we consider that it is interesting to exploit the underlying network topology to obtain a more compact representation of the trips. 
As future work, we are interested in exploiting the underlying network topology to obtain a more compact representation of the trips in $\repres$. In this promising line \cite{Han:2017:CCF:3086510.3015457, Koide:18} we are working on a succinct  representation for the context of public transportation networks.  Also, we want to explore ways to improve the compression of the temporal component of $\repres$. We consider that  an inverted-index based representation can be promising.

%\marginpar{\tiny \OJOFARI{Daniil, aqui podemos quizas dar una pincelada (en tres lineas) de la opcion que habiamos barajado en diciembre para comprimir los 
%	tiempos y que finalmente Nieves descarto'.}}

%It
%is flexible enough to allow new adaptations and functionality improvements we plan
%to do as future work, such as the analysis of line changes in
%switching stops (that would require storing the network  topology)
%or providing compression for the time index.

%As better tracking mechanisms will be installed, the problem of
%storing and querying trips to support network analysis  will gain
%interest for network management administrations and even end-user
%applications. For instance, with enough data of vehicle trips from a
%significant amount of drivers over the network formed by the streets
%of a city, it would be possible to infer traffic rules by examining
%turns that nobody takes, their usual driving speed across the
%network, and other useful information.

%We showed that $\repres$ is a powerful structure to represent
%user trips. Using compact data structures to represent trips over a
%transportation network allows us not only to keep a much larger amount
%of data in main memory (compression ratio is around 30\%), but also to efficiently perform spatial and
%spatio-temporal queries oriented to understand the real usage of the
%network.

\vspace*{\floatsep}
%We have presented $\repres$ as a proof of concept development. It
%is flexible enough to allow new adaptations and functionality improvements we plan
%to do as future work, such as the analysis of line changes in
%switching stops (that would require storing the network  topology)
%or providing compression for the time index.

% We assigned a range of IDs ($[1,\delta_r]$) to regular stops and a
% different one for switching stops ($[\delta_r+1,\delta]$) to provide
% a simple mechanism to know how many times a stop was used to change lines.
% For example, let us assume two lines $L_x$
% and $L_y$ whose stops are respectively $\langle ..., 25, \underline{3662}, 26,
% ... \rangle$ and $\langle ..., 116, \underline{3662}, 117,.. \rangle$, and where
% stop $3662$ is a switching stop among them that does not appear in any
% other line. To obtain the number of times stop $3662$ was used to
% change line we would  look for patterns: $\langle 25, 3662, ?x
% \rangle$, $\langle 116, 3662, ?y \rangle$, $\langle 25+1, 3662,
% ?\overline{x} \rangle$, $\langle 116+1, 3662, ?\overline{y} \rangle$
% checking that: $x \neq 26$, $y \neq 117$, $\overline{x} \neq 25$,
% and $\overline{y} \neq 116$. Obviously this is a simplification but
% a real implementation will need a proper representation of the lines
% and the network topology.

%%%%%%%%%%%%%%%%%%%%%%%%%%%%%%%%%%%%%%%%%%%%%%%%%%%%%%%%%%%%%%%%%%%%%%%%%%%%%%%%%%
%%%%%%%%%%%%%%%%%%%%%%%%%%%%%%%%%%%%%%%%%%%%%%%%%%%%%%%%%%%%%%%%%%%%%%%%%%%%%%%%%%%%%%%

\bibliographystyle{elsarticle-num}
%\bibliographystyle{plain}
%\bibliography{refs,abbrev,trajectories,secoD,BIB}

\section*{References}
\bibliography{paper}

\end{document}